\documentstyle[10pt,epsf,epsfig,hangcaption,xspace,amssymb,amsfonts,amsmath,amsthm,cite,
dp_delphititle,lineno]{dp_delphi}
\makeindex
\def\DpPaperGroup{EP}
\def\DpPaperRef{2003-092}
\def\DpDate{23 September 2003}
\def\DpAuthors{DELPHI Collaboration}
\def\DpSubmit{(Accepted by Eur. Phys. J. C)}
\def\DpTitle{{ 
Search for supersymmetric particles 
assuming  \boldmath$R$--parity non-conservation 
     in e$^+$e$^-$ collisions at $\sqrt{s}$~=~192~to~208~GeV }}
\def\DpComment{ }
\def\DpEMail{ }

\newcommand{\dm}      {$\Delta$M}

\newcommand{\XO}      {$\widetilde{\chi}^0$}
\newcommand{\XOI}     {$\widetilde{\chi}_1^0$}

\newcommand{\XOII}    {$\widetilde{\chi}_2^0$}

\newcommand{\XOi}     {$\widetilde{\chi}_i^0$}
\newcommand{\XOj}     {$\widetilde{\chi}_j^0$}

\newcommand{\XPI}{$\widetilde{\chi}_1^+$}

\newcommand{\XMI}{$\widetilde{\chi}_1^-$}

\newcommand{\XPM}{$\widetilde{\chi}^{\pm}$}

\newcommand{\squ}{\mbox{$\widetilde{\rm u}$}}
\newcommand{\sqd}{\mbox{$\widetilde{\rm d}$}}
\newcommand{\SQ}{$\widetilde{\rm q}$}
\newcommand{\SQB}{${\widetilde{\bar {\rm q}}}$}

\newcommand{\SLEPP}{$\widetilde{\ell}^+$}
\newcommand{\SLEPM}{$\widetilde{\ell}^-$}
\newcommand{\slep}     {\mbox{$ \tilde{\ell}                       $}}

\newcommand{\lum}{$\cal L$}

\newcommand{\tanb}{tan$\beta$}

\newcommand{\Labb}{$\lambda_{122}$}
\newcommand{\Lacc}{$\lambda_{133}$}
\newcommand{\Lijk}{$\lambda_{ijk}$}
\newcommand{\Lppijk}{$\lambda''_{ijk}$}

\newcommand{\LiLjEk}    {$ {\rm{L}}_i {\rm{L}}_j {\rm\bar{E}}_k $}
\newcommand{\LiQjDk}    {$ {\rm{L}}_i {\rm{Q}}_j {\rm\bar{D}}_k $}
\newcommand{\UiDjDk}    {$ {\rm\bar{U}}_i {\rm\bar{D}}_j {\rm\bar{D}}_k $}

\newcommand{\LLE}      {$ {\rm {LL \bar E}}$}
\newcommand{\UDD}{$\rm {\bar{U} \bar{D} \bar{D}}$}
\newcommand{\BLLE}      {\boldmath$ {\rm {LL \bar E}}$}
\newcommand{\BUDD}{\boldmath$\rm {\bar{U} \bar{D} \bar{D}}$}

\newcommand{\Emiss}{$E$\hspace{-0.6em}{{/}}\ }
\newcommand{\Pt}{\mbox{p$_{\rm t}$}}

\newcommand{\Ra}{$\rightarrow$\ }

\newcommand{\Zn}      {\mbox{$ {\mathrm Z}                               $}}

\newcommand{\WW}      {\mbox{$ {\mathrm W}^+{\mathrm W}^-                  $}}
\newcommand{\ZZ}      {\mbox{$ {\mathrm Z}{\mathrm Z}                 $}}

\newcommand{\Zg}      {\mbox{$ {\mathrm f \bar{\mathrm f}}    \gamma       $}}
\newcommand{\ecms}    {\mbox{$ \sqrt{s}                                    $}}

\newcommand{\GeV}     {\mbox{$ {\mathrm{GeV}}                              $}}
\newcommand{\GeVc}    {\mbox{$ {\mathrm{GeV}}/c                            $}}
\newcommand{\GeVcc}   {\mbox{$ {\mathrm{GeV}}/c^2                          $}}

\newcommand{\qqg}     {\mbox{$ {\mathrm q}\bar{\mathrm q}\gamma            $}}

\newcommand{\sel}     {\mbox{$ \tilde{{\rm e}}                       $}}

\newcommand{\smu}     {\mbox{$ \tilde{\mu}                              $}}

\newcommand{\stau}    {\mbox{$ \tilde{\tau}                             $}}

\newcommand{\snu}      {\mbox{$ \tilde{\nu}                                $}}
\newcommand{\snue}     {\mbox{$ \widetilde{\nu}_{{\rm e}}                            $}}
\newcommand{\snum}     {\mbox{$ \widetilde{\nu}_{\mu}                          $}}
\newcommand{\snut}     {\mbox{$ \widetilde{\nu}_{\tau}                          $}}
\newcommand{\asnu}      {\mbox{$ \tilde{\bar \nu}                                $}}
\newcommand{\asnue}     {\mbox{$ \widetilde{\bar \nu}_{{\rm e}}                            $}}
\newcommand{\asnum}     {\mbox{$ \widetilde{\bar \nu}_{\mu}                          $}}
\newcommand{\asnut}     {\mbox{$ \widetilde{\bar \nu}_{\tau}                          $}}

\newcommand{\sfe}       {\mbox{$ \tilde{\mathrm f}                              $}}
\newcommand{\sfea}      {\mbox{$ \tilde{\mathrm f}_1                            $}}
\newcommand{\sfeb}      {\mbox{$ \tilde{\mathrm f}_2                            $}}
\newcommand{\sfeL}      {\mbox{$ \tilde{\mathrm f}_{L}                  $}}
\newcommand{\sfeR}      {\mbox{$ \tilde{\mathrm f}_{R}                  $}}
\newcommand{\phimixf}   {\mbox{$\Phi_{\rm mix}                         $}}

\newcommand{\achiap}    {\mbox{$ \tilde{\chi}^{+}_{1}                    $}}

\newcommand{\achia}    {\mbox{$ \tilde{\chi}^{0}_{1}                        $}}

\newcommand{\mydeg}   {\mbox{$ ^\circ                                      $}}

\newcommand {\rp } {${R}_{p}$}

\newcommand {\mtwo} {M$_{2}$}

\newcommand{\stp}     {\mbox{$ \tilde{\rm t} $}}
\newcommand{\stpb}     {\mbox{$ \tilde{\bar{\rm t}}                                $}}
\newcommand{\sbt}     {\mbox{$ \tilde{\rm b} $}}

\def    \missEt      {\ifmmode{/\mkern-11mu E_t}\else{${/\mkern-11mu E_t}$}\fi}
\def    \missE          {\ifmmode{/\mkern-11mu E}\else{${/\mkern-11mu E}$}\fi}

\def\strutl{\protect\rule{0ex}{1.4em}}

\newenvironment{malist}
         {\begin{list}{\tiny{$\bullet$}}
         {\setlength{\parsep}{0pt}
         \addtolength{\leftmargin}{0mm}
         \setlength{\topsep}{0pt}
         \setlength{\itemsep}{0pt}}}{\end{list}}

\newcommand{\stitre}[1]{\noindent {\underline {#1}}\par \vskip 0.2cm}

\def\ap#1#2#3   {{\em Ann. Phys. (NY)} {\bf#1} (#2) #3}
\def\apj#1#2#3  {{\em Astrophys. J.} {\bf#1} (#2) #3}
\def\apjl#1#2#3 {{\em Astrophys. J. Lett.} {\bf#1} (#2) #3}
\def\app#1#2#3  {{\em Acta. Phys. Pol.} {\bf#1} (#2) #3}
\def\ar#1#2#3   {{\em Ann. Rev. Nucl. Part. Sci.} {\bf#1} (#2) #3}
\def\cpc#1#2#3  {{\em Computer Phys. Comm.} {\bf#1} (#2) #3}
\def\err#1#2#3  {{\it Erratum} {\bf#1} (#2) #3}
\def\epj#1#2#3  {\mbox{{\em Eur. Phys. J.~}{\bf#1} (#2) #3}}
\def\ib#1#2#3   {{\it ibid.} {\bf#1} (#2) #3}
\def\jmp#1#2#3  {{\em J. Math. Phys.} {\bf#1} (#2) #3}
\def\ijmp#1#2#3 {{\em Int. J. Mod. Phys.} {\bf#1} (#2) #3}
\def\jetp#1#2#3 {{\em JETP Lett.} {\bf#1} (#2) #3}
\def\jpg#1#2#3  {{\em J. Phys. G.} {\bf#1} (#2) #3}
\def\mpl#1#2#3  {{\em Mod. Phys. Lett.} {\bf#1} (#2) #3}
\def\nat#1#2#3  {{\em Nature (London)} {\bf#1} (#2) #3}
\def\nc#1#2#3   {{\em Nuovo Cim.} {\bf#1} (#2) #3}
\def\nim#1#2#3  {{\em Nucl. Instr. Meth.} {\bf#1} (#2) #3}
\def\np#1#2#3   {{\em Nucl. Phys.} {\bf#1} (#2) #3}
\def\npsup#1#2#3#4   {{\em Nucl. Phys.} {\bf #1} {\em (Proc. Suppl.)} 
{\bf#2} (#3) #4.}
\def\pcps#1#2#3 {{\em Proc. Cam. Phil. Soc.} {\bf#1} (#2) #3}
\def\pl#1#2#3   {{\em Phys. Lett.} {\bf#1} (#2) #3}
\def\prep#1#2#3 {{\em Phys. Rep.} {\bf#1} (#2) #3}
\def\prev#1#2#3 {{\em Phys. Rev.} {\bf#1} (#2) #3}
\def\prl#1#2#3  {{\em Phys. Rev. Lett.} {\bf#1} (#2) #3}
\def\prs#1#2#3  {{\em Proc. Roy. Soc.} {\bf#1} (#2) #3}
\def\ptp#1#2#3  {{\em Prog. Th. Phys.} {\bf#1} (#2) #3}
\def\ps#1#2#3   {{\em Physica Scripta} {\bf#1} (#2) #3}
\def\rmp#1#2#3  {{\em Rev. Mod. Phys.} {\bf#1} (#2) #3}
\def\rpp#1#2#3  {{\em Rep. Prog. Phys.} {\bf#1} (#2) #3}
\def\sjnp#1#2#3 {{\em Sov. J. Nucl. Phys.} {\bf#1} (#2) #3}
\def\spj#1#2#3  {{\em Sov. Phys. JEPT} {\bf#1} (#2) #3}
\def\spu#1#2#3  {{\em Sov. Phys.-Usp.} {\bf#1} (#2) #3}
\def\zp#1#2#3   {{\em Zeit. Phys.} {\bf#1} (#2) #3}

\def\delnote#1#2#3#4 {{DELPHI} {#1}-{#2}~{\sc #3}~#4}
\def\opalnote#1#2 {{OPAL} {\it#1} {#2}}
\def\alnote#1#2#3#4 {{ALEPH} {#1}-{#2}~{\sc #3}~#4}

\newcommand{\XPMk}{$\widetilde{\chi}_k^\pm$}

\newcommand{\slepr}    {\mbox{$ \tilde{\ell}_R            $}}

\newcommand{\Lbcc}{$\lambda_{233}$}

\newcommand{\ffgamma}{\mbox{$\rm f \bar{\rm f}(\gamma)$}}

\begin{document}
\makeatletter
\newcount\@tempcntc
\def\@citex[#1]#2{\if@filesw\immediate\write\@auxout{\string\citation{#2}}\fi
  \@tempcnta\z@\@tempcntb\m@ne\def\@citea{}\@cite{\@for\@citeb:=#2\do
    {\@ifundefined
       {b@\@citeb}{\@citeo\@tempcntb\m@ne\@citea\def\@citea{,}{\bf ?}\@warning
       {Citation `\@citeb' on page \thepage \space undefined}}%
    {\setbox\z@\hbox{\global\@tempcntc0\csname b@\@citeb\endcsname\relax}%
     \ifnum\@tempcntc=\z@ \@citeo\@tempcntb\m@ne
       \@citea\def\@citea{,}\hbox{\csname b@\@citeb\endcsname}%
     \else
      \advance\@tempcntb\@ne
      \ifnum\@tempcntb=\@tempcntc
      \else\advance\@tempcntb\m@ne\@citeo
      \@tempcnta\@tempcntc\@tempcntb\@tempcntc\fi\fi}}\@citeo}{#1}}
\def\@citeo{\ifnum\@tempcnta>\@tempcntb\else\@citea\def\@citea{,}%
  \ifnum\@tempcnta=\@tempcntb\the\@tempcnta\else
   {\advance\@tempcnta\@ne\ifnum\@tempcnta=\@tempcntb \else \def\@citea{--}\fi
    \advance\@tempcnta\m@ne\the\@tempcnta\@citea\the\@tempcntb}\fi\fi}
 
\makeatother
\begin{titlepage}
\pagenumbering{roman}
\CERNpreprint{\DpPaperGroup}{\DpPaperRef} 
\date{{\small\DpDate}} 
\title{\DpTitle} 
\address{\DpAuthors} 
\begin{shortabs} 
\noindent
\noindent
Searches for pair-production of supersymmetric particles under the assumption
of non-conservation of \mbox{$R$--parity} with a dominant \LLE~or
\UDD~term have been performed using the
data collected by the DELPHI experiment at LEP in e$^{+}$e$^{-}$ collisions
at centre-of-mass energies from 192 up to 208 GeV. No excess of data
above Standard Model expectations was observed.  The results were used to
constrain the MSSM parameter space and to derive limits on the masses of
supersymmetric particles.
\end{shortabs}
\vfill
\begin{center}
\DpSubmit \ \\ 
\DpComment \ \\
\DpEMail \ \\
\end{center}
\vfill
\clearpage
\headsep 10.0pt
\addtolength{\textheight}{10mm}
\addtolength{\footskip}{-5mm}
\begingroup
%
\newcommand{\DpName}[2]{\hbox{#1$^{\ref{#2}}$},\hfill}
\newcommand{\DpNameTwo}[3]{\hbox{#1$^{\ref{#2},\ref{#3}}$},\hfill}
\newcommand{\DpNameThree}[4]{\hbox{#1$^{\ref{#2},\ref{#3},\ref{#4}}$},\hfill}
\newskip\Bigfill \Bigfill = 0pt plus 1000fill
\newcommand{\DpNameLast}[2]{\hbox{#1$^{\ref{#2}}$}\hspace{\Bigfill}}
%
\footnotesize
\noindent
\DpName{J.Abdallah}{LPNHE}
\DpName{P.Abreu}{LIP}
\DpName{W.Adam}{VIENNA}
\DpName{P.Adzic}{DEMOKRITOS}
\DpName{T.Albrecht}{KARLSRUHE}
\DpName{T.Alderweireld}{AIM}
\DpName{R.Alemany-Fernandez}{CERN}
\DpName{T.Allmendinger}{KARLSRUHE}
\DpName{P.P.Allport}{LIVERPOOL}
\DpName{U.Amaldi}{MILANO2}
% ...
\normalsize
\endgroup
\titlefoot{Department of Physics and Astronomy, Iowa State
     University, Ames IA 50011-3160, USA
    \label{AMES}}
\titlefoot{Physics Department, Universiteit Antwerpen,
     Universiteitsplein 1, B-2610 Antwerpen, Belgium \\
     \indent~~and IIHE, ULB-VUB,
     Pleinlaan 2, B-1050 Brussels, Belgium \\
     \indent~~and Facult\'e des Sciences,
     Univ. de l'Etat Mons, Av. Maistriau 19, B-7000 Mons, Belgium
    \label{AIM}}
\titlefoot{Physics Laboratory, University of Athens, Solonos Str.
     104, GR-10680 Athens, Greece
    \label{ATHENS}}
\titlefoot{Department of Physics, University of Bergen,
     All\'egaten 55, NO-5007 Bergen, Norway
    \label{BERGEN}}
\titlefoot{Dipartimento di Fisica, Universit\`a di Bologna and INFN,
     Via Irnerio 46, IT-40126 Bologna, Italy
    \label{BOLOGNA}}
\titlefoot{Centro Brasileiro de Pesquisas F\'{\i}sicas, rua Xavier Sigaud 150,
     BR-22290 Rio de Janeiro, Brazil \\
     \indent~~and Depto. de F\'{\i}sica, Pont. Univ. Cat\'olica,
     C.P. 38071 BR-22453 Rio de Janeiro, Brazil \\
     \indent~~and Inst. de F\'{\i}sica, Univ. Estadual do Rio de Janeiro,
     rua S\~{a}o Francisco Xavier 524, Rio de Janeiro, Brazil
    \label{BRASIL}}
\titlefoot{Coll\`ege de France, Lab. de Physique Corpusculaire, IN2P3-CNRS,
     FR-75231 Paris Cedex 05, France
    \label{CDF}}
\titlefoot{CERN, CH-1211 Geneva 23, Switzerland
    \label{CERN}}
\titlefoot{Institut de Recherches Subatomiques, IN2P3 - CNRS/ULP - BP20,
     FR-67037 Strasbourg Cedex, France
    \label{CRN}}
\titlefoot{Now at DESY-Zeuthen, Platanenallee 6, D-15735 Zeuthen, Germany
    \label{DESY}}
\titlefoot{Institute of Nuclear Physics, N.C.S.R. Demokritos,
     P.O. Box 60228, GR-15310 Athens, Greece
    \label{DEMOKRITOS}}
\titlefoot{FZU, Inst. of Phys. of the C.A.S. High Energy Physics Division,
     Na Slovance 2, CZ-180 40, Praha 8, Czech Republic
    \label{FZU}}
\titlefoot{Dipartimento di Fisica, Universit\`a di Genova and INFN,
     Via Dodecaneso 33, IT-16146 Genova, Italy
    \label{GENOVA}}
\titlefoot{Institut des Sciences Nucl\'eaires, IN2P3-CNRS, Universit\'e
     de Grenoble 1, FR-38026 Grenoble Cedex, France
    \label{GRENOBLE}}
\titlefoot{Helsinki Institute of Physics, P.O. Box 64,
     FIN-00014 University of Helsinki, Finland
    \label{HELSINKI}}
\titlefoot{Joint Institute for Nuclear Research, Dubna, Head Post
     Office, P.O. Box 79, RU-101 000 Moscow, Russian Federation
    \label{JINR}}
\titlefoot{Institut f\"ur Experimentelle Kernphysik,
     Universit\"at Karlsruhe, Postfach 6980, DE-76128 Karlsruhe,
     Germany
    \label{KARLSRUHE}}
\titlefoot{Institute of Nuclear Physics PAN,Ul. Radzikowskiego 152,
     PL-31142 Krakow, Poland
    \label{KRAKOW1}}
\titlefoot{Faculty of Physics and Nuclear Techniques, University of Mining
     and Metallurgy, PL-30055 Krakow, Poland
    \label{KRAKOW2}}
\titlefoot{Universit\'e de Paris-Sud, Lab. de l'Acc\'el\'erateur
     Lin\'eaire, IN2P3-CNRS, B\^{a}t. 200, FR-91405 Orsay Cedex, France
    \label{LAL}}
\titlefoot{School of Physics and Chemistry, University of Lancaster,
     Lancaster LA1 4YB, UK
    \label{LANCASTER}}
\titlefoot{LIP, IST, FCUL - Av. Elias Garcia, 14-$1^{o}$,
     PT-1000 Lisboa Codex, Portugal
    \label{LIP}}
\titlefoot{Department of Physics, University of Liverpool, P.O.
     Box 147, Liverpool L69 3BX, UK
    \label{LIVERPOOL}}
\titlefoot{Dept. of Physics and Astronomy, Kelvin Building,
     University of Glasgow, Glasgow G12 8QQ
    \label{GLASGOW}}
\titlefoot{LPNHE, IN2P3-CNRS, Univ.~Paris VI et VII, Tour 33 (RdC),
     4 place Jussieu, FR-75252 Paris Cedex 05, France
    \label{LPNHE}}
\titlefoot{Department of Physics, University of Lund,
     S\"olvegatan 14, SE-223 63 Lund, Sweden
    \label{LUND}}
\titlefoot{Universit\'e Claude Bernard de Lyon, IPNL, IN2P3-CNRS,
     FR-69622 Villeurbanne Cedex, France
    \label{LYON}}
\titlefoot{Dipartimento di Fisica, Universit\`a di Milano and INFN-MILANO,
     Via Celoria 16, IT-20133 Milan, Italy
    \label{MILANO}}
\titlefoot{Dipartimento di Fisica, Univ. di Milano-Bicocca and
     INFN-MILANO, Piazza della Scienza 2, IT-20126 Milan, Italy
    \label{MILANO2}}
\titlefoot{IPNP of MFF, Charles Univ., Areal MFF,
     V Holesovickach 2, CZ-180 00, Praha 8, Czech Republic
    \label{NC}}
\titlefoot{NIKHEF, Postbus 41882, NL-1009 DB
     Amsterdam, The Netherlands
    \label{NIKHEF}}
\titlefoot{National Technical University, Physics Department,
     Zografou Campus, GR-15773 Athens, Greece
    \label{NTU-ATHENS}}
\titlefoot{Physics Department, University of Oslo, Blindern,
     NO-0316 Oslo, Norway
    \label{OSLO}}
\titlefoot{Dpto. Fisica, Univ. Oviedo, Avda. Calvo Sotelo
     s/n, ES-33007 Oviedo, Spain
    \label{OVIEDO}}
\titlefoot{Department of Physics, University of Oxford,
     Keble Road, Oxford OX1 3RH, UK
    \label{OXFORD}}
\titlefoot{Dipartimento di Fisica, Universit\`a di Padova and
     INFN, Via Marzolo 8, IT-35131 Padua, Italy
    \label{PADOVA}}
\titlefoot{Rutherford Appleton Laboratory, Chilton, Didcot
     OX11 OQX, UK
    \label{RAL}}
\titlefoot{Dipartimento di Fisica, Universit\`a di Roma II and
     INFN, Tor Vergata, IT-00173 Rome, Italy
    \label{ROMA2}}
\titlefoot{Dipartimento di Fisica, Universit\`a di Roma III and
     INFN, Via della Vasca Navale 84, IT-00146 Rome, Italy
    \label{ROMA3}}
\titlefoot{DAPNIA/Service de Physique des Particules,
     CEA-Saclay, FR-91191 Gif-sur-Yvette Cedex, France
    \label{SACLAY}}
\titlefoot{Instituto de Fisica de Cantabria (CSIC-UC), Avda.
     los Castros s/n, ES-39006 Santander, Spain
    \label{SANTANDER}}
\titlefoot{Inst. for High Energy Physics, Serpukov
     P.O. Box 35, Protvino, (Moscow Region), Russian Federation
    \label{SERPUKHOV}}
\titlefoot{J. Stefan Institute, Jamova 39, SI-1000 Ljubljana, Slovenia
     and Laboratory for Astroparticle Physics,\\
     \indent~~Nova Gorica Polytechnic, Kostanjeviska 16a, SI-5000 Nova Gorica, Slovenia, \\
     \indent~~and Department of Physics, University of Ljubljana,
     SI-1000 Ljubljana, Slovenia
    \label{SLOVENIJA}}
\titlefoot{Fysikum, Stockholm University,
     Box 6730, SE-113 85 Stockholm, Sweden
    \label{STOCKHOLM}}
\titlefoot{Dipartimento di Fisica Sperimentale, Universit\`a di
     Torino and INFN, Via P. Giuria 1, IT-10125 Turin, Italy
    \label{TORINO}}
\titlefoot{INFN,Sezione di Torino, and Dipartimento di Fisica Teorica,
     Universit\`a di Torino, Via P. Giuria 1,\\
     \indent~~IT-10125 Turin, Italy
    \label{TORINOTH}}
\titlefoot{Dipartimento di Fisica, Universit\`a di Trieste and
     INFN, Via A. Valerio 2, IT-34127 Trieste, Italy \\
     \indent~~and Istituto di Fisica, Universit\`a di Udine,
     IT-33100 Udine, Italy
    \label{TU}}
\titlefoot{Univ. Federal do Rio de Janeiro, C.P. 68528
     Cidade Univ., Ilha do Fund\~ao
     BR-21945-970 Rio de Janeiro, Brazil
    \label{UFRJ}}
\titlefoot{Department of Radiation Sciences, University of
     Uppsala, P.O. Box 535, SE-751 21 Uppsala, Sweden
    \label{UPPSALA}}
\titlefoot{IFIC, Valencia-CSIC, and D.F.A.M.N., U. de Valencia,
     Avda. Dr. Moliner 50, ES-46100 Burjassot (Valencia), Spain
    \label{VALENCIA}}
\titlefoot{Institut f\"ur Hochenergiephysik, \"Osterr. Akad.
     d. Wissensch., Nikolsdorfergasse 18, AT-1050 Vienna, Austria
    \label{VIENNA}}
\titlefoot{Inst. Nuclear Studies and University of Warsaw, Ul.
     Hoza 69, PL-00681 Warsaw, Poland
    \label{WARSZAWA}}
\titlefoot{Fachbereich Physik, University of Wuppertal, Postfach
     100 127, DE-42097 Wuppertal, Germany
    \label{WUPPERTAL}}
\addtolength{\textheight}{-10mm}
\addtolength{\footskip}{5mm}
\clearpage
\headsep 30.0pt
\end{titlepage}
%
\pagenumbering{arabic} 
\setcounter{footnote}{0} %
\large
\section{Introduction}

The \mbox{$R$--parity} (\rp) symmetry plays an essential role in the construction
of supersymmetric theories, such as the  Minimal 
Supersymmetric extension of the Standard Model (MSSM)~\cite{mssm}.
The conservation  of {\rp} is closely
related to the conservation of lepton ($L$) and baryon ($B$) numbers
and the multiplicative quantum number associated to the {\rp}
symmetry is defined by $R_p=(-1)^{3B+L+2S}$ for a particle with
spin~$S$~\cite{fayet}. Standard model particles have even {\rp},
whereas the corresponding superpartners have odd {\rp.}
The conservation of {\rp} guarantees that 
the spin--0 sfermions cannot
be directly exchanged between standard fermions. It also implies 
that the sparticles (\mbox{$R_p=-1$}) can only be produced in pairs, 
and that the decay of a sparticle leads to another sparticle,
or an odd number of them. Therefore, it ensures the stability of
the Lightest Supersymmetric Particle (LSP). 
In the MSSM, the  conservation of {\rp} is assumed: this is phenomenologically
justified by proton decay constraints, and by the fact that a neutral
LSP could be a good dark matter candidate. 

From a theoretical point of view, the conservation of
{\rp} is not mandatory in supersymmetric extensions
of the Standard Model (SM). Nevertheless,  to be in agreement with the
present experimental limit on proton lifetime, 
{\rp} violation can be introduced in MSSM either
via the non-conservation of $L$ or the non-conservation of $B$.
One of the major consequences of the non-conservation
of {\rp} is the allowed decay 
of the LSP into fermions;
this modifies the signatures of supersymmetric particle
production compared to the expected signatures in the case
of {\rp} conservation.

In this paper, searches for  pair-produced supersymmetric particles 
in the hypothesis of {\rp} violation 
via one dominant sparticle-particle coupling
are presented. The data recorded in 1999 and 2000 
by the DELPHI experiment have been analyzed, and no signal
of \rp-violating decays
was found in any of the channels. 
Previous results published by DELPHI on this subject
can be found in references~\cite{lle189,udd189}.
Similar searches performed by the other three LEP experiments
have also shown no evidence for \rp-violating effects~\cite{lesautres}.

The  paper is organized as follows. Section 2 is dedicated to the
{\rp} violation phenomenology considered in the present search. 
The data samples  and simulated
sets  are presented in Section 3. Section 4 is devoted to the
description of the analyses, and
in Section 5 the search results are given and interpreted in order to 
constrain the mass spectrum of
SUSY particles. A brief summary is given in the last section.        

\section{{\rp} non-conservation framework \label{rpvframe}}

In the presence of {\rp} violation the superpotential~\cite{weinberg} contains three trilinear terms, 
two violating $L$ conservation, and one violating $B$ conservation.  
We consider here the  
$\lambda_{ijk}$\LiLjEk\ (non-conservation of~$L$) 
and $\lambda''_{ijk}$\UiDjDk\ (non-conservation of $B$) 
terms~\footnote{$i, j, k$ are generation indices, 
$L$ denotes the lepton  doublet superfields, 
 ${\bar E}$ (${\bar U}$, ${\bar D}$) denote the lepton (up and down quark)
 singlet superfields, 
${\lambda}_{ijk}$ and ${\lambda}^{\prime \prime}_{ijk}$ are  Yukawa
couplings.}, which couple the sleptons to
the leptons  and the squarks to the quarks,  respectively.
Since \Lijk~=~-- $\lambda_{jik}$ and
$\lambda''_{ijk}$~=~--$\lambda''_{ikj}$,
due to SU(2) and SU(3) symmetries, 
there are only 9 \Lijk\ and only 9 $\lambda''_{ijk}$ free couplings.
In the present work, it is assumed that only one $\lambda_{ijk}$ or 
$\lambda''_{ijk}$  is dominant at a time. 
In the following, searches  assuming \rp-violation via one
dominant  $\lambda_{ijk}$\LiLjEk\ term are referred as ``\LLE'', and those
via one 
 $\lambda''_{ijk}$\UiDjDk\ term as ``\UDD''.   
Searches assuming  \rp-violation  via one
 $\lambda'_{ijk}$\LiQjDk\ term (non-conservation of $L$) 
were not performed in DELPHI for data collected in 1999 and 2000.

In the pair-production of supersymmetric particles studied here, 
{\rp} is not conserved in the decay 
of the sparticles, but is conserved at the production vertex.
The production cross-sections behave as in the MSSM with
$R_p$ conservation (see section \ref{sec:pairprod}).

\subsection{\rp-violating decays of sparticles via \BLLE\ or \BUDD\ terms}

Two types of supersymmetric particle decays are considered: 
{\it direct decay} and {\it indirect decay}.

\subsubsection{Direct decays}

\begin{figure}[p]
\begin{center}
\epsfig{file=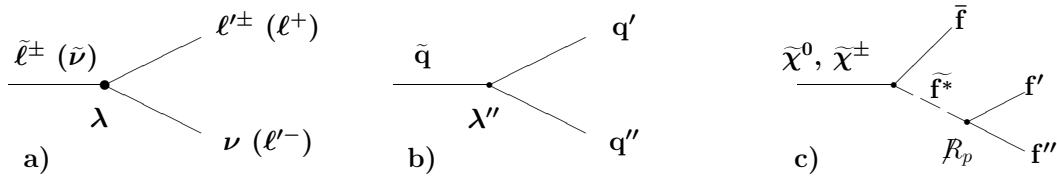,width=14cm} 
\caption{Diagrams of sparticle direct decays. 
a: slepton direct decay via \LLE\ term; 
b: squark direct decay via \UDD\ term;
c: neutralino/chargino direct decay via any \rp-violating trilinear term.}
\label{dirdec}
\end{center}
\end{figure}

{\rp} violation 
allows the direct decay of a sfermion into two conventional
fermions (\mbox{Fig.~\ref{dirdec}--a,~b}), or the direct decay of 
a neutralino or a chargino into a fermion and a virtual sfermion which then 
decays into two conventional fermions (\mbox{Fig.~\ref{dirdec}--c}).
A direct decay is the only possibility for the LSP.\\

\stitre{Decays through \LLE~terms}
\noindent Sleptons are coupled to leptons through the
 \Lijk \LiLjEk\ term.
In four-component Dirac notation, the $\rm{LL} \bar{\rm E}$
Yukawa interaction terms are\footnote{Here $\nu$ (\snu) 
refer to neutrino (sneutrino) fields, $\ell$ (\slep) refer to 
charged lepton (slepton) fields, $i$,$j$,$k$ are generation indices and the 
superfix $c$ refers to a charge-conjugate field.}~\cite{barger89}:
$$
\lambda_{ijk} \left( \snu_{iL}\bar{\ell}_{kR}\ell_{jL} +
  \slep_{jL}\bar{\ell}_{kR}\nu_{iL} + \slep_{kR}^*({\overline{ {\nu}}}_{iL})^c
  \slep_{jL} - i \leftrightarrow j \right) + {\rm{h.c.}}
$$
Considering the  above expression,
it can be deduced that the \rp-violating decay of a sfermion 
is possible only with specific indices $i,j,k$ of
the coupling which is considered to be dominant.
The possible sparticle decays with such a dominant \Lijk\
coupling are listed below.
\begin{malist}
\item The sneutrino direct decay gives two charged leptons: 
via \Lijk\, only the
\snu$_i$ and \snu$_j$ are allowed to decay directly:  
\mbox{\snu$_i$ \Ra $\ell^{\pm}_{jL}\ell^{\mp}_{kR}$} and 
\mbox{\snu$_j$ \Ra $\ell^{\pm}_{iL}\ell^{\mp}_{kR}$} 
respectively.

\item The charged slepton direct decay gives one neutrino and one
charged lepton (the lepton flavour may be different from the slepton one).
Among the supersymmetric partners of the right-handed leptons, only
the one belonging to the $k^{th}$ generation can decay directly:  
\mbox{$\slep^{-}_{kR}$ \Ra  $\nu_{iL} \ell^{-}_{jL}$ , $\ell^-_{iL}
\nu_{jL}$}.  For the supersymmetric partners of the left-handed
leptons, the allowed direct decays are: 
\mbox{$\slep^-_{iL}$ \Ra  $\bar{\nu}_{jL} \ell^-_{kR}$}~and  
\mbox{$\slep^-_{jL}$ \Ra  $\bar{\nu}_{iL} \ell^-_{kR}$}.  

\item The neutralino decays via a virtual slepton and a lepton, 
and subsequently gives three-lepton final states
(two charged leptons and one neutrino): \\
\mbox{\XO  \Ra $\ell^+_i \bar{\nu}_j \ell^-_k~,~\ell^-_i \nu_j \ell^+_k~,~ 
    \bar{\nu}_i \ell^+_j \ell^-_k~,~\nu_i \ell^-_j \ell^+_k $  }.
    
\item The chargino decays 
via a virtual slepton and gives either three charged leptons, or
two neutrinos and one charged lepton: \\
\mbox{\XPI  \Ra $\ell^+_i  \ell^+_j \ell^-_k~,~\ell^+_i \bar{\nu}_j  \nu_k~, 
    \bar{\nu}_i \ell^+_j \nu_k~,~\nu_i \nu_j \ell^+_k $}.\\
\end{malist}

\stitre{Decays through \UDD~terms} 
\noindent The squarks are coupled to the quarks through the
 \Lppijk \UiDjDk\  term.
The decays allowed via  this term 
can be inferred  by considering the Lagrangian 
for the trilinear Yukawa interactions written in expanded notation:
$$
\lambda_{ijk}'' \left( \overline{ (\rm u_i)^c} ~\overline{ (\rm d_j)^c }
    \sqd^*_k +  \overline{ (\rm u_i)^c }~ \sqd^*_j ~\overline{
    (\rm d_k)^c } + \squ^*_i ~\overline{ (\rm d_j)^c } ~\overline{
    (\rm d_k)^c } \right)  +  {\rm{h.c.}}
$$
From this Lagrangian, we can derive the following rules:
\begin{malist}
\item The direct decays of squarks into two
quarks are given by:
\mbox{$\squ_{i,R}$ \Ra  $\bar{\rm d}_{j,R} \bar{\rm d}_{k,R}$},  
\mbox{$\sqd_{j,R}$ \Ra  $\bar{\rm u}_{i,R} \bar{\rm d}_{k,R}$} and 
\mbox{$\sqd_{k,R}$ \Ra  $\bar{\rm u}_{i,R} \bar{\rm d}_{j,R}$.}  
\item The neutralino decays via a virtual 
squark and a quark and subsequently 
gives a three-quark final state:

\mbox{\XO  \Ra  $\bar{\rm u}_{j} \bar{\rm d}_{j}\bar{\rm d}_{k}~,~
                 {\rm u}_{j} {\rm d}_{j} {\rm d}_{k}$}
\item Chargino decay is similar to the neutralino one, and 
then gives also a three quarks final state:
\mbox{\XPI  \Ra  ${\rm u}_{j} {\rm d}_{j} {\rm u}_{k}~,~
                  {\rm u}_{i} {\rm u}_{j} {\rm d}_{k}~,~
           \bar{\rm d}_{i} \bar{\rm d}_{j}\bar{\rm d}_{k}$}
\end{malist}

\subsubsection{Indirect decays}

\begin{figure}[htb!]
\begin{center}
\epsfig{file=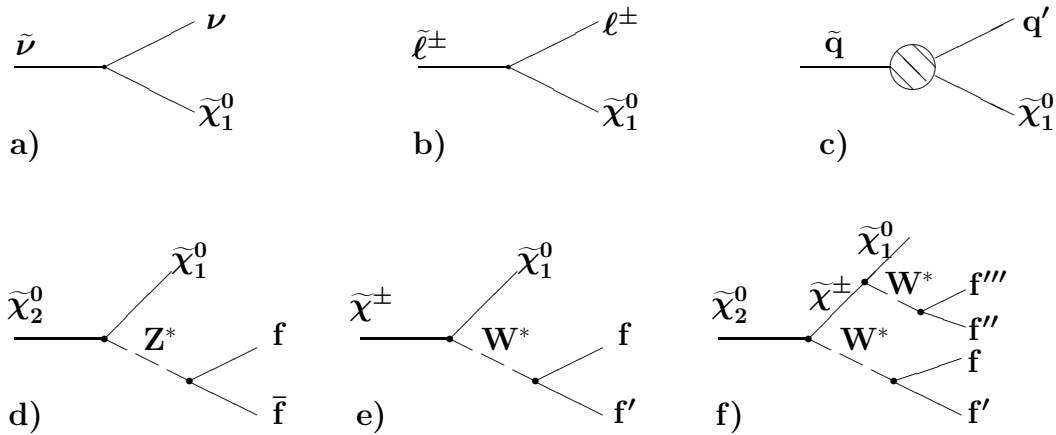,width=14cm} 
\caption{Diagrams of sparticle $R_p$-conserving decays. To get 
the whole chain of the sparticle indirect decay, the LSP (lightest
neutralino) has to undergo a direct \rp-violating decay. 
a, b: slepton  decays into a lepton and the lightest neutralino;
c: squark  decay into a quark and the lightest neutralino,
the hatched disk means decay beyond tree level for the stop case; 
d, e, f: examples of gaugino  decays.}
\label{inddec}
\end{center}
\end{figure}

Indirect decays are cascade decays through  
{\rp}-conserving
vertices to  on-shell supersymmetric particles, down to the 
lightest supersymmetric particle, which then
decays via one \LLE\ or \UDD\ term~(\mbox{Fig.~\ref{inddec}}).
A typical example is the
\rp-conserving decay 
\mbox{\achiap $\rightarrow$ \achia + W$^{*+}$} (see
\mbox{Fig.~\ref{inddec}--e}) and the subsequent decay of
\achia \ through the \rp-violating couplings.
The indirect decay mode usually dominates when there is enough phase space 
available in the decay between
``mother'' and ``daughter'' sparticles.
For example, when the difference 
of masses between these two sparticles is larger than 
5--10~\GeVcc. Regions of the parameter
space where there is a ``dynamic'' suppression of the  \rp-conserving modes
also exist. 
In this case, even if the sparticle is not the LSP, it decays 
through an \rp-violating  mode. For example, if 
the field component of the two lightest
neutralinos is mainly the photino, then the decay
\mbox{\XOII\ \Ra \XOI\ Z$^*$} is suppressed. 

The sfermion indirect decay studied here is the
decay through the lightest neutralino considered as the LSP 
(\mbox{\sfe \Ra f$^{\prime}$ \XOI}), followed by the \rp-violating decay of 
the LSP.
With \LLE, the indirect decay of a sneutrino 
(charged slepton)  through a neutralino
and  a neutrino (charged lepton) leads to 
two charged leptons and two neutrinos (three charged leptons  and one neutrino). 
The squark decay into a quark and a gaugino leads to one quark and three
leptons.
With \UDD, the indirect decay of a squark (slepton) 
leads to four quarks (three quarks and one lepton).

\subsection{\rp-violating coupling upper limits and LSP lifetime}

Upper limits on the \Lijk\ and \Lppijk\ 
couplings can  be derived mainly from indirect
searches of \rp-violating effects~\cite{barger89,dreiner99}, assuming that only one
coupling is dominant at a time. 
They  are dependent on the sfermion mass, and  usually given 
for m$_{\widetilde{\rm f}} =$~100~\GeVcc.
The  upper bounds on  \Lijk\ are obtained  from charged-current
universality, lepton universality, $\nu_{\mu}-$e scattering, 
forward-backward asymmetry in e$^{+}$e$^{-}$ collisions, and bounds on 
$\nu_e$--Majorana mass. 
Most present indirect limits are in the range
of 10$^{-3}$ to 10$^{-1}$;
the most stringent upper limit is given for \Lacc\ ($\simeq 6 \cdot 10^{-3}$).
Upper limits on \Lppijk\ couplings come from experimental
measurements of
double nucleon decays for $\lambda''_{112}$ (10$^{-6}$),  
$n-\bar{n}$ oscillations for $\lambda''_{113}$ (10$^{-5}$) and of
$R_\ell =\Gamma_{had}(\rm Z^0)/\Gamma_\ell (\rm Z^0)$  in e$^{+}$e$^{-}$ collisions
for $\lambda''_{312},\lambda''_{313}, 
\lambda''_{323}$ (0.43).
The upper limits on the other \Lppijk\ couplings 
are obtained from the requirement of perturbative unification 
at the Grand Unified Theory (GUT) 
scale of $10^{16}$ GeV. This gives a limit of 1.25.\\

In the present searches, the LSP lifetime
was a crucial parameter since the analyses 
were valid only if the \rp-violating decays were
close to the production vertex, which means a LSP flight path
 shorter than a few centimetres. 

The LSP mean decay length is given by ~\cite{dreiner-ross,dawson}:
\begin{equation}
\mathrm{L(cm)} = 0.3 \ (\beta \gamma) 
\left({\rm m_{\tilde{\rm f}}\over{100~\mathrm{GeV}/{c}^2}} \right)^4 
\left({1~\mathrm{GeV}/{c}^2 \over{\rm m_{\widetilde{\chi}}}}\right)^5 
{1\over{\Lambda^2}}
\label{gaulife}
\end{equation}
if the neutralino or the chargino is the  LSP with 
$\beta \gamma =  \rm P_{\widetilde{\chi}}/\rm m_{\widetilde{\chi}}$
and with $\Lambda$~=~\Lijk\ or $\Lambda$~=~$\sqrt{3}$\Lppijk.
Considering the upper limits on
the couplings described above and according
to  equation~(\ref{gaulife}), the analyses are not sensitive to
a light neutralino (m$_{\chi} \leq $~15~GeV/$c^2$), 
due to the terms 
m${_{\widetilde{\chi}}}^{-5}$  and  $(\beta \gamma)$.
Moreover, when studying  neutralino
decays, for the typical masses considered in the present study,
the analyses are sensitivite to 
\rp-violating couplings  greater than $10^{-4}$ to $10^{-5}$,
where the \rp-violating decay has a negligible decay length. 
For much lower values of the coupling strength, the
LSP escapes the tracking devices before decaying and the results of
the searches performed under the assumption of  $R_p$ conservation are
recovered~\cite{rpc}. Between these two extreme cases,
the LSP decay produces a displaced vertex 
topology~\footnote{This particuliar topology, not considered
in the present searches, has been studied in other
searches performed by the DELPHI collaboration~\cite{displaced}.}. 

\subsection{Pair-production of supersymmetric particles}\label{sec:pairprod}

Pair-production of supersymmetric particles in the MSSM 
assuming  {\rp} violation is
identical to pair-production in the case of 
{\rp} conservation, since the trilinear couplings are not present 
at the production vertex. The production of single supersymmetric particles 
via trilinear couplings   has been studied in other
searches performed by the DELPHI collaboration~\cite{singlerpv}. \\

In the constrained MSSM scheme~\cite{mssm} considered
in the present searches, the mass spectrum of neutralinos 
and charginos is determined by three parameters,
with the  assumption that both  the gaugino 
and the sfermion masses are unified at the 
GUT scale. The relevant
parameters  
are then:
M$_2$, the SU(2) gaugino mass 
at the electroweak scale (it is assumed that
\mbox{M$_1={ 5\over3}{\mathrm{tan}}^2\theta_{\rm W} \rm M_2$}),
m$_0$, the common sfermion mass at the GUT scale, 
$\mu$, the mass-mixing term of the Higgs doublets at the electroweak
scale and \tanb, the ratio of the
vacuum expectation values of the two Higgs doublets. 
It is assumed that the running of the \Lijk\ and  \Lppijk\  couplings from
the GUT  to the electroweak scales does not have a significant
effect on the ``running'' of the gaugino and sfermion masses.

The charginos are produced in pairs in the $s$-channel 
via $\gamma$ or $\mathrm{Z}$ exchange, or in the 
$t$-channel via \snue\  exchange  
if the charginos have a gaugino component;
the neutralinos  are produced in pairs  
via $s$-channel $\mathrm{Z}$ exchange  provided they have a higgsino
component, or 
via $t$-channel \sel\ exchange if they have a gaugino component.
The  $t$-channel contribution
is suppressed when  the slepton masses (depending on m$_0$) are high enough. 
When the  \sel\  mass is
sufficiently small (less than 100~\GeVcc),
neutralino production can be enhanced, because of the
$t$-channel contribution. 
On the contrary, if the \snue\  mass is in the same range, 
the chargino cross-section can decrease due to
destructive interference between the $s$- and $t$-channel amplitudes. \\

The pair-production  cross-section of sfermions 
mainly depends on the sfermion masses. 
The \sel\ and \snue\ cross-sections are also
very sensitive to the neutralino and chargino compositions  
(which are function of $\mu$,  M$_2$ and tan$\beta$) 
via the $t$-channel exchange.    
The sfermion mass-eigenstates, \sfea\ and
\sfeb\ (where f is  a quark  or lepton and \sfea\ is lighter than \sfeb), 
are obtained from
the two supersymmetric scalar partners \sfeL\ and 
\sfeR\ of the corresponding left and right-handed 
fermion~\cite{ellis,bartl}: 
\begin{center}
\begin{tabular}{lcl}
\sfea\ &=& ~~\sfeL\ cos\phimixf\ + \sfeR\ sin\phimixf \\
\sfeb\ &=& --\sfeL\ sin\phimixf\ + \sfeR\ cos\phimixf 
\end{tabular}
\end{center}
where \phimixf\ is the mixing angle with 0~$\leq$~\phimixf~$\leq \pi$.
The supersymmetric partner of the left-handed fermions are likely 
to be heavier than their  right-handed
counterparts.  The \sfeL--\sfeR\ mixing is related to the off-diagonal
terms of  the scalar squared-mass matrix. It is
proportional to the fermion mass, and is small compared to the
diagonal terms, with the possible exception of the third family
sfermion~\cite{drees}.  
The lighter stop, $\tilde{t}_1$, is then probably the lightest
squark. This is not only due to the mixing effect but also to  
the effect of the large Yukawa coupling of the top; both tend to decrease the mass of
$\tilde{t}_1$~\cite{dreesmartin}.  
Similarly the lightest charged slepton is probably
the $\tilde\tau_1$. For small values of tan$\beta$, $\tilde\tau_1$ is
predominantly a $\tilde\tau_R$, and it is not so much lighter than
$\tilde{e}_1$ and $\tilde\mu_1$.
In the present slepton search, a no-mixing scenario is assumed. 
In the third squark generation searches
two left-right mixing angle cases have been considered. The first one with
mixing angle equal to zero and the second one with the mixing angle 
$\Phi_{\rm mix}= 56 ^{\circ}$ ($\Phi_{\rm mix}= 68 ^{\circ}$) 
corresponding to the minimum production cross-section of the stop (sbottom)
via \Zn\ exchange~\cite{dreeshikasa}.

\section{Data and generated samples}

\subsection{Data samples} 

The data recorded in 1999 and 2000 by the DELPHI experiment 
at  centre-of-mass energies from
$\sqrt{s}$~=~192~GeV to 208.8 GeV,
correspond to a total integrated
luminosity of around 450~pb$^{-1}$.
The DELPHI detector has been
described elsewhere~\cite{delphidet}.  
An integrated luminosity of 386~pb$^{-1}$~(Table~\ref{lumitable}) 
has been analysed, corresponding to
high quality data, with the tracking detectors and the electromagnetic
calorimeters in good working conditions. 
At the end of the data taking period in 2000, one sector (among twelve)
of the Time Projection Chamber (TPC) failed beyond repair. 
This required modifications 
in the data treatment (pattern recognition), and a specific
simulation of the detector with one TPC sector off 
has been performed. 
An integrated luminosity of 51.8~pb$^{-1}$ recorded
with one TPC sector off have been analysed.

\subsection{Event generators}

To evaluate background contaminations,
different contributions coming from the SM processes  
were considered. The SM events were produced by  
the following generators:
\begin{malist}
\item \underline{$\gamma\gamma$ events:}
{\tt BDK}~\cite{bdk} for $\gamma\gamma\to \ell^+\ell^-$ processes,
including radiative corrections for the e$^+$e$^- \mu^+ \mu^-$
and e$^+$e$^- \tau^+ \tau^-$ final states,
and {\tt TWOGAM}
for $\gamma\gamma\to$ hadron processes.
\item \underline{two-fermion processes:}
 {\tt BHWIDE}~\cite{bhwd} for Bhabha scattering 
($\rm e^+ \rm e^- \to \rm e^+ \rm e^- (\gamma)$),
{\tt KORALZ}~\cite{koralz} for e$^+$e$^-\to \mu^+\mu^-(\gamma )$ and
for e$^+$e$^-\to \tau^+\tau^-(\gamma)$ 
and {\tt PYTHIA 6.143}~\cite{pythia} for e$^+$e$^-\to \rm q \rm\bar q(\gamma)$ 
events.
\item \underline{four-fermion processes:}
{\tt EXCALIBUR}~\cite{excal} and {\tt GRC4F}~\cite{grc4f} for all types of 
four-fermion processes: non resonant ($\rm{f\bar ff^\prime\bar f^\prime}$), 
singly resonant 
(Z$\rm f \rm\bar f$, W$\rm f \rm\bar f^\prime$) and doubly resonant (ZZ, WW)
({\tt PYTHIA} was used also for cross-checks on the final results).
\end{malist}
Signal events were generated for all analyses with
the {\tt SUSYGEN 3.00} program~\cite{susygen}.

All generated background and signal events were passed through 
the full \mbox{DELPHI} simulation and reconstruction
chain~\cite{delphidet} and then processed in the same way as the 
real data. 
To treat the data taken with one sector of the TPC off, special
background and signal event samples were generated, and the same
treatment applied to them as to the real data.

\subsection{Signal samples}
\stitre{Choice of the \rp-violating couplings} 
\noindent Among the nine \Lijk\ couplings, 
\Labb\ (which leads to several muons
in the final states) and \Lacc\ (which leads to several taus
in the final states) 
have been chosen for most of the signal generation.
Their values were set for m$_{\tilde\ell} = 100$~\GeVcc\ 
at 0.04 and 0.003 respectively, below 
their upper bound derived from indirect searches of \rp-violating effects.
Any value between $10^{-3}$ and  $10^{-1}$
would not change the neutralino decay topologies.
Simulations with other couplings have been also performed in order to check
that the analyses developed for  \Labb\ or \Lacc\
were able to select the corresponding signal with an equal
or better efficiency. 

For the generation of all \UDD\ signals, a $\lambda''_{212}$ coupling of 
strength 0.1 (for m$_{\tilde{\rm q}} = 100$~\GeVcc)
was used. Any value between $10^{-2}$ and 0.5 
would not change the neutralino decay topologies.   

Searches for decays through specific \Lppijk\ couplings, 
leading to the production of one or several b quarks, can use 
b--tagging techniques to reach higher sensitivities, but at 
the cost of losing generality. \\

\stitre{Generated signal sets} 
\noindent Two different procedures were applied to 
the signal generation for gaugino pair-production 
and subsequent decays through either \LLE\ or \UDD\ terms
in order to cover the MSSM parameter space.

For the \LLE\ term, 
the \XOi\ and \XPMk\ pair-production processes were considered for
different values of \tanb\ (from 1 to 30), m$_0$ 
(between 90~\GeVcc\ and 500~\GeVcc),
$\mu$ (between --200~\GeVcc\ and 200~\GeVcc) 
and M$_2$ (between 5 and 400~\GeVcc), 
for centre-of-mass energies of 200 and 206~\GeV.
For the \UDD\ term,
pair-production of neutralinos  was generated for several masses. 
The simulated masses started from 10~\GeVcc \ and were  increased
in steps of 10~\GeVcc, as long as the mass of the chargino remained kinematically
accessible. Masses corresponding to the kinematic limit were also simulated. 
To generate chargino pairs,
the mass of the chargino was varied from 45~\GeVcc \ to 95~\GeVcc \ with a
10~\GeVcc \ step. Chargino masses were also simulated at the kinematic
limit. The neutralino mass was varied from
10~\GeVcc \ to a mass difference with the chargino of 5~\GeVcc\ with a 10~\GeVcc \ step. 
For each mass pair, a set of the variables $\mu$,  M$_2$ and  tan$\beta$ was
found for the chosen simulation. \\

Sfermion indirect decay signals were simulated at different
masses with  steps of 10~\GeVcc\ at centre-of-mass energies of 200 and
206 GeV,  with tan$\beta$ and 
$\mu$ fixed at 1.5 and --200~\GeVcc\ respectively. M$_2$  
was used to fix the neutralino mass at the required value.
 The points were simulated from 45
to 100~\GeVcc\ for the 
sfermion  masses and from 15 to 95~\GeVcc\ for the \XOI~masses up
to a mass difference between the sfermion and the LSP of ~5~\GeVcc. \\

Among the sfermions, only the sneutrino direct decay 
via \LLE\ terms was studied.
Specific signal sets have been produced 
with Br(\snu~\Ra~$\ell^+\ell^-$) = 100\%. The
processes \mbox{\snue\asnue \Ra 4$\mu$} (\Labb), 
\mbox{\snue\asnue \Ra 4$\tau$} (\Lacc),
\mbox{\snum\asnum \Ra 4$\tau$} (\Lbcc) 
and \mbox{\snut\asnut \Ra 2e2$\tau$} (\Lacc) have been generated for
different values of the sneutrino mass up to 98~\GeVcc,  with tan$\beta$ and 
$\mu$ fixed at 1.5 and --200~\GeVcc\ respectively. 
In order to check that all final states from \snu\asnu\ decay
were covered, signals
obtained for other \Lijk\ couplings and for sneutrino masses around 90~\GeVcc\
were also generated. 

\section{Description of the analyses}

The analyses covering the decay of pair-produced sparticles
were designed to cover multi-lepton final states 
for \LLE\ coupling and multi-jet final states for \UDD\ coupling.
Different preselections were applied, one for the
multi-lepton channels and one for the multi-jet channels.
In each case, dedicated analyses were necessary
to take into account the specific characteristics of the sparticle decay.
The  multi-jet analyses required
a specific treatment based on neural network techniques.

The sensitivity of the searches for sparticle indirect decays
depended on the mass difference ($\Delta$M) between the sparticle
being searched for and
the LSP. The analyses were designed to be efficient for 
$\Delta$M~$\geq$~5~\GeVcc.
The multi-jet analyses required different signal selection optimisations
to cover efficiently all  $\Delta$M regions; therefore they were
divided into windows according to the value of $\Delta$M.

No excess in the data appeared in these searches, therefore a  
working point optimization on the selection criteria was performed 
minimizing  the expected excluded cross-section 
as a function of the average 
signal efficiency.

\subsection{Description of the final states}

\subsubsection{Decays via \BLLE}
Direct and indirect decays of gauginos, 
direct and indirect decays of sneutrinos
and indirect decays of charged sleptons and squarks were studied.

The direct decay of a pair of lightest neutralinos
leads to two neutrinos and four
charged leptons. For an indirect decay of  chargino  or heavier
neutralino pairs the final state may contain some jets and/or leptons 
in addition to the four leptons and the missing energy 
from the decay of the LSP.
The direct decay of a sneutrino pair 
gives final states with four charged leptons, in which the leptons
can be of two different flavours. The direct decay
of a charged slepton pair gives 
final states with two charged leptons, in which the leptons
can be of two different flavours, and missing energy. This
final state has not been covered by the present 
analyses.
In the indirect decay of any sfermion pair,
the final states are composed of two fermions plus the decay 
products of the neutralinos. 

Compared to other couplings the highest efficiencies and background reduction 
were obtained
in analyses performed on the signal with a dominant \Labb\ coupling.
For analyses dedicated to a  \Lacc\ coupling, 
due to the presence of several taus in the decay channels,  
the efficiencies and the rejection power were low.
For final states produced by  other $\lambda_{ijk}$, the detection
efficiencies lay between these two limiting cases.
Therefore conservative limits can be derived by considering the results of the
analyses  performed assuming a dominant \Lacc\  coupling, and
only these analyses will be described in section~\ref{multilepton}.

The decay of pair-produced sparticles via a \Lacc\ coupling
leads to different types of final states, depending on the
produced sparticles.
The \XOI\XOI\ decay via \Lacc\ leads to 2$\tau + \ell + \ell' +
$\Emiss, where $\ell, \ell' = $~e or  $\tau$, and \Emiss\
means missing energy. 
In addition, jets and/or leptons from the W or Z decays show up in
the final state from \XOi\XOj\ and \XPI\XMI\ indirect decays.
The indirect decay of a slepton pair gives 
2$\tau + \ell + \ell' +$\Emiss, 
with two additional charged leptons 
(same flavour, opposite charge) in the case of charged sleptons, and
additional missing energy in the case of sneutrinos.  A 4$\tau$ final
state is produced by the direct decay of \snue\asnue\ via  \Lacc.
The direct decay of \snut\asnut\ gives 2e2$\tau$ and then,
there is finally less missing energy comming from the taus decay
than in the previous cases.
The indirect decay of squarks adds exactly two jets to the 
2$\tau + \ell + \ell' + $\Emiss.

Four analyses have been performed to search for all these topologies.  
They are summarized in the first part of Table~\ref{topos}.  

\subsubsection{Decays via \BUDD}
Direct and indirect decays of gauginos, 
and indirect decays of charginos,
charged sleptons and squarks were studied.

For each indirect decay of a chargino, squark or
slepton pair there are at least six quarks in the final state.    
Therefore the most important feature of these decays is
the number of quarks  
produced,  which can be up to ten for the indirect decay of
two charginos with the hadronic decays of the W bosons.  
The indirect decay channel presents the only possibility for the sleptons to decay through a \UDD~term. 
In this case, two leptons are produced in the
$R_p$ conserving decay of the slepton pair, and they add to the 
six jets coming from the decay of the two  neutralinos: 
a \mbox{6 jets + 2$\ell$, $\ell =$~e, $\mu$} 
final state is the signature of these signals.   
The indirect decay of a
stop or sbottom pair produces 8 jets in the final state. 
Two b quarks are produced from the sbottom decay.  
The analysis of the different decay channels was 
organized on the basis of the number of hadronic jets in the final state (see
Table~\ref{topos}).

\subsection{Analysis tools and techniques}

\subsubsection{Lepton identification}
The identification of  a muon or an electron, 
used in all $\lambda_{ijk}$ analyses
and several $\lambda''_{ijk}$ ones,   was based on standard DELPHI 
algorithms~\cite{delphidet}. The identification  
could be ``tight'' if it was an 
unambiguous one, or ``loose'' otherwise. 
In the multi-lepton analyses described in this section  
a particle was considered as a well identified 
electron if it satisfied the tight conditions from the DELPHI electron
identification algorithm,  its 
momentum was greater than 8~\GeVc\ and  there was no other charged 
particle in a  cone of half-angle 2$^\circ$ around it.
A particle was considered as a well identified muon if its momentum 
was greater than 
5~\GeVc\ and  it was tagged as a tight muon candidate by the 
DELPHI algorithm.\\

\subsubsection{Jet reconstruction algorithms}

Two different jet reconstruction algorithms have been used.   
The {\tt DURHAM} algorithm~\cite{Durham} was used for the multi-lepton  
(\LLE\ coupling) analyses, where jets were expected from $\tau$ or W
boson decays. In case of  multi-jet
analyses, the {\tt CAMBRIDGE}
 clustering algorithm~\cite{camjet} implemented in the {\tt CKERN} 
package~\cite{ckern}
was used. 
 
The {\tt CAMBRIDGE} algorithm was introduced to select soft jets,
coming from quark-jets with gluon emission. The specific 
procedure of clusterization which extracts 
soft jets from the list of objects to be clustered, was
particularly interesting for multi-jet analyses, 
where the jets (more than six) 
may not be well separated in momentum space.   
For each event, the two algorithms provided all possible
configurations of jets between 2 and 10.  
They have the same definition 
of y$_{\rm cut}$ distance,
but mainly differ in the iterative procedure of clustering. 
In this paper, the transition value of the y$_{\rm cut}$ in the  {\tt DURHAM} or 
{\tt CAMBRIDGE} algorithm at which the event changes from a clustering with
$n$ jets, called $n$-jet configuration,  
to a clustering with ($n-$1)-jets, is denoted y$_{nn-1}$. In other words, the
y$_{nn-1}$  value is the y$_{\rm cut}$ value for which 
the number of particle clusters flips from $n$ to $n-$1 for increasing
y$_{\rm cut}$ distances. For example, the y$_{43}$ value of one 
event is the highest value of y$_{\rm cut}$
 to obtain 4 separated clusters of particles.   

\subsubsection{Neural networks \label{nnw_techno}}

A neural network method was applied in order to distinguish signals from 
SM background events for all multi-jet analyses. 
The trainings of the neural networks were done in the standard 
back-propagation manner with one hidden layer on samples of simulated 
background ($ \rm q \rm\bar q$  and four-fermion) and signal events.  
A feed-forward algorithm has been implemented to compute from the input
discriminating variables a  single discriminant
variable (signal output node) which was used first to validate the training 
with different signal and background samples and then to 
select the final number of candidate events for each analysis. 
The exact configuration and the input discriminating variables of 
each neural network depended on the search channel. 
The working point on the signal ouput node value has been chosen to 
minimize the expected excluded cross-section at 95\% confidence level 
(CL) when there is no signal.

\subsection{Multi-lepton \BLLE\ channels \label{multilepton}}

\subsubsection{Preselection \label{presel-LLE}}
The selections were based on the criteria already
presented in~\cite{lle189},
using mainly missing momentum, lepton identification
and kinematic properties. The  preselection
requirements were:
 \begin{malist}
\item more than three charged particles
and at least one of them with a polar angle between 40$^{\circ}$
and 140$^{\circ}$;
\item at least one identified lepton (e or $\mu$); 
\item a total energy greater than 0.1$ \cdot \sqrt{s}$; 
\item a missing momentum component transverse to the beam
(\Pt) greater than  5~\GeVc;  
\item a polar angle of the missing momentum ($\theta_{\rm miss}$)  
between 27\mydeg\ and 153\mydeg;
\item a thrust axis not close to the beam pipe, viz. $|\cos\theta_{\rm th}|$ less than 0.9;
\item an acollinearity\footnote{The acollinearity is computed 
between the two vectors corresponding to the sum of the particle
momenta  in each hemisphere of the event. The two hemispheres are defined by the
 plane orthogonal to the thrust axis.} greater than 2$^\circ$,
and greater than 7$^\circ$
for events  with a charged particle multiplicity greater than 6.
\end{malist}

\begin{figure}[htb!]
\begin{center}
\epsfig{file=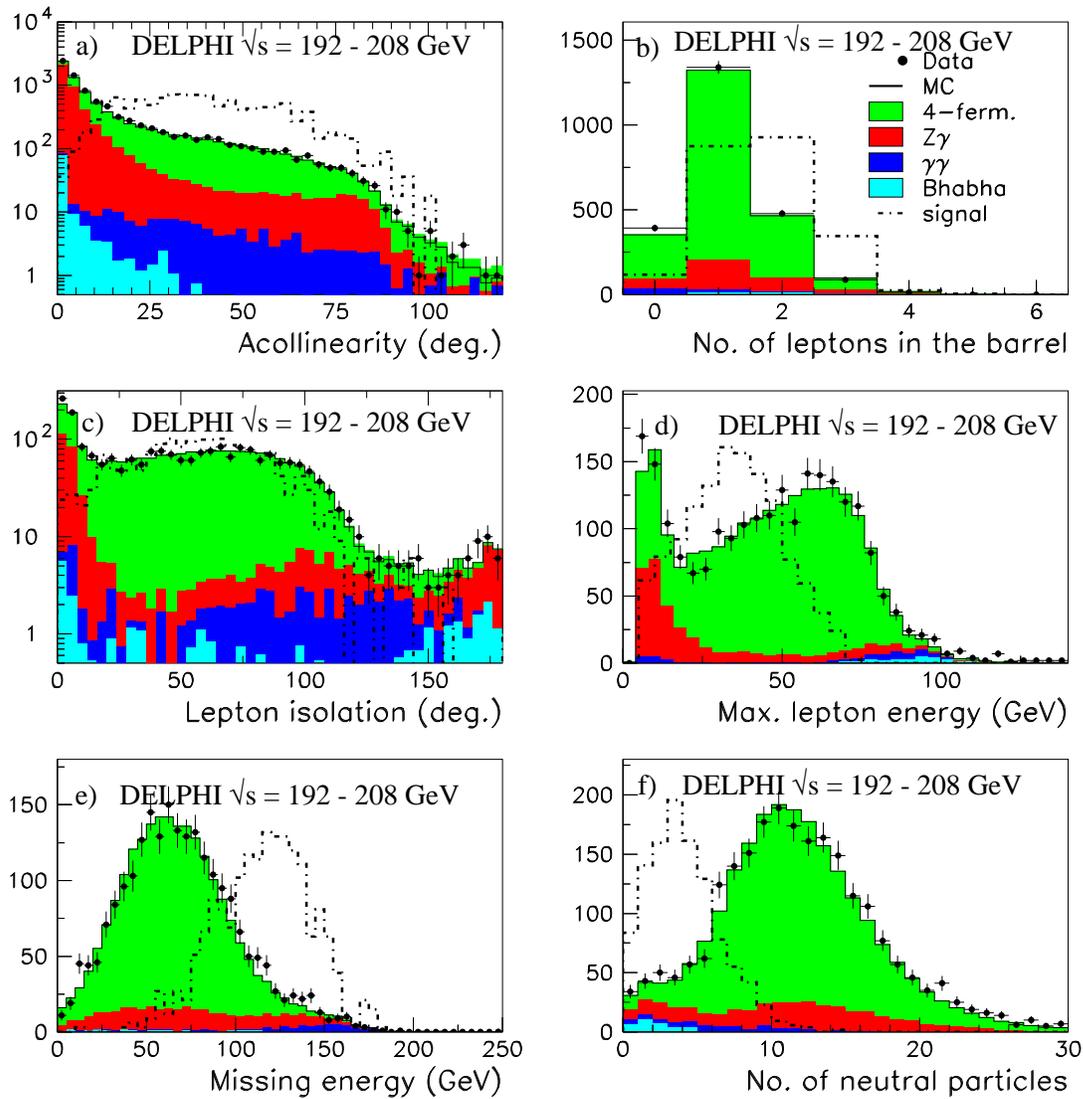,width=16cm} 
\caption{\LLE: sparticle pair search in multi-lepton channels -- 
Event variable (see text) distributions before the criteria
applied on the acollinearity (preselection)
 (a) and after (b--f) the preselection.
The simulated signal corresponds to \XOI\XOI\ 
production, with \mbox{m$_{\chi_1^0} = 90$~\GeVcc} 
(the normalisation is arbitrary).}  
\label{l133dist0}
\end{center}
\end{figure}

The preselection 
was efficient in suppressing 99.9\% of the backgrounds coming from Bhabha 
scattering and two-photon processes while removing 97\% of the 
\ffgamma\ contribution. The preselection also reduced the four-fermion 
contamination by 75\%. 
After this preselection stage, 2310~events (1220 for data
 at  centre-of-mass energies between 192 and 202~\GeV, and
1090 for those collected above 202~\GeV)
were selected to be compared to 2254~$\pm$~6  expected from the background 
sources (1189~$\pm$~5 at centre-of-mass energies between 192 and
202~\GeV, 1065~$\pm$~4  above 202~\GeV). The corresponding
efficiencies for the large majority
of \LLE\ signals lay between 60\% and 80\%.
The distributions of several event variables  
at the hadronic preselection stage are shown in Figure~\ref{l133dist0}.

The above requirements had to be 
slightly modified for the stop
analysis (see Section \ref{squa_lle}),
in particular to take into
account the fact that the final state always contains 
two jets. 
A minimum multiplicity of eight charged particles was
required. No requirement was applied on the thrust axis.
On the other hand, a stronger cut 
was applied on the polar angle of the missing momentum  
(30$^\circ  \leq \theta_{\rm miss} \leq 150^\circ$).
After this preselection stage, 2197~events
were selected to be compared to 2208~$\pm$~6  expected from the
background sources.

\subsubsection{Gaugino search \label{gaug_lle}}

The gaugino analysis was designed to cover the 
$2\tau + {\rm n}\ell + {\rm m}j +$~\Emiss  
($\rm n \geq 2$,~$\rm m  \geq 0$) final states, from
direct or indirect decays of gauginos,
and to be efficient for both low and high 
multiplicity cases. 

The thrust had to be less than 0.9 
and a lower limit on the missing energy was applied
 viz. E$_{\rm miss}$ greater than 0.3$\cdot\sqrt{s}$. 
The number of neutral (charged) particles  had 
to be less than~20 (25) and
the polar angle of at least one lepton had to be between 
40$^\circ$  and 140$^\circ$. 

The events were then separated in to two classes, according to
their charged particle multiplicity.
\begin{itemize}
\item For events with a charged particle multiplicity from four to six, 
(mainly for neutralino direct decay topologies), the
following criteria were applied:
\begin{malist}
\item the energy in a cone of 30$^\circ$ around the beam axis was
required to be less than 50\% of the total visible energy;
\item the energy of the most energetic lepton (e or $\mu$)
 had to be between 2 and 70~GeV;
\item there should be no other charged particle in a cone of half angle
of 20\mydeg\ (6\mydeg) 
around any identified lepton for a charged particle multiplicity 
equal to 4 (5 or 6).
\end{malist}
\item For events  with a charged particle multiplicity greater than six,
the previous criteria became:
\begin{malist}
\item the energy in a cone of 30$^\circ$ around the beam axis was
required to be less than 40\% of the total visible energy;
\item the energy of the most energetic lepton had to be between 5 and 60~GeV,
\item if there was only one identified lepton;
there should be no other charged particle in a cone with
a half angle of  6\mydeg\  around it;
if there was more than one identified lepton there should be no other charged 
particle in  a cone with
a half angle of  10\mydeg\  around at least two of them; 
\item at least one electron (loose identification) was required.
\end{malist}
\end{itemize}
These criteria removed 95\% of \Zg, \ZZ\ and \WW\ events.

\begin{figure}[htb!]
\begin{center}
\epsfig{file=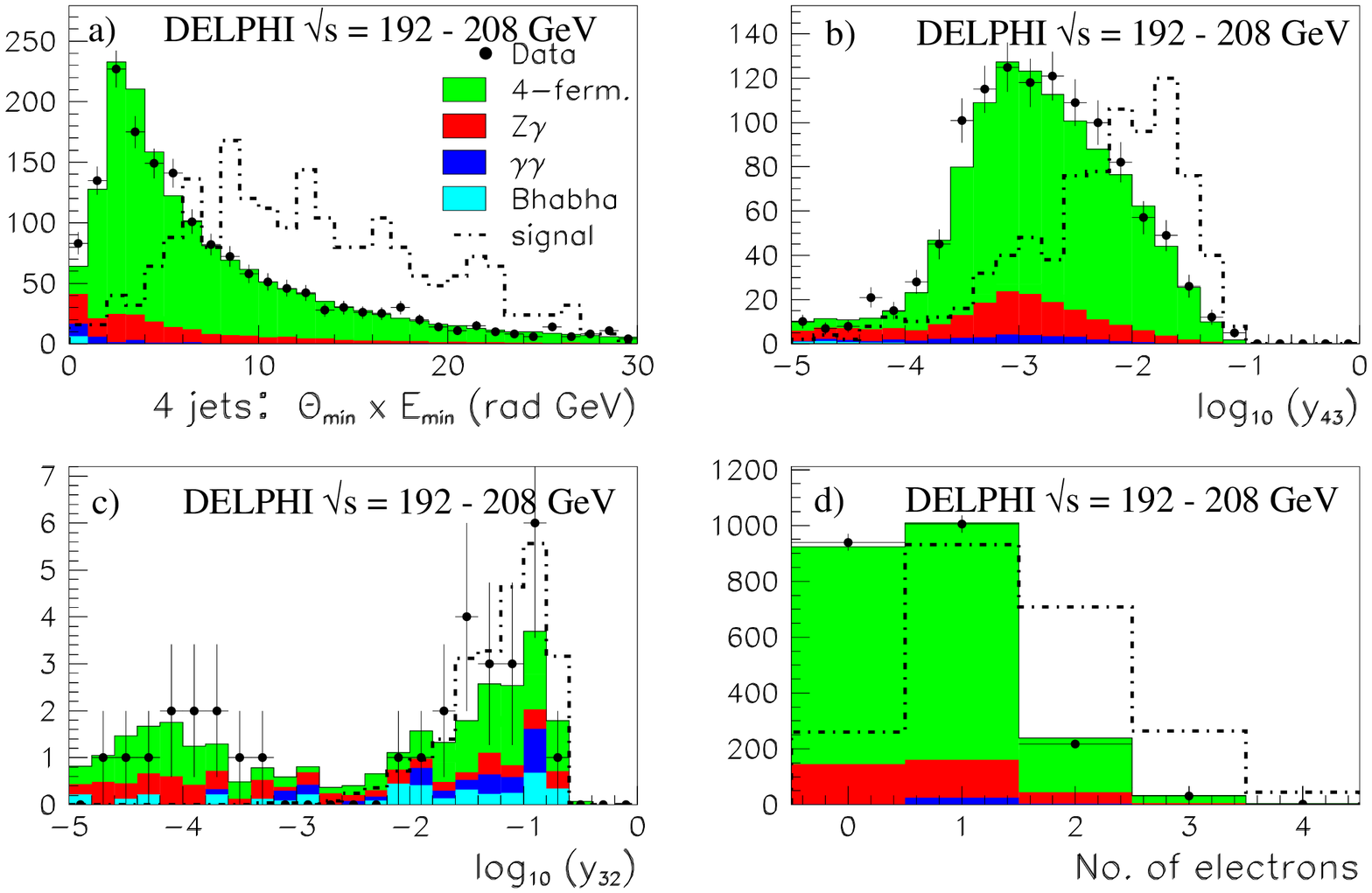,width=16cm} 
\caption{\LLE: sparticle pair search in multi-lepton channels --
Distributions of variables used in the four analyses, as described 
in the text: a) gaugino, b) slepton, c) sneutrino tau, d) squark.
The simulated signals correspond to production  of a) \XOII\XOI\ and \XPI\XMI\
(m$_{\widetilde{\chi}^{0}_{1}}$~=~50~\GeVcc,
m$_{\widetilde{\chi}^{0}_{2}}$~=~85~\GeVcc, 
m$_{\widetilde{\chi}^{\pm}_{1}}$~=~95~\GeVcc), 
b) \snue\asnue\ (m$_{\tilde{\nu}_{\rm e}}$~=~100~\GeVcc), 
c) \snut\asnut\ (m$_{\tilde{\nu}_{\tau}}$~=~95~\GeVcc), 
d) \stp\stpb\ (m$_{\tilde{\rm t}}$~=~95~\GeVcc)
(the signal normalisation is arbitrary).}  
\label{l133anal}
\end{center}
\end{figure}

A selection based on the jet characteristics and topologies was then
applied, depending on the charged particle multiplicity, as mentioned
above. First, constraints were imposed on the
y$_{32}$ and  y$_{43}$ values: they had to be greater
than 0.002 and 0.0001 respectively for events with low  charged particle 
multiplicity, and greater than 0.016 and 0.005  respectively for
events whose charged particle 
multiplicity was above 7; these criteria eliminated 99\% 
of the remaining \Zg\ contribution.  
In events with more than six charged particles, it was required
that at least one jet had no more than two  charged particles. 
In four or five-jet 
configurations, a minimum number of 4 jets with at least one charged particle
was required. For
a four-jet topology, a cut was applied on the value of 
E$^j_{{\rm min}} \cdot \theta_{{\rm min}}$
where  E$^j_{{\rm min}}$ is the energy of the least energetic jet, and
$\theta_{{\rm min}}$ is the minimum di-jet angle~(Fig.~\ref{l133anal}--a).
It had to be greater than 1~GeV$\cdot$rad for events with low  charged 
particle multiplicity, and greater then 5~GeV$\cdot$rad for events with a
charged particle multiplicity above 7. These requirements reduced 
the background from 4-fermion processes.\\

\subsubsection{Slepton search \label{slep_lle}} 
  
A slepton analysis, aimed to search for 
$2\tau + 2\ell +$~\Emiss\ (+~2$\ell$)
 was performed in order to study the following three channels:
\begin{malist}
\item 
$\tilde{\ell}_R^+ \tilde{\ell}_R^-$ \Ra $\ell$\XOI $\ell$\XOI;
\item  \snu\asnu \Ra $\nu$\XOI $\nu$\XOI; 
\item \snue\asnue\ \Ra 4$\tau$. 
\end{malist}
These final states have several taus, and most of them have missing energy.
After the tau decays, all channels  present a large amount of missing energy. 
The criteria used 
to eliminate almost all remaining \Zg\ events and most of the
4-fermion events
were the following:
\begin{malist}
\item the missing energy had to be greater than 0.3$\cdot\sqrt{s}$; 
\item the energy in a cone of half-angle of 30$^\circ$ around the beam axis was
required to be less than 40\% of the total visible energy;
\item the number of charged (neutral) particles  had 
to be less than~8 (10);
\item the energy of the most energetic lepton had to be between 2 and 70~GeV;
\item at least one lepton should have a polar angle between 40$^{\circ}$
and 140$^{\circ}$;
\item  there should be no other charged particle in a cone 
with a half-angle of 6\mydeg\ around at least one lepton;
\item y$_{32}$ and y$_{43}$, computed with the {\tt DURHAM} algorithm,
(Fig.~\ref{l133anal}--b) had to be 
greater than $2\cdot 10^{-3}$ and $4 \cdot 10^{-4}$ respectively;
\item in a four-jet topology
a minimum angle of 20\mydeg\ between any pair of jets was required.
\end{malist}

\subsubsection{\snut\ search \label{snut_lle}}
An analysis searching for 2e2$\tau$ final states
produced in the direct decay of \snut\asnut\ was performed.
Compared to the selection described in~\ref{slep_lle}, 
the most important change was the
suppression of the criterion on the missing energy, and the
introduction of the requirement of having at least one well identified
electron:
\begin{malist}
\item the energy in a cone of 30$^\circ$ around the beam axis was
required to be less than 50\% of the total visible energy;
\item the number of charged (neutral) particles  had 
to be less than~7 (10);
\item there should be at least one electron;
\item the energy of the most energetic lepton had to be between 25 and 80~GeV;
\item at least one lepton should have a polar angle between 40$^{\circ}$
and 140$^{\circ}$;
\item  there should be no other charged particle in a cone of
half-angle of 6\mydeg\ around at least one lepton;
\item y$_{32}$ (Fig.~\ref{l133anal}--c), computed with the
{\tt DURHAM} algorithm, had to be 
greater than $2\cdot 10^{-3}$;
\item in a four-jet topology,
a minimum angle of 20\mydeg\ between any pair of jets
was required.
\end{malist}
This selection removed \Zg\ background and most of the 4-fermion 
events.

\subsubsection{Stop search \label{squa_lle}}

In stop pair-production, each of the stops decays into a 
charm quark and a neutralino. The 
subsequent \rp-violating  decay via the \Lacc\ coupling 
of the neutralino into leptons, the
final state:~$2\tau + 2\ell$~+~\Emiss~+~2$j$, $\ell = $e or $\tau$.

After the preselection described in Section~\ref{presel-LLE},
the criteria used to select the final states of the stop pair 
indirect decay were:
\begin{malist}
\item the missing energy greater than 0.3$ \cdot \sqrt{s}$;
\item  the charged and neutral particle multiplicities below 
25 and 20 respectively;
\item the polar angle of at least one lepton  
between 40$^\circ$  and 140$^\circ$;  
\item the energy of the most energetic identified 
lepton between 5 and 50~GeV;
\item  no other charged particle  in a cone of half-angle of 6\mydeg\ 
around at least one lepton;
\item at least one well identified electron~(Fig.~\ref{l133anal}--d);
\item the y$_{32}$ and  y$_{54}$ values  constrained to
be less than 0.016  and  10$^{-3}$   respectively;
\item in the four-jet configuration, at least one
jet with less than three charged particles.
\end{malist}

\subsection{Multi-jet \BUDD\ channels}

For all \UDD\ channels the main SM backgrounds come from 
four-fermion processes except for the low neutralino mass channel where 
the hadronic Z decay is the dominant background. The following \UDD\ analyses 
were based on neural network techniques since the optimisation 
of the signal selection over 
the four-fermion background was performed on topological variables, 
such as jet resolution parameters, which are extensively correlated.  

\subsubsection{Preselection \label{presel-UDD}}

The multi-jet \UDD\ signals have final states with a large hadronic
activity, independent of the produced sparticles. Therefore a 
general hadronic preselection was performed with the aim of a high 
efficiency for the signal (especially for the gaugino analysis) and at the 
same time a good rejection of low particle multiplicity hadronic background events:

\begin{malist}
\item the number of charged particles had to be greater than 15; 
\item the total energy was required to be greater than
      {\mbox{0.6$\cdot$\ecms}};
\item the energy associated to
    charged particles was required to be greater than
      {\mbox{0.3$\cdot$\ecms}};
\item the effective centre-of-mass energy~\footnote{the effective
centre-of-mass energy is the centre-of-mass energy after the emission of one
or more photons from the initial state.} had to be greater than 150~GeV;
\item the discriminating  variable d$_{\alpha}$ =
$\alpha_{\rm min} \cdot  \rm{E}_{\rm min} - 0.5 \cdot \beta_{\rm min}
\cdot \rm{E}_{\rm max} / \rm{E}_{\rm min} $ (where the 0.5 energy factor is in GeV)
had to be greater than --10~GeV.rad~\footnote{$\alpha_{\rm min}$ 
is the minimum angle between two jets,
 $\beta_{\rm min}$ is 
the minimum angle between the most energetic jet and any other,
E$_{\rm min}$
(E$_{\rm max}$) is the minimum (maximum) jet energy from the four-jet
topology of the event.};
\item the minimum jet invariant mass had to be greater than {\mbox{500
MeV/$c^2$}} when forcing the event into four jets;
\item the  ln(y$_{32}$) had to be greater than --6.9;
\item the  ln(y$_{43}$) had to be greater than --8.
\end{malist}

\begin{figure}[htb!]
\begin{center}
\epsfig{file=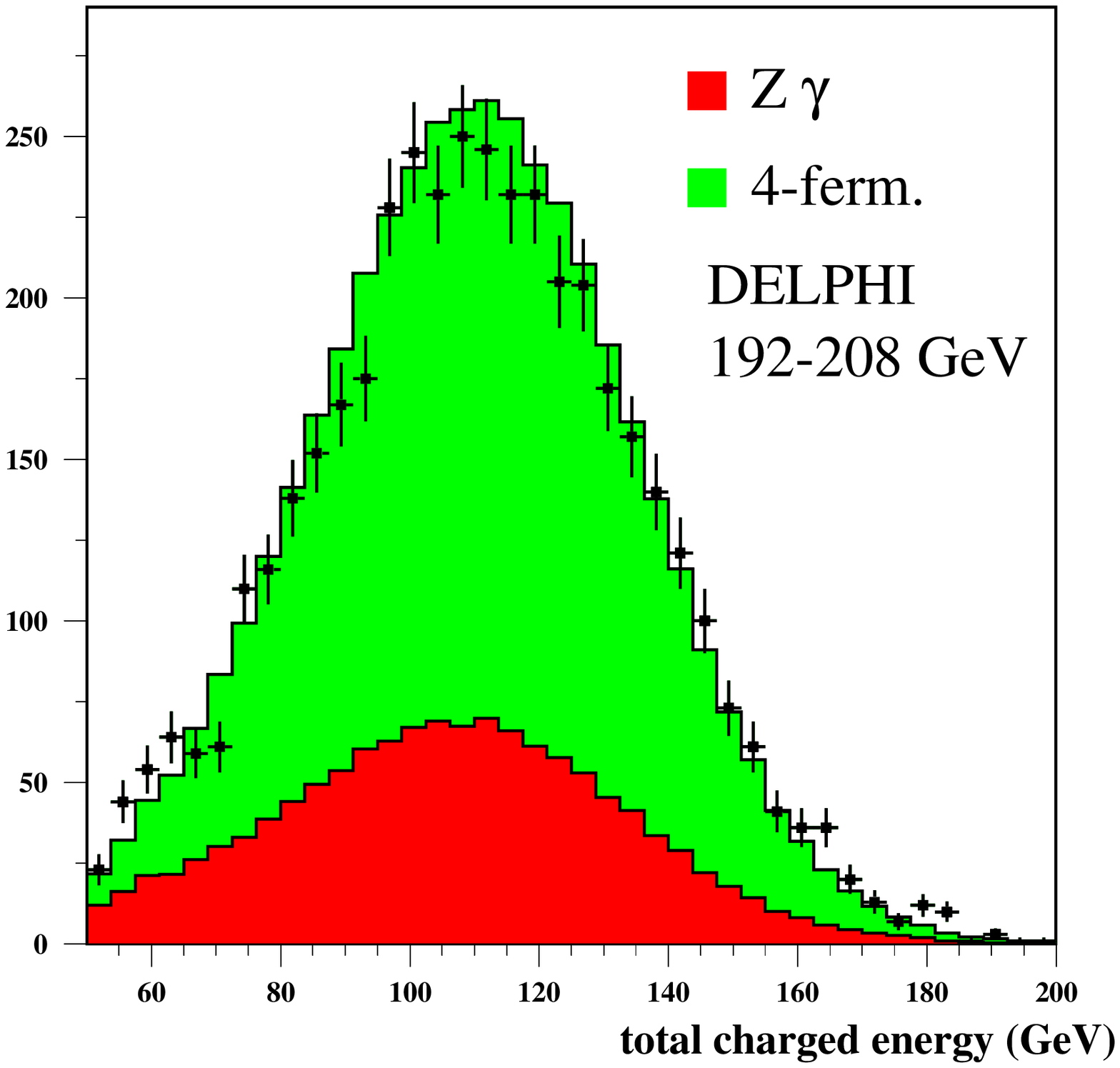,width=.47\linewidth,}  
\epsfig{file=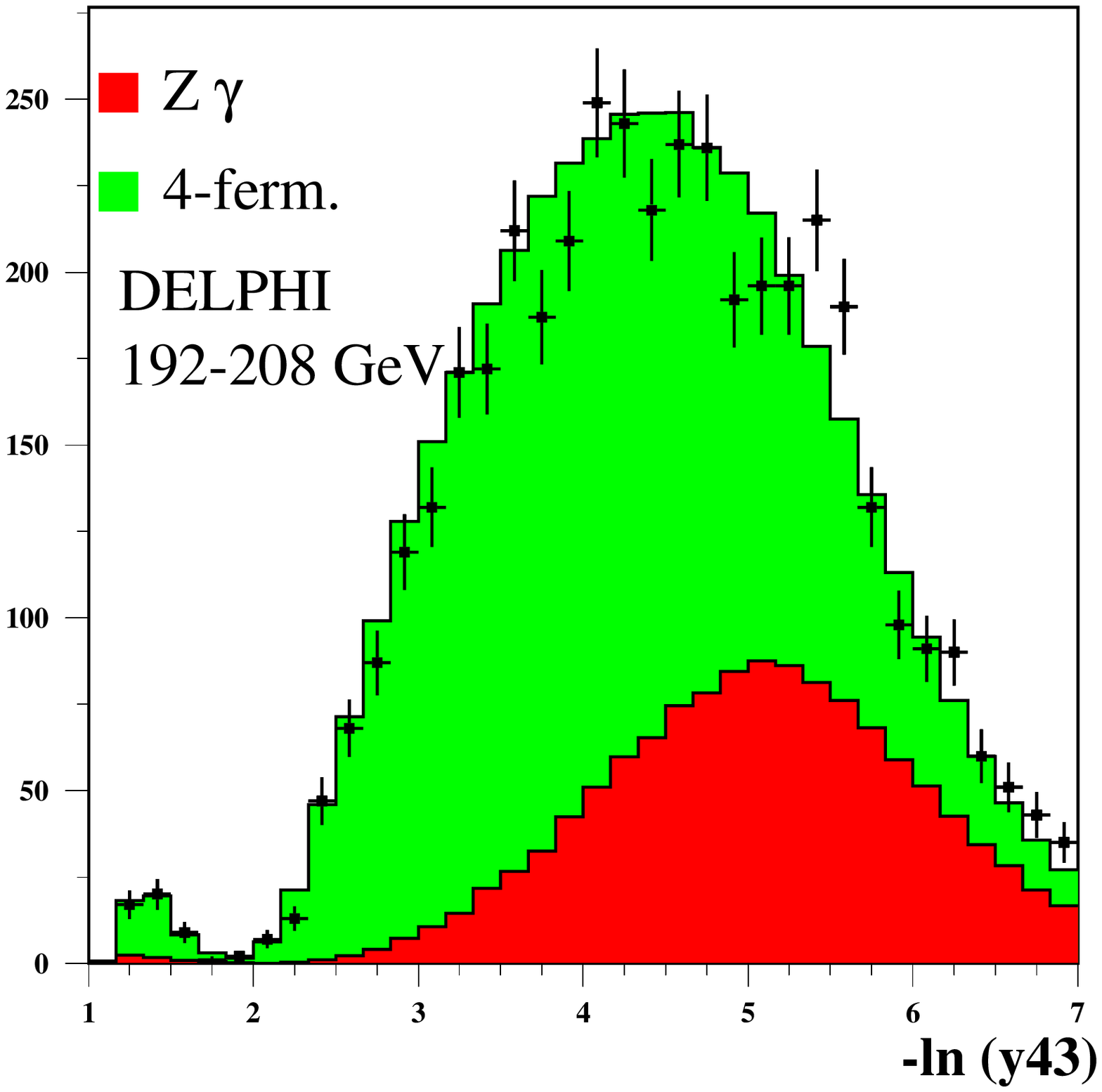,width=.47\linewidth}  
\vspace{-0.4cm}
\caption{\UDD: the total energy associated to
charged particles (left) and -ln(y$_{43}$) (right) distributions in the
multi-jet slepton analyses at the hadronic preselection level.} 
\label{udd_presel}
\vspace{1.5cm}
\epsfig{file=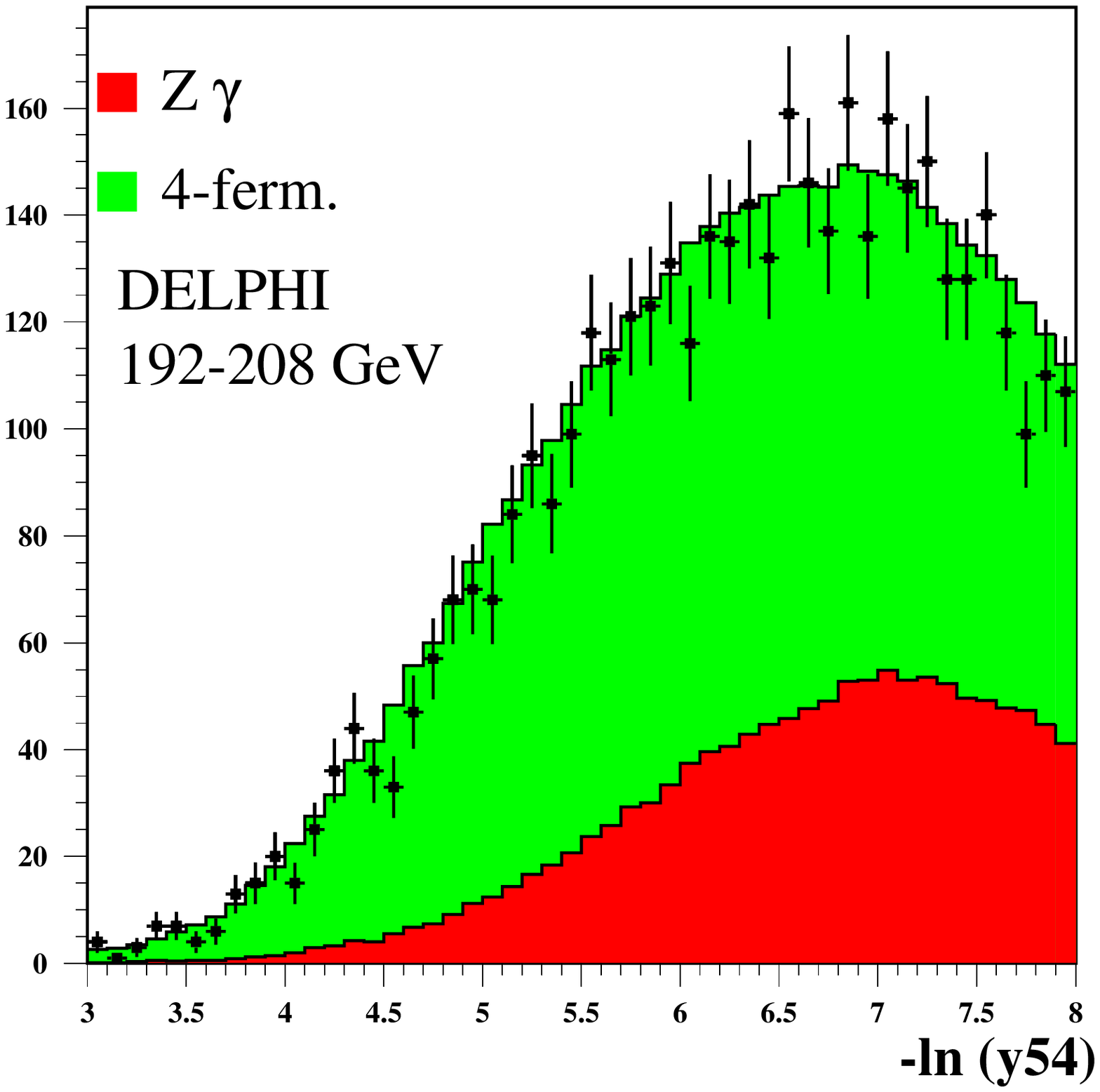,width=.47\linewidth,}  
\epsfig{file=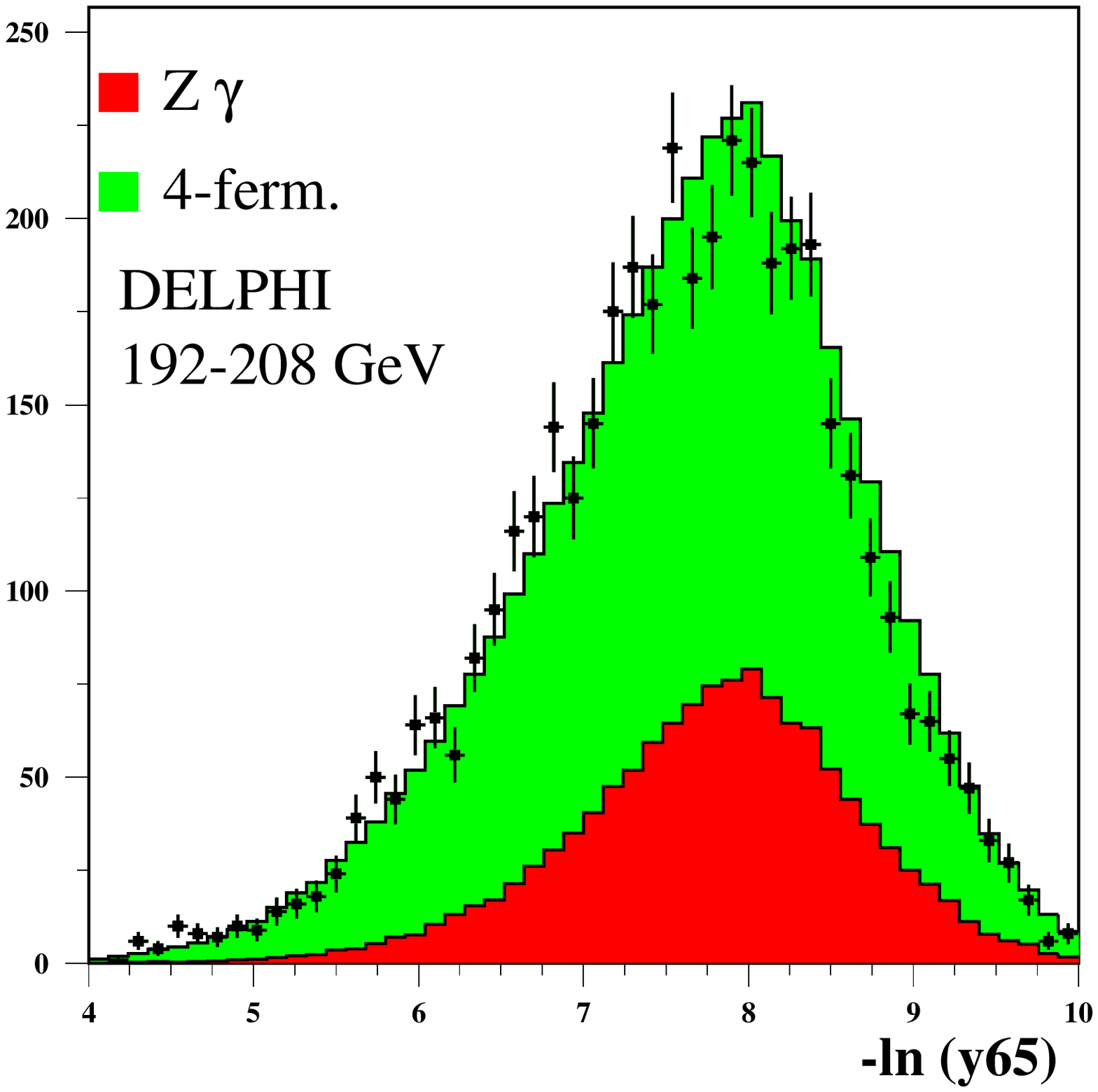,width=.47\linewidth}  
\vspace{-0.4cm}
\caption{\UDD: the -ln(y$_{54}$) (left) and -ln(y$_{65}$) (right) distributions
in the multi-jet sfermion analyses at preselection level.} 
\label{udd_sle_ymn}
\end{center}
\end{figure}

After the hadronic preselection, the main remaining background events were the
four-fermion events and the \qqg\ events with hard gluon
radiation. We observed 3844 events in the data 
with
 3869~$\pm$~4 events expected from background processes for the year 2000
(4180 events in the data to be compared to 4096~$\pm$~7 events 
in the simulation for the
year 1999). 
Examples of the distributions of several event variables 
at the hadronic preselection level 
are shown in Figure~\ref{udd_presel}.

The efficiencies for the \UDD\ signals
varied  from 60\% to 99\% depending on the simulated 
masses and $\Delta$M.
This preselection (sometimes with slight modifications, described 
in the following) was used for all the \UDD\ analyses.

These requirements had to be slightly 
modified to be better optimised for slepton searches. 
The discriminating  variable d$_{\alpha}$ was not used 
in the preselection, and the effective centre-of-mass energy 
was required to be above {\mbox{0.6$\cdot$\ecms}},
because a tighter cut was set 
on y$_{43}$ and a cut on y$_{54}$ was applied. 

Additional criteria to the basic hadronic preselection have been applied
 before the optimal neural network selection:
\begin{malist}
\item the charged particle multiplicity had to be greater than  16; 
\item the total energy was required to be greater than
      {\mbox{0.6$\cdot$\ecms}}; 
\item the energy of charged particles was required to be greater than
      {\mbox{0.3$\cdot$\ecms}};
\item the  ln(y$_{43}$) had to be greater than --7;
\item the  ln(y$_{54}$) had to be greater than --8;
\item the  thrust had to be lower than 0.94;
\item the maximum di-jet mass in a four-jet configuration had to be greater
than 10~\GeVcc.
\end{malist}
This selection was applied for the
\UDD \ sleptons ans squark analyses.  
After this selection the remaining number of events for all energies was 
4245 for the data and 4378~$\pm$~8 for the expected background from 
SM processes.\\
The distributions of the variables ln(y$_{54}$) and ln(y$_{65}$) 
at this stage of the preselection level 
are shown in Figure~\ref{udd_sle_ymn}.\\

\subsubsection{Neutralino search\label{neut_udd}}
The analysis described here was mainly designed to
search for neutralino direct decays; it was also
efficient in the search for chargino direct decays.
The six-jet analyses were based on three different  neural networks for the
optimization of the background and signal discrimination. The neural
network method used has
been presented  in Section~\ref{nnw_techno}. \\

Events with low gaugino mass have a large boost and 
look like di-jet events.
On the contrary,  events with heavy gauginos 
are almost spherical with six well separated jets. 
Therefore, we distinguished 3 mass windows to increase the sensitivity of
each signal configuration: 
\begin{malist}
\item low mass window N1:    $ 10 \le {\rm m}_{\tilde{\chi}} \le 45$~\GeVcc; 
\item medium mass window N2: $ 45 <   {\rm m}_{\tilde{\chi}} \le 75$~\GeVcc;
\item high mass window N3:   $  {\rm m}_{\tilde{\chi}} > 75$~\GeVcc.
\end{malist}

A mass reconstruction was performed using a method depending on the
mass window. For the N1 analysis, the events were forced into two jets 
and the average of the two-jet masses was computed. 
For the other analyses, the events
were forced into six jets and criteria on di-jet angles were
applied to choose the optimum three-jet combinations corresponding to the
decays of two neutralinos with the same mass. The minimum and maximum angles
between the jets belonging to the same three-jet cluster (same neutralino)
had to be in the intervals $[20^\circ, 80^\circ]$ and $[50^\circ, 165^\circ]$ for
the medium mass  window N2 ($[40^\circ, 110^\circ]$ and $[100^\circ,
175^\circ]$ for the high mass  window N3).  
If more than one combination was selected, the combination with the minimum
difference between the two energies of the three-jet clusters was chosen
to compute the neutralino mass.

Three neural networks were used, one for each mass window,
with the following variables as inputs:
\begin{malist} 
\item the thrust;
\item  dist$_{\rm WW}$= 
\( \sqrt{\frac{(\rm M_{1}-\rm M_{2})^{2}}{\sigma _{-}^{2}
}+\frac{(\rm M_{1}+\rm M_{2}-2\rm M_{\rm W})^{2}}{\sigma _{+}^{2}}} \),  
where M$_{1}$ and M$_{2}$ are  the di-jet masses of the jet combination 
which minimized this variable (after forcing the event into 4 jets); 
we took M$_{\rm W}=80.4$~\GeVcc\ for
the W mass,  $\sigma _{-}=9.5$~\GeVcc\ and $\sigma _{+}=4.8$~\GeVcc\ 
for the mass
resolutions of the difference and the sum of the reconstructed di-jet
masses respectively; this variable is peaked at 0 for WW events,
allowing a good discrimination against this background;
\item the energy of the least energetic jet multiplied by the
minimum di-jet angle in four and five-jet configurations;
\item the difference between the energies of
the two combinations of three jets, 
after the mass reconstruction;
\item the reconstructed neutralino mass;
\item y$_{n~n-1}$ with $n$=3 to 10.
\end{malist}

It was observed that the modelling of the gluon emission was unable to
describe the event distributions of y$_{n~n-1}$ correctly 
for $n$ greater than 5.  To take into account
this imperfect description, corrections were applied
to the background distributions  of the y$_{n~n-1}$ variables (for $n$
between 6 and 10) \cite{these_VP}.

\subsubsection{Chargino search \label{char_udd}}

To take into account the effect of the mass difference between
chargino and neutralino, \dm, on the topology of the
event, the ten-jet analysis was divided into two windows: 
\begin{malist}
\item low \dm\ window C1:  5$\leq$~\dm$ \leq 10$~\GeVcc; 
\item high \dm\ window C2: \dm$ > 10$~\GeVcc.
\end{malist} 

Two neural networks were trained 
with the following discriminating variables as inputs:
\begin{malist} 
\item the thrust;
\item the variable dist$_{\rm{WW}}$ described above;
\item the energy of the least energetic jet multiplied by
the minimum di-jet angle in four and five-jet configurations;
\item y$_{n~n-1}$ with $n$=3 to 10 (for $n$=6 to 10, the corrected
y$_{n~n-1}$ were used); for the C1 analysis, the variables
y$_{87}$, y$_{98}$ and y$_{10\,9}$ were not used.
\end{malist}

\subsubsection{Slepton search \label{slep_udd}}

Three mass windows were defined to take into account the mass
difference, \dm, between the sfermion and the neutralino
considered as the LSP: 
\begin{malist}
\item low \dm\ window 1:  5~$\le \Delta$M~$ \le 10$~\GeVcc\ with m$_{\tilde{\chi}^0} > $ 55~\GeVcc;
\item high \dm \ window 2:  $\Delta$M~$ > 10$~\GeVcc\ with m$_{\tilde{\chi}^0} > $ 55~\GeVcc; 
\item low neutralino mass window 3:  m$_{\tilde{\chi}^0} \le $ 55~\GeVcc.
\end{malist}
Different selection criteria on the momentum of the tagged leptons were applied 
depending on the mass window.\\

\stitre{Electron and muon momentum selection}
\noindent
In addition to the high rejection power of the topological jet variables, lepton
identification has been used since two opposite sign leptons of the same 
flavour 
are produced in the final state (see Table~\ref{topos}). Therefore, 
an electron and positron, or two muons with opposite sign were required, with
thresholds on the momentum which depended on \dm,
in order to discriminate 
the selectron or the smuon pair-production signal  from the
SM  background:  
\begin{malist}
\item the momentum of the less energetic tagged lepton (electron or muon) had
to be lower than 30~\GeVc\ (window 1), 70~\GeVc \ (windows 2 and 3); 
\item the momentum of the more energetic tagged electron had to be in the
intervals [2,40]~\GeVc\ (window 1), [10,70]~\GeVc \ (window 2) and  [10,90]~\GeVc \ (window 3);
\item the momentum of the more energetic tagged muon had to be in the
intervals [2,40]~\GeVc\ (window 1), [30,70]~\GeVc \ (window 2) and  
[30,90]~\GeVc \ (window 3). \\
\end{malist}

\stitre{Neural network signal selection optimisation}
\noindent
The following variables
have been used as inputs to the neural networks:
\begin{malist}
\item the clustering variables y$_{43}$, y$_{54}$, y$_{65}$, computed with
the {\tt CAMBRIDGE} algorithm;
\item the minimum di-jet mass in the four, five and six-jet configurations;  
\item the energy of the least energetic jet $\cdot$ minimum di-jet angle
in four and five-jet configurations;
\item the thrust (only for window 3);
\item the energy of the most energetic electromagnetic cluster (only for window 3).
\end{malist}

The training was performed on signal samples of 
selectrons and smuons at centre-of-mass energies of 200 and 206~\GeV\
for each  analysis, and  with the same statistics for two samples of
the most important 
expected  SM backgrounds (two and four-fermion
events separately). \\

\subsubsection{Squark search \label{squa_udd}}

Searches for stop and sbottom were performed in the case of indirect
decays. The eight quarks event topology depends strongly on \dm, the
mass difference
between the squark and the \XOI.
The same mass windows as those defined for slepton analysis (Section 4.4.4)
were used. 
After the preselection 
and before training the neural networks,  additional criteria were  
applied to select high jet multiplicity events:   
\begin{malist}
\item the total  multiplicity had to be lower than  40;
\item the effective centre-of-mass energy had to be greater than  {\mbox{0.7$\cdot$\ecms}};
\item the total electromagnetic energy had to be lower than 20~\GeV \ (window 3 only);
\item the  ln(y$_{43}$) had to be greater than -6;
\item the  ln(y$_{54}$) had to be greater than -6.5;
\item the momentum of the less energetic tagged electron had to be lower
than 16~\GeVc \ (windows 1 and 2), 20~\GeVc \ (window 3); 
\item the momentum of the most energetic tagged electron had to be lower
than 40~\GeVc;
\item the energy of the most energetic electromagnetic cluster had to be lower
than 40~\GeV\  (window 3 only).
\end{malist}
Sbottom decays produce b--quarks in the final state which may be 
identified with the impact parameter information provided by the 
micro-vertex detector.  
The event tagging obtained with the DELPHI algorithm for
tagging events containing a b--quark~\cite{aabtag} was 
therefore added as a sequential cut for
the  sbottom analysis. 

The same input variables as in the selectron and smuon searches
(Section~\ref{slep_udd}) were used in the neural network, except for the low
 \dm\ analysis, where the energy of the most energetic electromagnetic
cluster was suppressed.    

\section{Results and limits}
In this section, the number of selected and expected events
after the final event selection,
and the signal
efficiencies obtained  
for each channel under study are presented.
The results are
in agreement with the SM expectation. 
Together with the signal efficiencies they were used to exclude at 95\% CL
 possible  regions of the MSSM parameter space.
Unless otherwise stated, the limits were derived using the results
from the centre-of-mass energies between 192 and 208~\GeV.

As already mentioned, to obtain the most
conservative constraints on the MSSM parameter values from
\LLE\ searches, only the analyses performed considering the \Lacc\ 
coupling as the dominant one were used:
in fact, if a different \Lijk\ coupling
is dominant, the exclusions would be at least as large as those from a
dominating \Lacc\ coupling.

After a presentation of the methods applied to derive limits,
the efficiencies and the number of  selected events are 
given for each channel, as well as the derived limits.

\subsection{Limit computation \label{lim_comp}}
\stitre{Limits on gaugino masses}
\noindent An upper limit to the number of signal
events, N$_{95}$, at 95\%~CL, was calculated  according to the monochannel
Bayesian method~\cite{bayesian} from the number of events 
remaining in the data and those expected 
in the SM, summed over all centre-of-mass  energies
from 192 to 208~\GeV.

The gaugino pair-production was considered for
different values of \tanb\ (from 0.5 to 30), m$_0$ 
(between 90~\GeVcc\ and 500~\GeVcc),
$\mu$ (between --200~\GeVcc\ and 200~\GeVcc) 
and M$_2$ (between 5 and 400~\GeVcc); 
for a given set of tan$\beta$ and  m$_0$ values 
the ($\mu$,~{\mtwo}) point 
was excluded at 95\% CL if the expected number of
signal, N$_{\rm exp}$ at this point was greater than  N$_{95}$.
The computations of N$_{\rm exp}$ were slightly
different for \LLE\ and \UDD\ searches, as detailed 
below. \\

To obtain the limits on the gaugino masses with a good precision,
special studies were performed to scan the regions of the
parameter space from which the limits were determined: the steps in
M$_2$ and $\mu$ were of 0.25~\GeVcc\ and 1~\GeVcc, respectively.  \\

\stitre{Limits on sfermion masses}
\noindent For all the sfermion searches
the limits at 95\%~CL were derived using the modified
frequentist likelihood ratio method~\cite{aread}. 
Expected exclusion limits
were obtained with the same algorithm where the number of observed events 
was set to the number of expected background events (absence of signal).    
To extract the mass limits,
a branching ratio of 100\% was assumed for the $R_{p}$-conserving
decay of the sfermion into a neutralino and a fermion. 
The MSSM values chosen to present
 the exclusion plots were tan$\beta$~=~1.5 and 
$\mu$~=~--200~\GeVcc.\\

The statistical errors on the efficiencies, which were between $\pm$1\% 
and $\pm$3\%, and on the expected background were used in the limit
computation, for gauginos and sfermions. The systematic uncertainties on the
signal selection efficiencies were negligeable compared to statistical errors
in the \LLE \ analyses. In the case of \UDD \ analyses, the systematic uncertainties
on the signal efficiences were larger. Indeed, the  hard gluon
radiation in  the parton shower of the Monte Carlo \UDD \ signal simulation 
is not implemented. Therefore this generates systematically events with
background-like y$_{mn}$ distributions. This is the reason why the \UDD \
results of the present search are conservative.     

\subsection{Gaugino searches} 
\subsubsection{\BLLE\ scenario}

\stitre{Efficiencies and selected events}
\noindent The efficiency of the selection described 
in Section~\ref{gaug_lle}
was computed from simulated samples 
at different points of the MSSM parameter space. 
In order to benefit from the high centre-of-mass energies and
luminosities, all \mbox{$\rm e^+ \rm e^-$ \Ra \XOi\XOj } and 
 \mbox{ $\rm e^+ \rm e^-$ \Ra \XPI\XMI }
processes which contribute
significantly have been simulated, at each \mbox{MSSM} point of this
study. Then a global event selection efficiency was determined for each
point.
The efficiencies lay between 11\% and 38\%.

At each selection step of the gaugino analysis,
good agreement between the number of observed and expected
background events 
was obtained, and no excess was observed in the data; at the end, 
24 candidates remained in the data from 192 to 208 GeV, 
compared to 23.7$\,\pm\,0.6$ expected from SM 
background processes (see Table~\ref{results}), mainly from \WW\ 
events and the rest from other four-fermion 
processes. \\

\begin{figure}[htb!]
\begin{center}
\epsfig{file=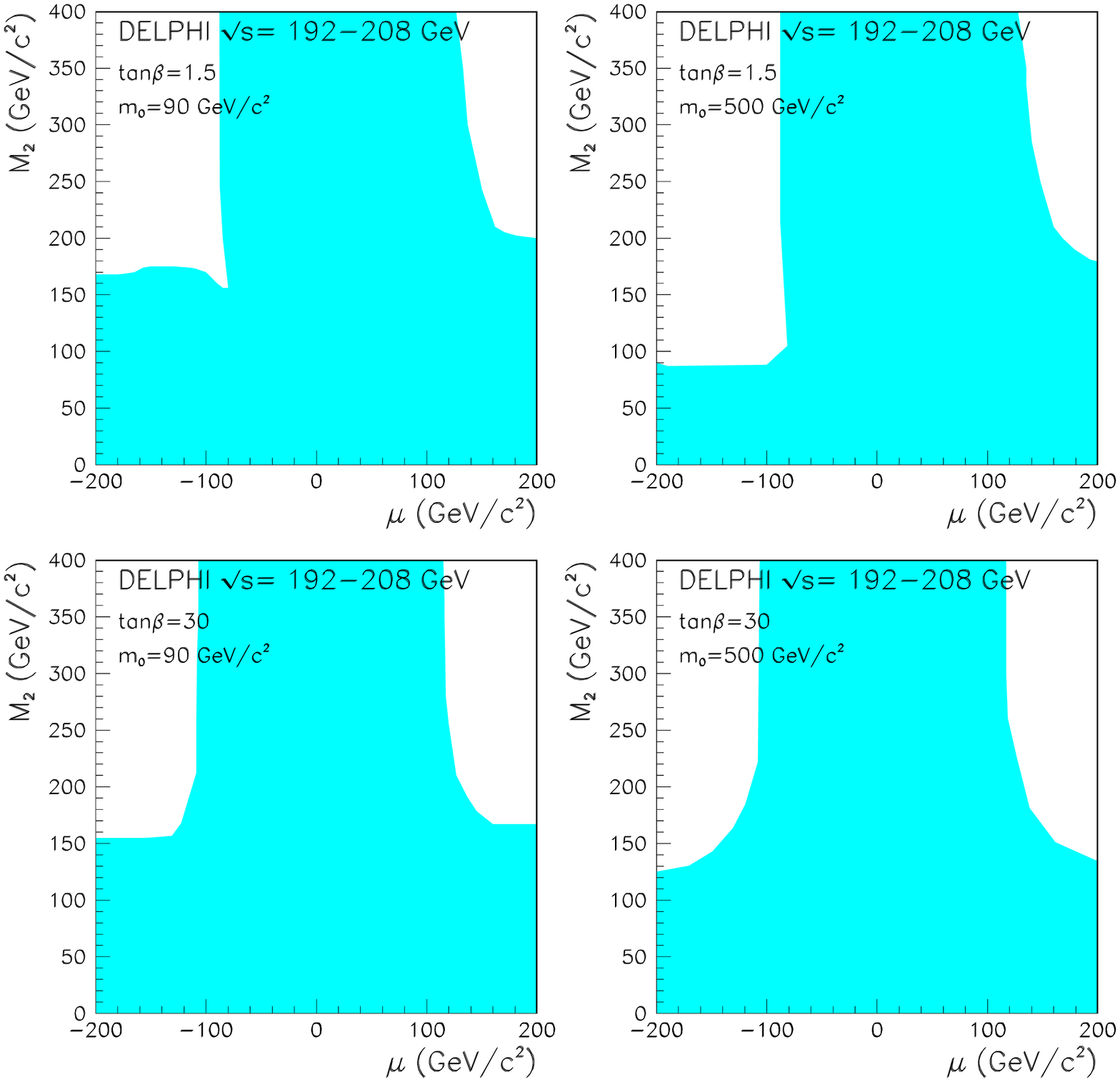,width=15cm}
\caption{\LLE: regions in $\mu$, M$_{\rm 2}$ parameter 
space excluded at 95\%~CL by the neutralino and chargino searches
for two values of \tanb\ and two values of m$_{0}$.}
\label{rp_lle_gau_exc}
\end{center}
\end{figure}

\stitre{Limits}
\noindent 
The number of expected events corresponding to gaugino pair-production
at each point of the explored MSSM parameter space was obtained by:
\begin{center}
N$_{\rm exp}=
{\large \boldmath \epsilon_{200}} \cdot \rm
\ \sum_{E_{cm}=192}^{E_{cm}={202}} $ \lum$_{\rm E_{cm}} \cdot \sigma_{\chi\chi}
+ {\large \boldmath \epsilon_{206}} \cdot \rm
\ \sum_{E_{cm}=203}^{E_{cm}={208}} $ \lum$_{\rm E_{cm}} \cdot \sigma_{\chi\chi}$
\end{center}
where $\sigma_{\chi\chi}=
 \sum_{i,j=1}^4 \sigma( \rm e^+ \rm e^-$ \Ra \XOi\XOj
$) + \sigma( \rm e^+ \rm e^-$ \Ra \XPI\XMI $)   $,
\lum$_{\rm E_{\rm cm}}$ is the integrated luminosity collected at
the centre-of-mass energy E$_{\rm cm}$, and
$\large \boldmath \epsilon_{200}$ and $\large \boldmath
\epsilon_{206}$ are the
global efficiencies determined as explained above 
at 200 and 206~\GeV\ respectively.
All points which satisfied N$_{\rm exp} > $N$_{95}$
were excluded at 95\% CL, and
the corresponding excluded area in ($\mu$,~M$_2$) planes obtained with the 
present searches are presented  in Figure~\ref{rp_lle_gau_exc}, for
m$_0$~=~90,~500~\GeVcc\ and tan$\beta$~=1.5, 30. 

\begin{figure}[htb!]
\begin{center}
\epsfig{file=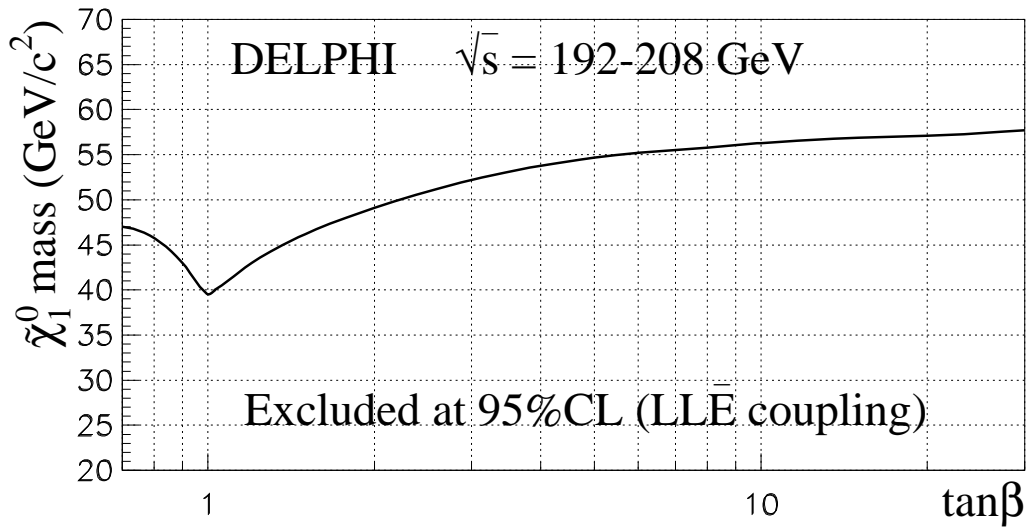,width=15cm}
\vspace{-0.4cm}
\caption{\LLE: excluded lightest neutralino mass as a function of
\tanb\ at 95\%~CL.
This limit is independent of the choice of the generation indices
$i$,$j$,$k$ of the $\lambda_{ijk}$ coupling and  
is for values of m$_0$ between 90 and 500~\GeVcc.}
\label{rp_lle_gau_lim}
\end{center}
\end{figure}

For each tan$\beta$, the highest value of the  mass of the lightest 
neutralino which can be excluded has been determined in the
($\mu$,~M$_2$) plane for several m$_0$~values from 90 to 500~\GeVcc; the most
conservative mass limit was obtained for high
m$_0$~values. The corresponding limit on neutralino mass as a function
of tan$\beta$ is shown in Figure~\ref{rp_lle_gau_lim}.\\
The same procedure has been applied to determine the 
most conservative lower limit on the chargino mass. The result is less 
dependent on tan$\beta$, and almost reaches the kinematic limit
for any value of tan$\beta$. 
The lower limit obtained on the neutralino mass is 39.5~\GeVcc, and
the one on the chargino mass is 103.0~\GeVcc. \\

\subsubsection{\BUDD~scenario} 
\begin{figure}[htb!]
\begin{center}
\epsfig{file=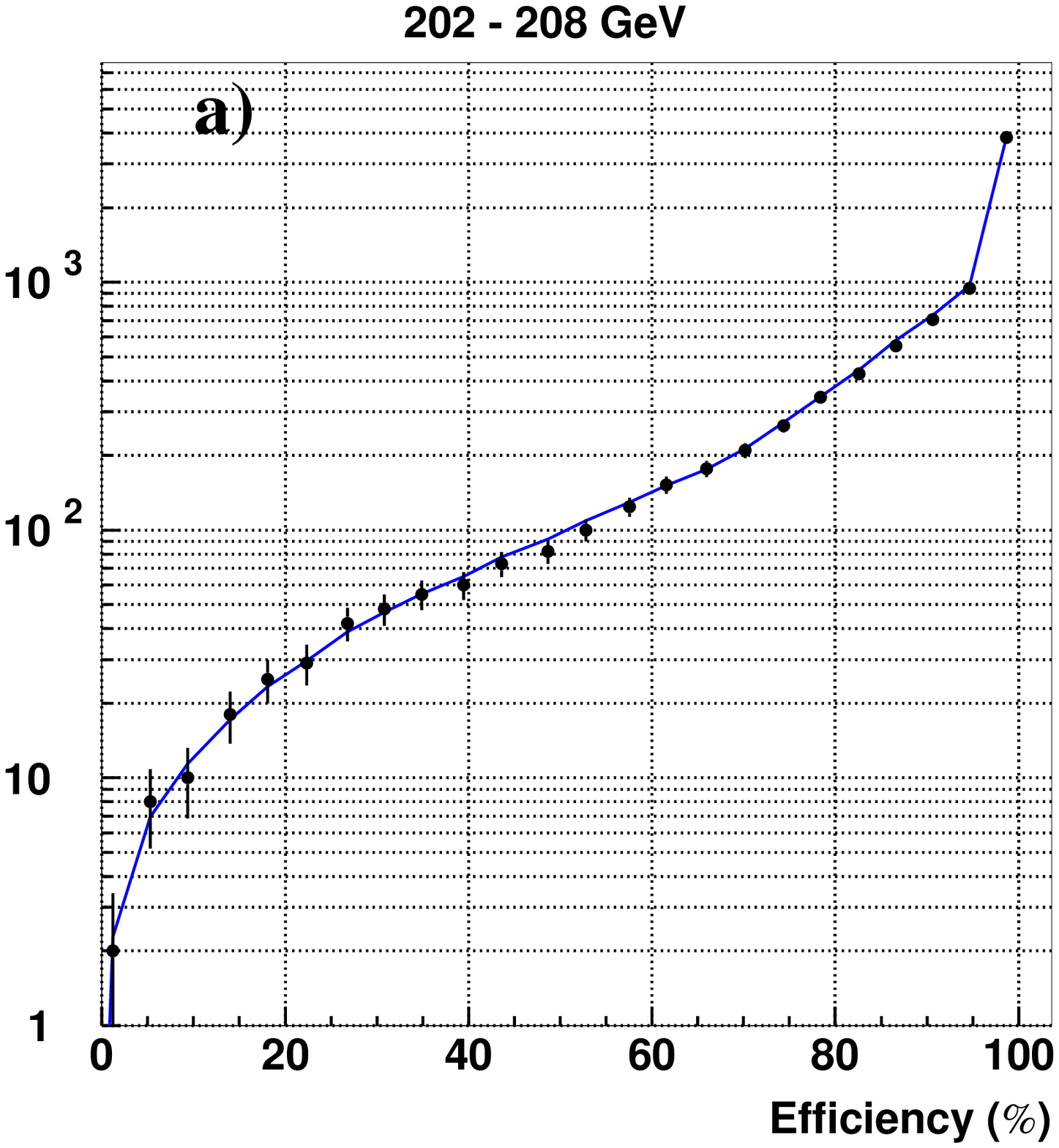,width=.47\linewidth,}  
\epsfig{file=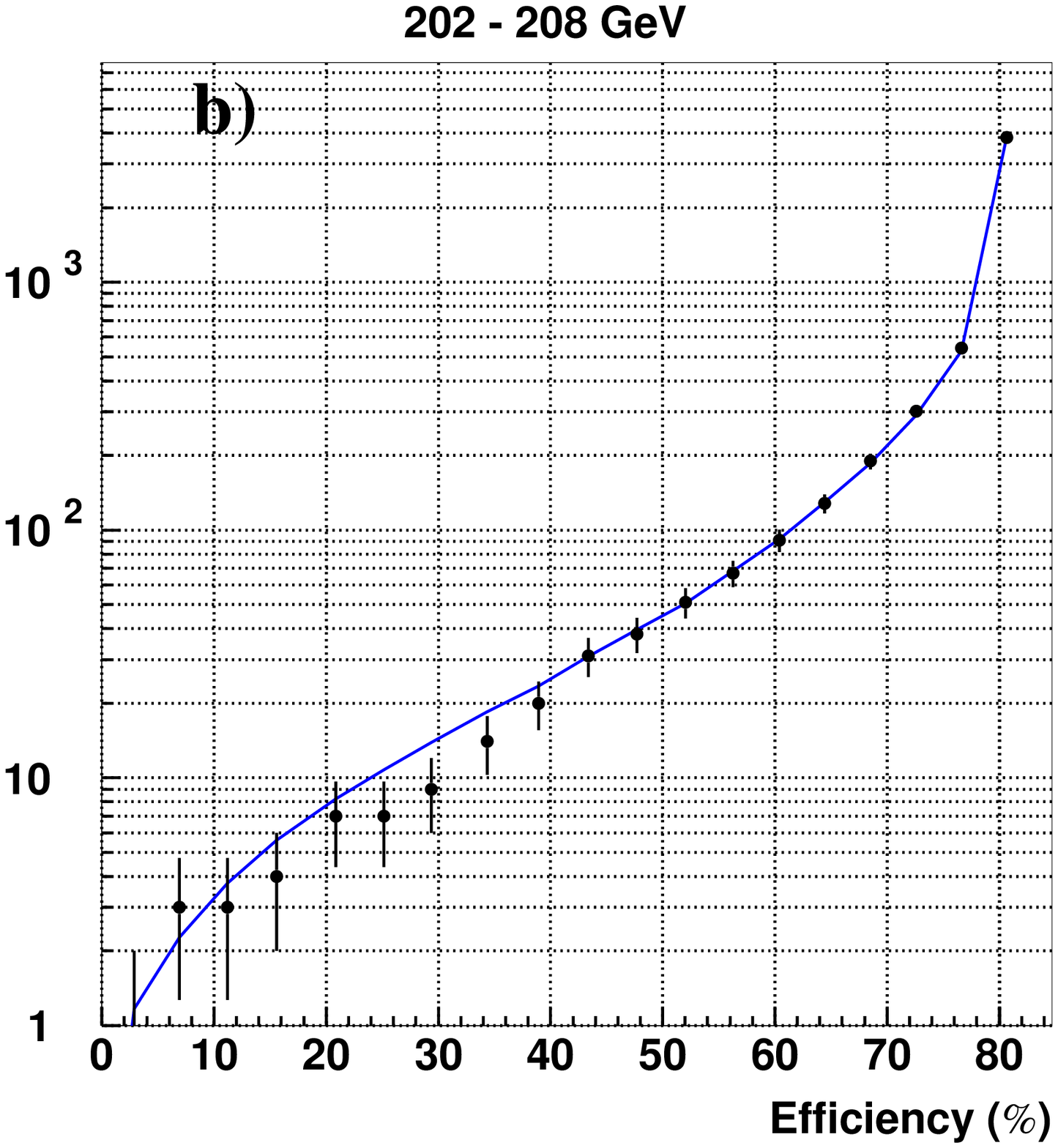,width=.47\linewidth,}  
\vspace{-0.4cm}
\caption{\UDD: number of expected events (continuous line) and data events (black dots)
versus average signal efficiency for the high neutralino mass search
N3 (plot a) and for the large \dm\ chargino search C2 (plot b)
for centre-of-mass energies between 202 and 208~GeV.}
\label{sortie_reseau}
\end{center}
\end{figure}

\stitre{Efficiencies and selected events}
\noindent At the end of the analysis procedure  to search
for gauginos described in Sections~\ref{neut_udd} and~\ref{char_udd},  
no significant excess of events was seen in the 
data with respect to the SM expectations.
Figure~\ref{sortie_reseau} shows the number of expected events from the 
SM and the number of
observed events as a function of the average signal 
efficiency obtained with all simulated masses for the N3 and C2 analyses
after a step-by-step cut on the neural network output.

For neutralino pair-production,
the efficiencies were typically 
around \mbox{30--60\%} at the values of the optimized neural network outputs, 
depending on the simulated masses. 
For chargino pair-production, 
the signal efficiencies were between 10\% and 70\%.
The expected and observed numbers of events for both analyses
are reported  in Table~\ref{results} for each mass window. \\

\stitre{Limits}
\noindent The signal efficiency for any values of the \XOI~and \XPM~masses
was interpolated using an efficiency grid determined with signal samples 
produced with the full DELPHI detector simulation. 
The number of expected events N$_{\rm exp} $ has been computed 
separatly for neutralino and chargino pair productions.  
\begin{center}
N$_{\rm exp}=
{\large \boldmath \epsilon_{200}} \cdot \rm
\ \sum_{E_{cm}=192}^{E_{cm}={202}} $ \lum$_{\rm E_{cm}} \cdot \sigma_{\chi\chi}
+ {\large \boldmath \epsilon_{206}} \cdot \rm
\ \sum_{E_{cm}=203}^{E_{cm}={208}} $ \lum$_{\rm E_{cm}} \cdot \sigma_{\chi\chi}$
\end{center}
where $\sigma_{\chi\chi}=
 \sigma( \rm e^+ \rm e^-$ \Ra \XOI \XOI
$) $ or $ \sigma( \rm e^+ \rm e^-$ \Ra \XPI\XMI $)   $, and
$\large \boldmath \epsilon_{200}$ and $\large \boldmath
\epsilon_{206}$ are taken from the efficiency grids.

\begin{figure}[htb!]
\begin{center}
\epsfig{file=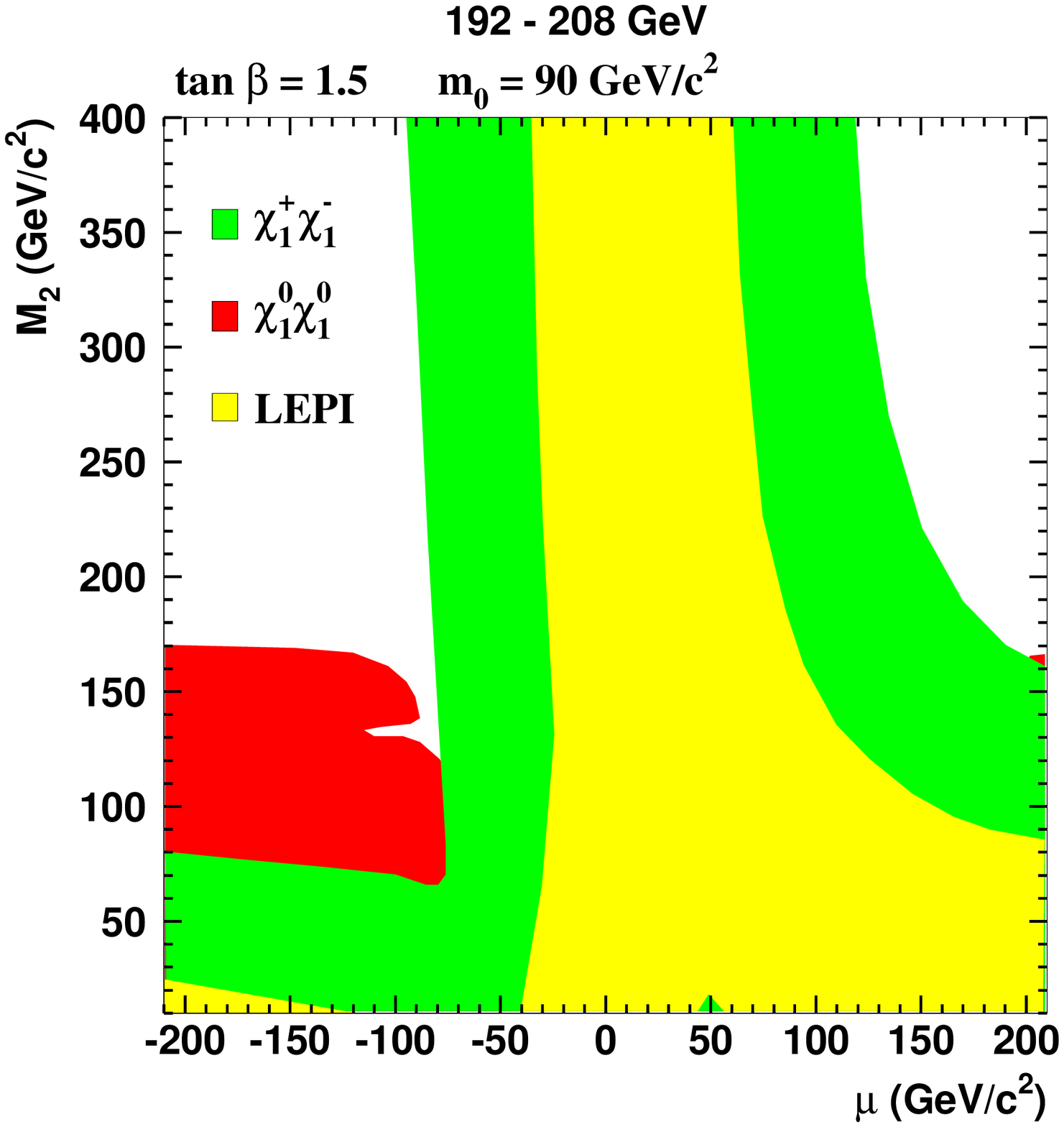,width=.4\linewidth}
\epsfig{file=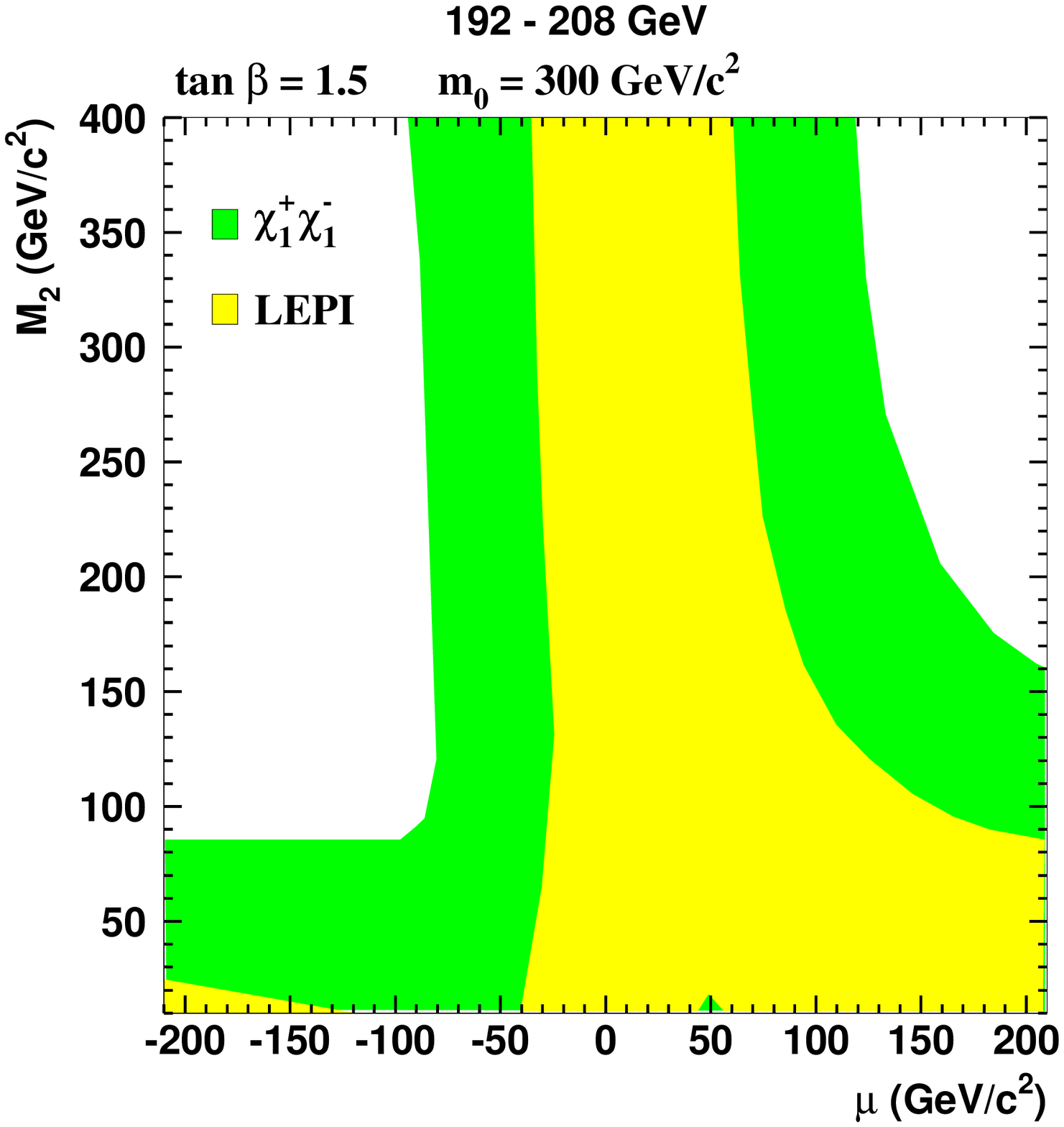,width=.4\linewidth}
\epsfig{file=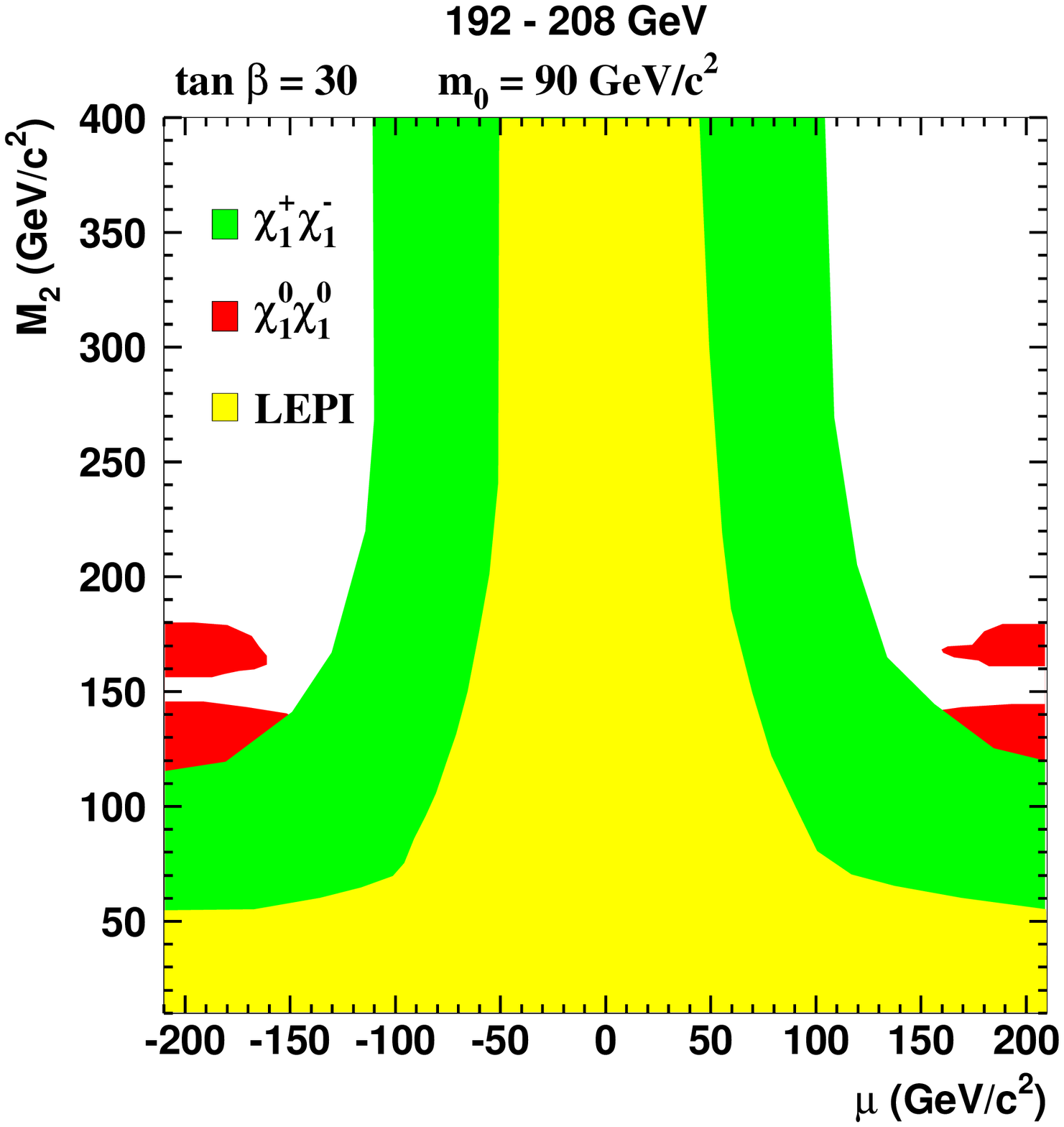,width=.4\linewidth}
\epsfig{file=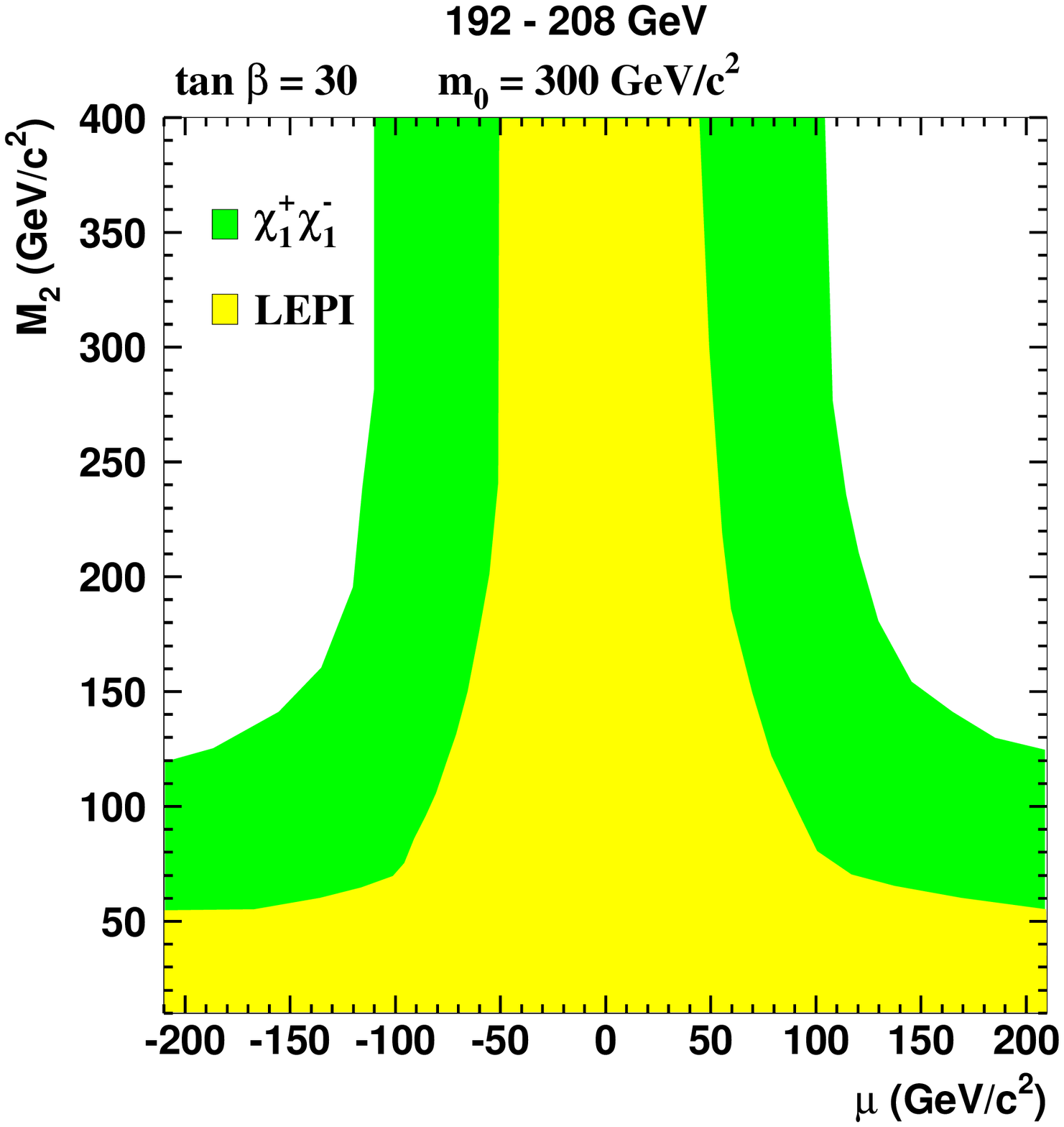,width=.4\linewidth}
\vspace{-0.4cm}
\caption{\UDD: regions in $\mu$, M$_{\rm 2}$ parameter 
space excluded at 95\%~CL by the neutralino and chargino searches
for two values of \tanb\ and two values of m$_{0}$.}
\label{exclu}
\end{center}
\end{figure}

\begin{figure}[htb!]
\begin{center}
\epsfig{file=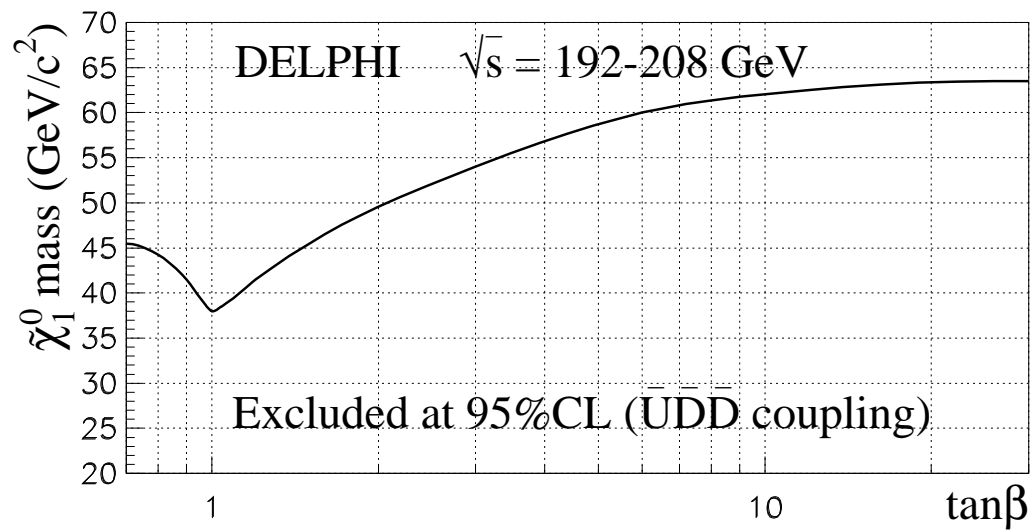,width=15cm}
\vspace{-0.4cm}
\caption{\UDD: lightest neutralino mass excluded  at 95\% CL 
as a function of tan$\beta$. This limit was obtained for m$_0$= 500~\GeVcc.}
\label{lowne1}  
\end{center}
\end{figure}

Using the six-jet  and the ten-jet analysis results,
an exclusion contours in the ($\mu$, M$_2$) plane 
at 95\% CL were derived for different values of 
m$_0$ (90 and 300~\GeVcc) and tan$\beta$ (1.5 and 30), as shown in
Figure~\ref{exclu}. 
In the exclusion plots the main contribution  comes 
from the study of the chargino indirect decays with the ten-jet analysis, 
due to the high cross-section. The six-jet analysis becomes crucial 
in the exclusion plot for low tan$\beta$, low m$_0$ and 
negative $\mu$. The  lower limits on the mass of the lightest
neutralino and chargino  are obtained from the scan on tan$\beta$  from
0.5 to 30. The lower limit on the neutralino mass of 38.0~\GeVcc\ is
obtained for tan$\beta$~=~1 and  m$_0$~=~500~\GeVcc\ 
(Figure~\ref{lowne1}). 
The chargino is mainly excluded up to the 
kinematic limit at 102.5~\GeVcc.

\subsection{Slepton searches} 

\subsubsection{\BLLE~scenario}
\stitre{Efficiencies and selected events}
\noindent  The efficiencies of the slepton analysis 
described in Section~\ref{slep_lle}
were between 18\% and 38\% 
for the sneutrino indirect decay channel, depending only on
the neutralino mass.
The efficiencies were higher for the final states obtained in indirect decay 
of charged 
slepton pairs, due to the presence of two additional leptons. They ranged 
from $\sim$~20\% (m$_{\widetilde{\chi}^0}$~=~15~\GeVcc) 
to 43\% for stau pairs; they were of the same order but up to $\sim$~5\%
higher  for selectron pairs, and ranged   
from $\sim$~25\% (m$_{\widetilde{\chi}^0}$~=~15~\GeVcc) 
to 64\% for smuon pairs.
For the direct decay of \snue, the analysis efficiencies lay 
in the  range 27--36\%, depending on the sneutrino mass. 

At the end of the selection,
11 events remained in the data compared to 8.1~$\pm$~0.3 expected from
the SM processes (Table~\ref{results}).
The background was mainly
composed of four--fermion events, in particular from 
W pair-production.

The efficiencies of the \asnut\snut\ direct decay analysis (see
Section~\ref{snut_lle})
varied with the $\tilde{\nu}_\tau$ mass and ranged from 
45\% to 51\%.
At the end of the selection, 6~candidates were 
obtained for 6.3~$\pm$~0.4 expected (see Table~\ref{results}).\\

\stitre{Limits}
\noindent To derive limits on slepton masses, 
the results of the search described  above
were combined with those obtained with data
at $\sqrt{s} = 189$~\GeV~\cite{lle189}.

\begin{figure}[htb!]
\begin{center}
\epsfig{file=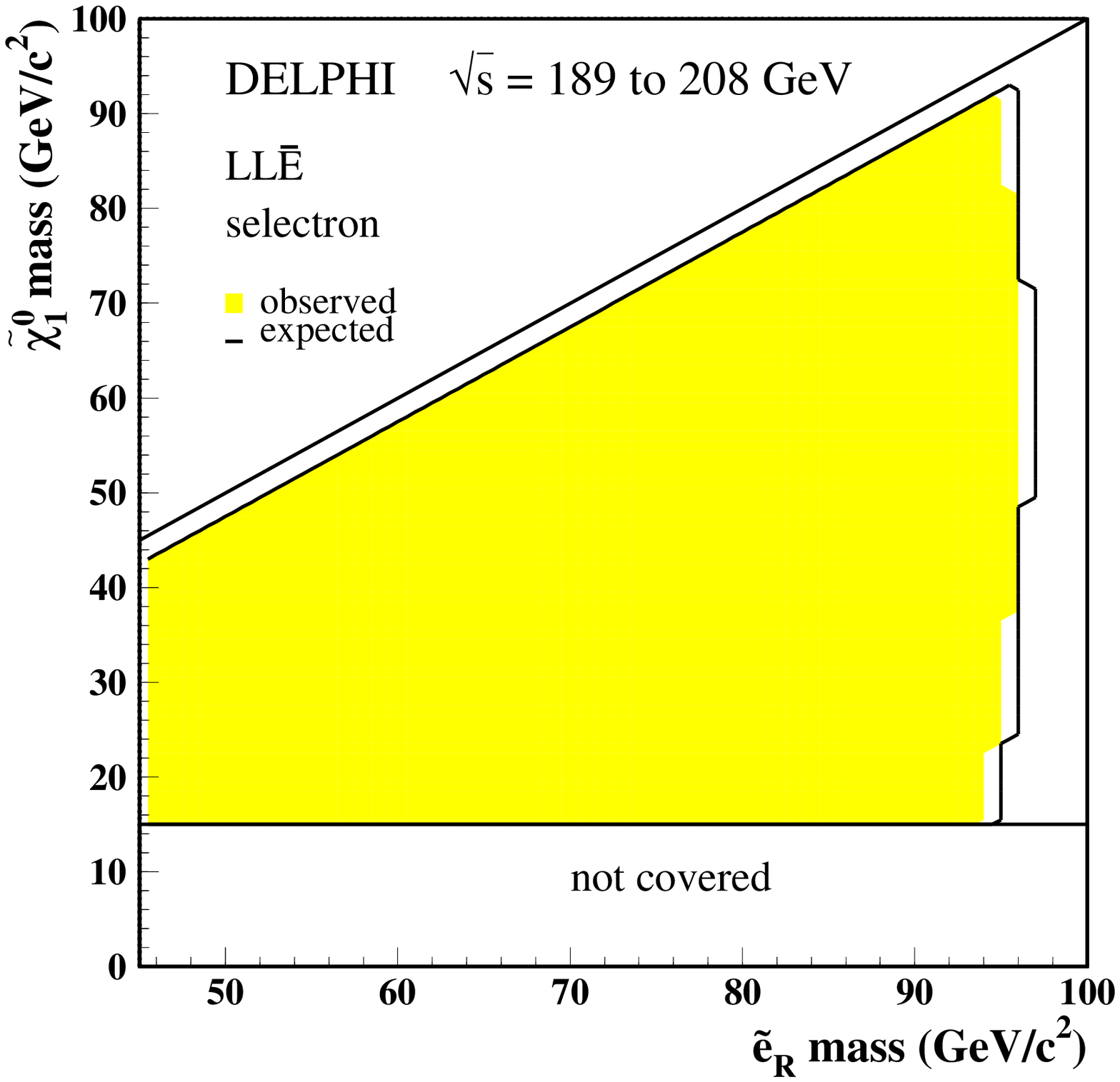,width=0.47\linewidth}
\epsfig{file=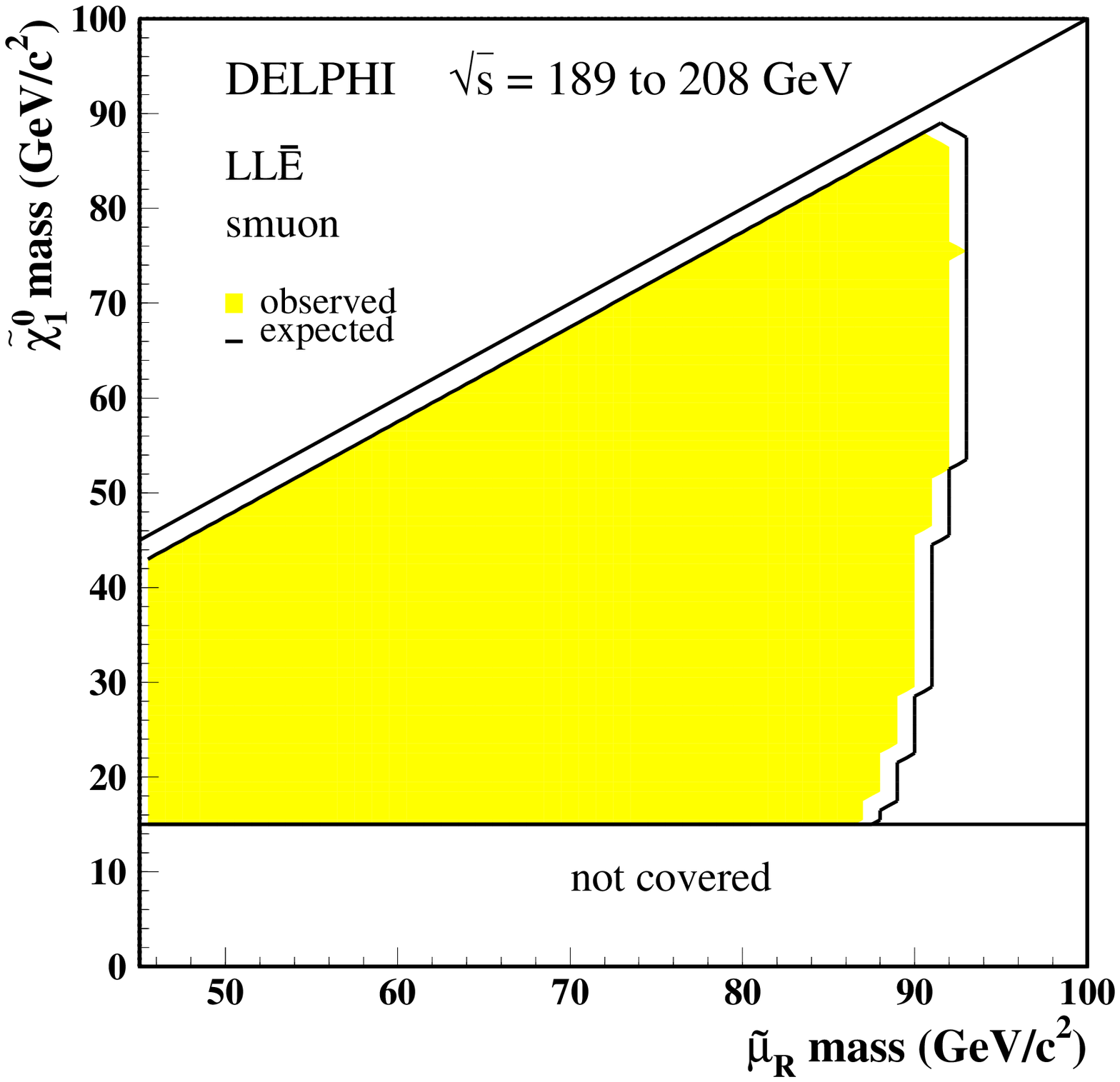,width=0.47\linewidth}
\epsfig{file=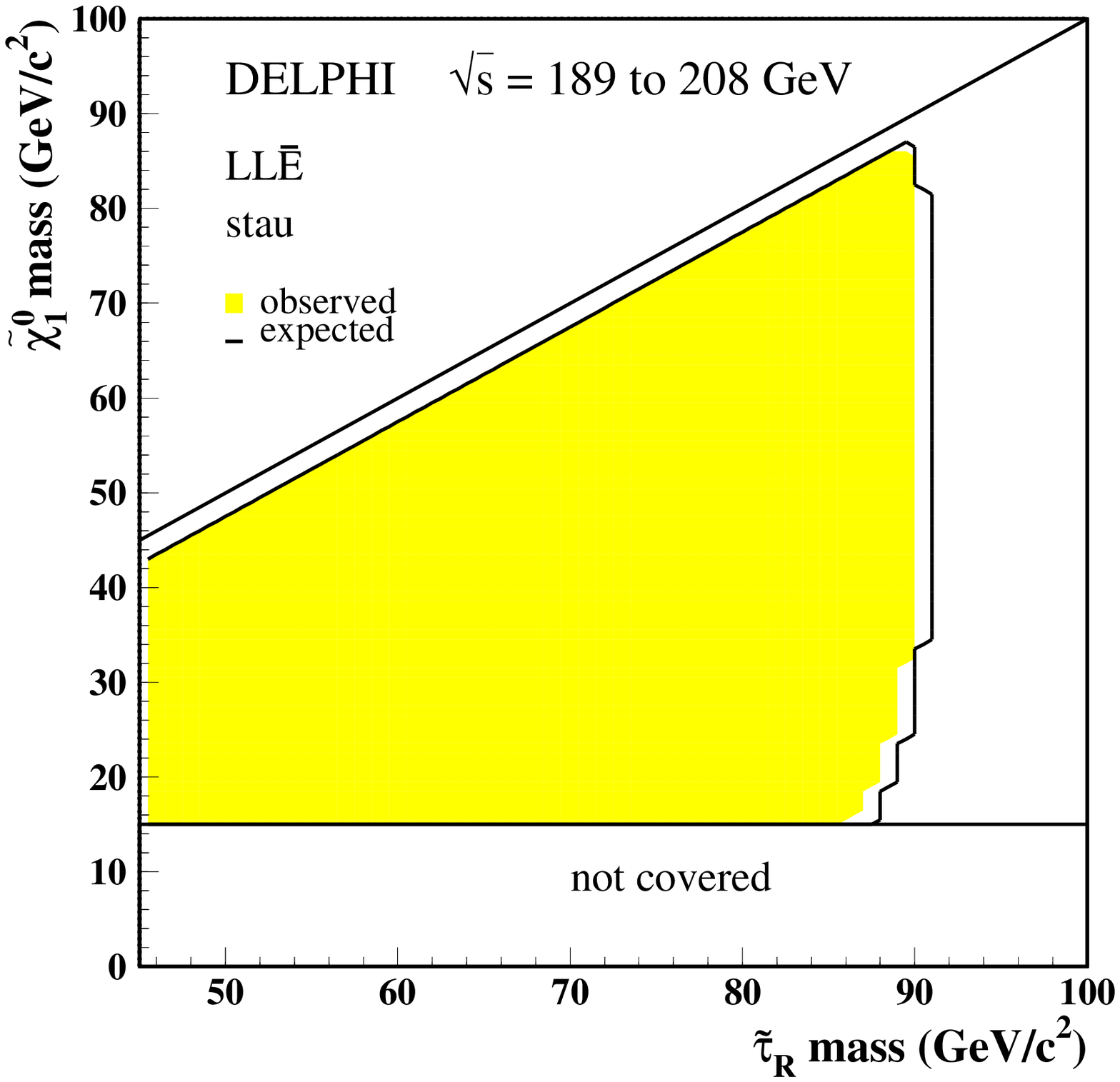,width=0.47\linewidth}
\caption{\LLE: excluded regions at 95\%~CL 
in the m$_{\tilde{\chi}^0}$ versus
m$_{\tilde{\ell}_R}$ planes
for \slepr\ pair-production, with
\mbox{BR(\slepr~\Ra \XOI~$\ell$) = 100\%},
and neutralino decay into leptons.
The  plots show the exclusion (filled
area) for selectron, smuon and stau. 
The black contour is the corresponding 
expected exclusion at 95\% CL.}
\label{rp_lle_slep}
\end{center}
\end{figure}

For charged slepton indirect decay, the areas  excluded  in the
m$_{\tilde{\chi}^0}$ versus m$_{\tilde{\ell}_R}$
planes are plotted in Figure~\ref{rp_lle_slep}. 

As was explained in Section~\ref{sec:pairprod}, a pair
of selectrons  can be produced in the $t$-channel via 
neutralino exchange. With the MSSM parameters fixed to derive 
limits, the \sel$^+$\sel$^-$ cross-section is higher than the 
\smu$^+$\smu$^-$ and \stau$^+$\stau$^-$
ones. So, though the analysis efficiencies for the smuon 
pair-production  were higher, 
the excluded area in case of the selectron pair search is the largest; 
the smallest is obtained for the \stau$^+$\stau$^-$ production.
For $\Delta$M~$\geq$~5~\GeVcc, the limits on the slepton mass are 
94~\GeVcc, 87~\GeVcc\ and 86~\GeVcc\ for the \sel, \smu\ and
\stau, respectively, and become 95~\GeVcc, 90~\GeVcc\ and 90~\GeVcc\
if the neutralino mass limit is taken into account.

\begin{figure}[htb!]
\begin{center}
\epsfig{file=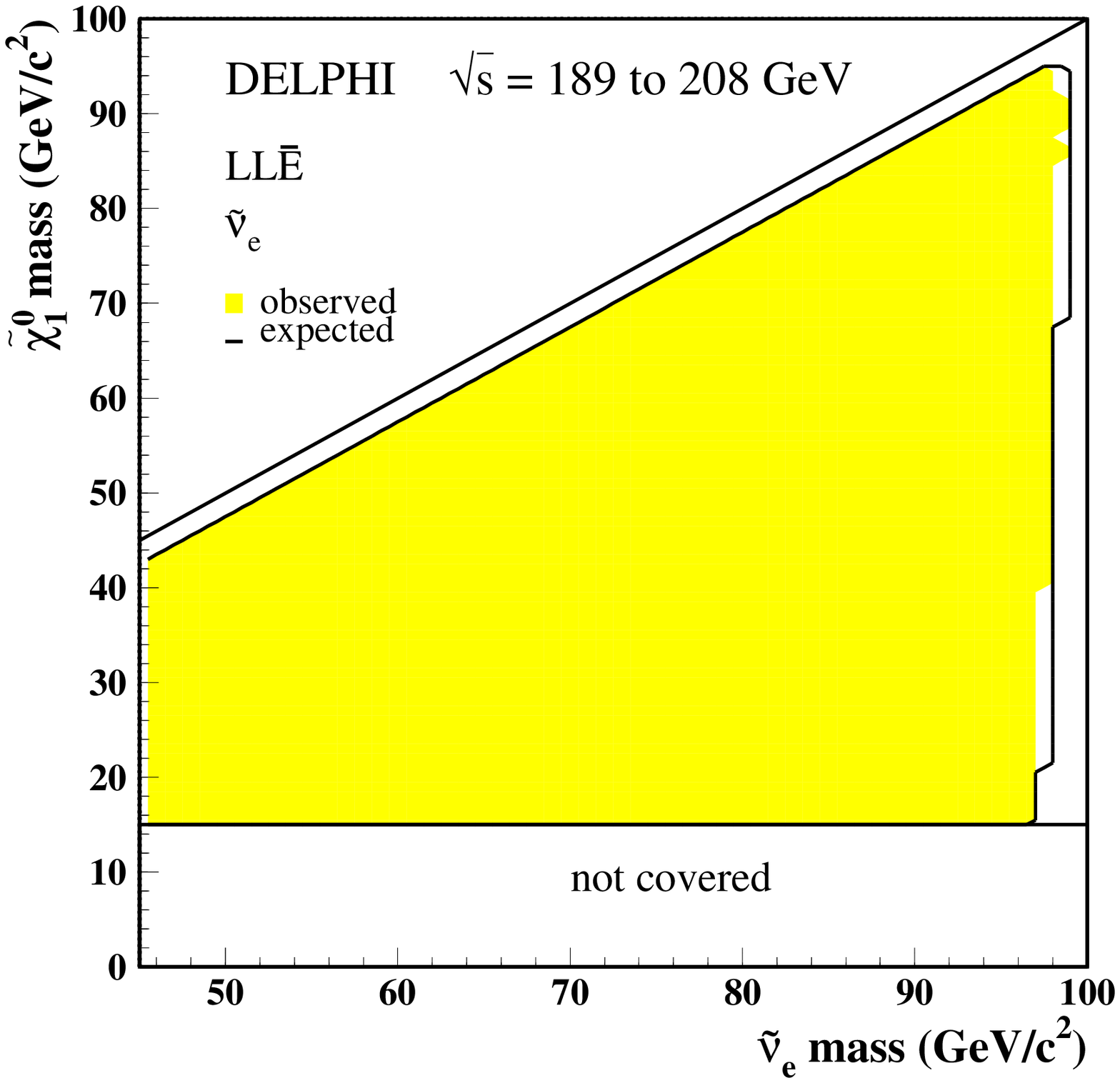,width=0.47\linewidth}
\epsfig{file=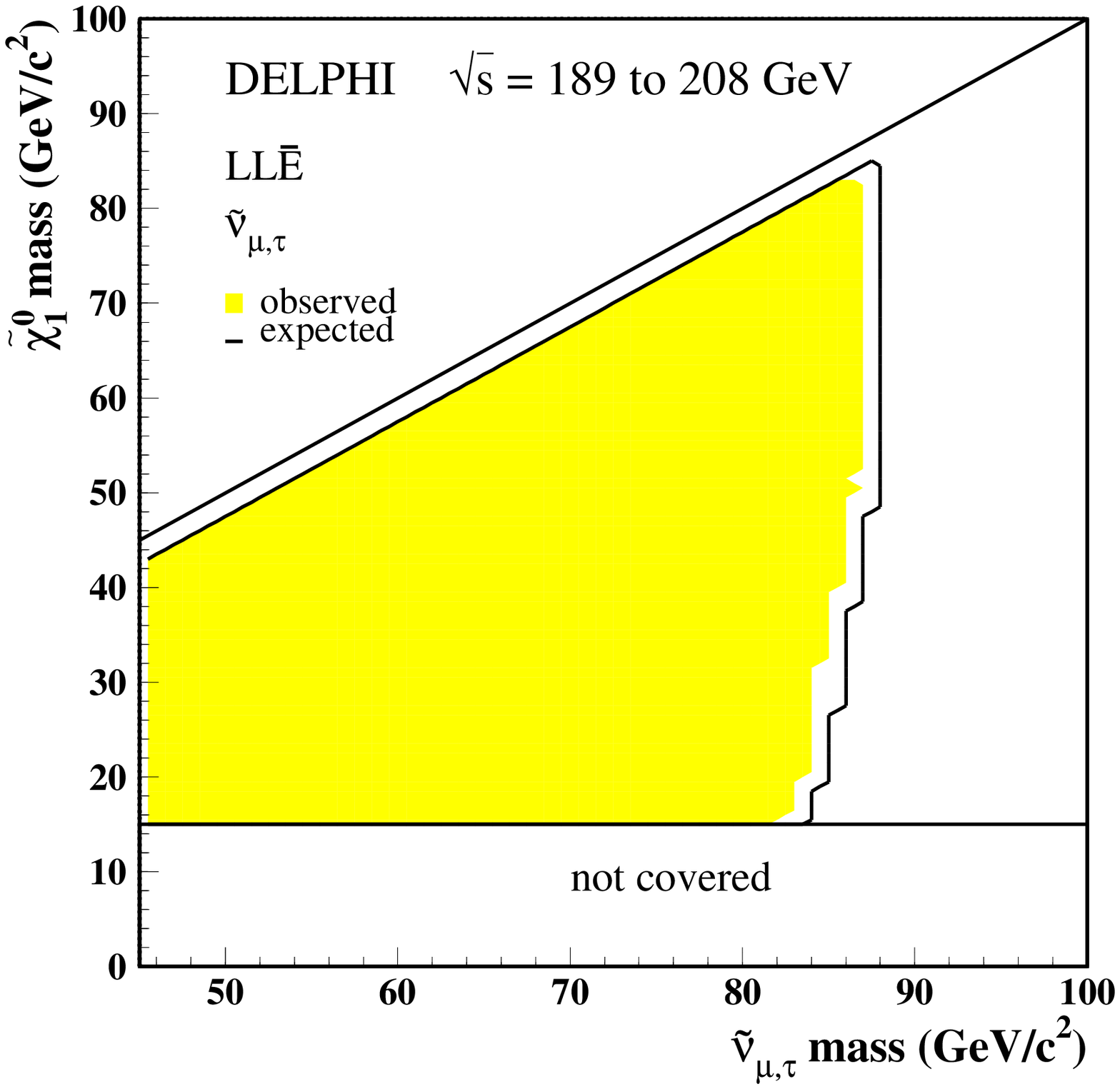,width=0.47\linewidth}
\caption{\LLE: excluded regions at 95\%~CL 
in the m$_{\tilde{\chi}^0}$ versus 
m$_{\tilde{\nu}}$ planes
for \snue\  (left) and \snum,\snut\ (right) pair-production, 
with \mbox{BR(\snu~\Ra \XOI~$\nu$) = 100\%},
and neutralino decay into leptons.
The black contour is the corresponding 
expected exclusion at 95\% CL.}
\label{rp_lle_snu}
\end{center}
\end{figure}

The results of the search for the indirect  decay of the \snu\ 
 were used to exclude  areas
in the m$_{\tilde{\chi}^0}$ versus m$_{\tilde{\nu}}$
planes, as shown in Figure~\ref{rp_lle_snu}. These exclusion areas 
are also valid for all the \Lijk\ couplings. 
As already mentioned in Section~\ref{sec:pairprod}, 
the \snue\asnue\ cross-section
can be enhanced compared to the \snum\asnum\ and \snut\asnut\
cross-sections if production via a chargino exchange
is possible and the excluded area depends on the chargino mass. 
 For $\Delta$M~$\geq$~5~\GeVcc,
the limit on the \snum\ and \snut\ mass is 82~\GeVcc, and 
is 85~\GeVcc\ if the limit on the neutralino mass is taken into
account. These limits are 96 and 98~\GeVcc\ respectively for \snue.

The results of the searches for 4$\tau$  and 2e2$\tau$ 
final states, from sneutrino pair direct decays,
were combined to obtain lower limits on the sneutrino mass. The results
from the 4$\tau$ search were used 
to derive limits on \snue\ 
and on \snum, those from the 2e2$\tau$ 
to derive limits on \snut. The limits obtained 
are respectively 96~\GeVcc, 83~\GeVcc\  and 91~\GeVcc.

\subsubsection{\BUDD~scenario} 

\begin{figure}[htb!]
\begin{center}
\vspace{-0.4cm}
\epsfig{file=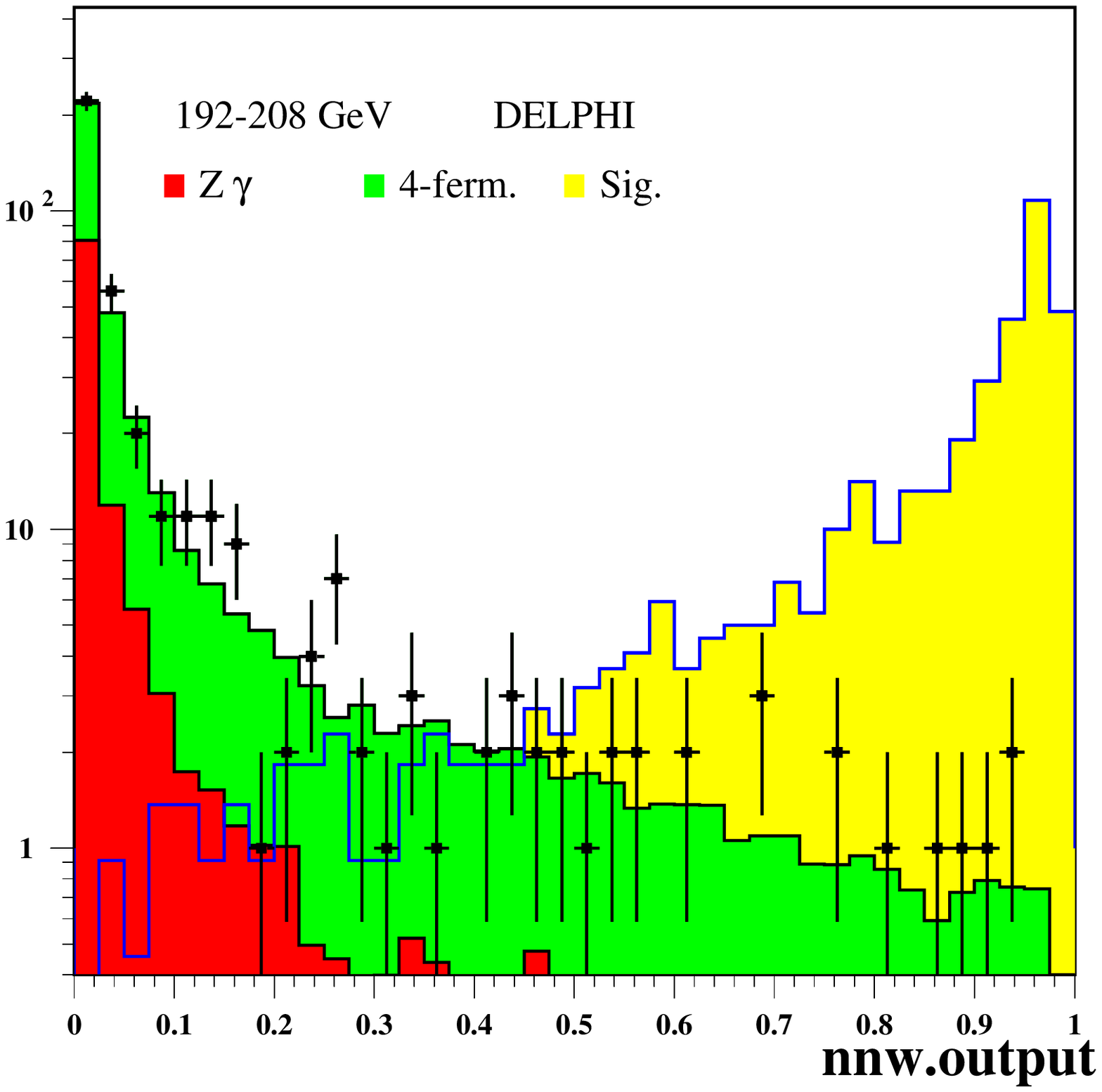,width=.47\linewidth}  
\epsfig{file=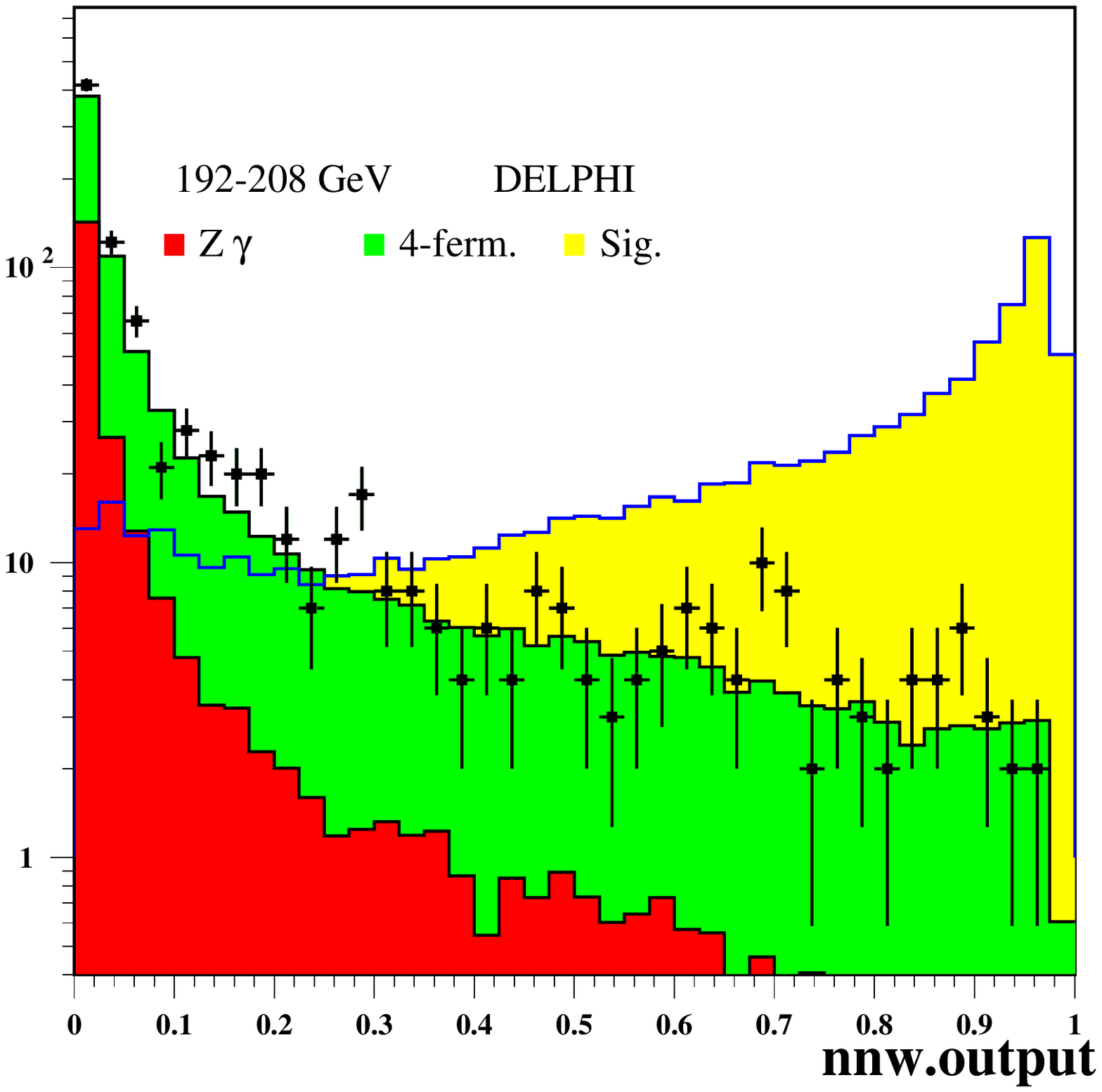,width=.47\linewidth}  
\caption{\UDD: the neural network signal output distributions 
for the selectron (left) and smuon (right) window 2 analyses.
The cuts on the neural network output variable
were chosen for the final selection at
0.83 (selectrons) and 0.92 (smuons).}
\label{udd_sle_nnw}


\vspace{1.5cm}
\epsfig{file=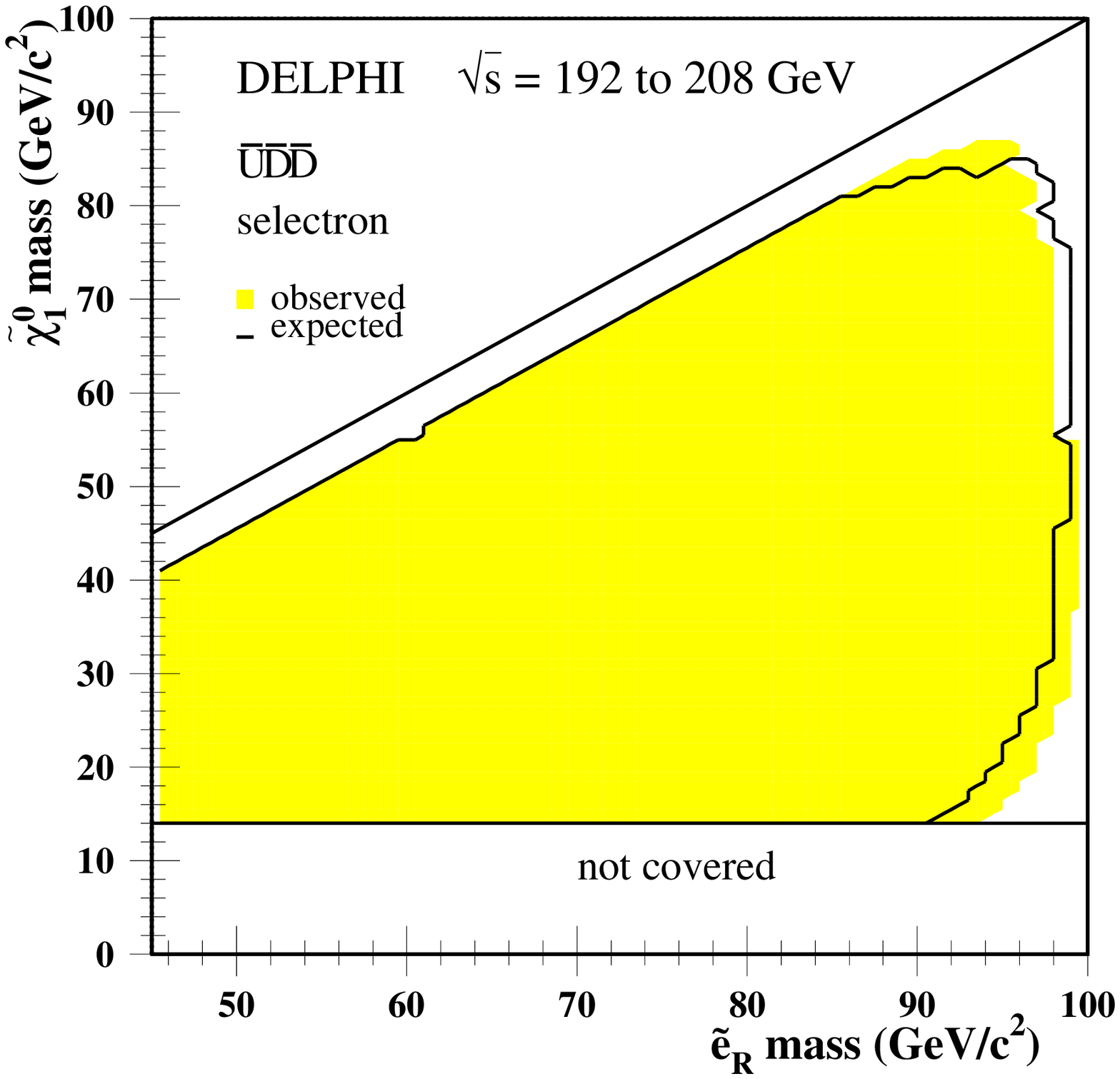,width=.47\linewidth,}
\epsfig{file=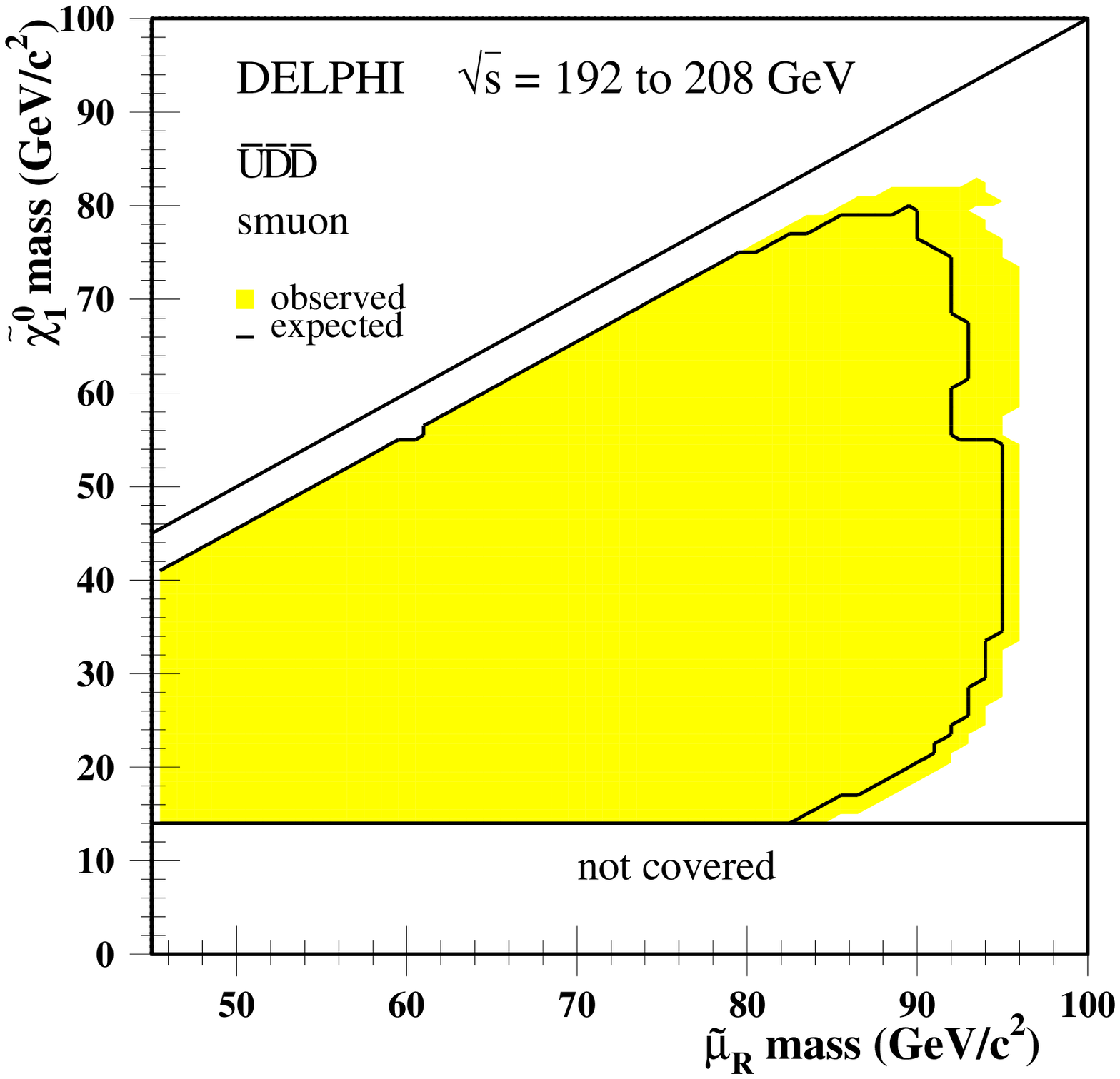,width=.47\linewidth,}
\caption{\UDD: excluded domains at 95\% CL in the 
m$_{\tilde{\chi}^0_1}$ versus
m$_{\tilde{\ell}}$ planes for  selectron 
(left) and for smuon (right) pair-production
 with \mbox{BR(\slepr\ \Ra \XOI\ $\ell$)~=~100\%} 
and  neutralino decay into jets (filled area).
The superimposed contours show
the expected exclusion at 95\% CL.}
\label{udd_sle_exc}
\end{center}
\end{figure}

\stitre{Efficiencies and selected events}
\noindent  
No significant excess has been observed in the output signal node
distributions for any analyses.
The  signal output distributions for the selectrons and smuons 
analyses are shown in Figure~\ref{udd_sle_nnw} for the medium
$\Delta$M analyses.
The signal efficiency of the slepton analyses 
described in Section~\ref{slep_udd}
was evaluated at each of the 
simulated points for the two centre-of-mass energies (200 and 206 \GeV). 
Efficiencies for the signal (selectron and smuon) 
were in the range from 5--40\%, for small mass differences and small
neutralino mass, 
and increased up to 60\% for medium \dm\ analyses. 
The 5\% efficiency was obtained for the \dm\ = 5 GeV and for a neutralino 
mass of 45 GeV. This efficiency increased rapidly with \dm \ and with the
neutralino mass. 

No excess of data with respect to the
SM expectations was observed for the selectron
and smuon analyses; the numbers of events observed and expected 
from background contributions are shown in Table~\ref{results}.
The  remaining background comes mainly from four-fermion processes.\\

\stitre{Limits}
\noindent From the  selectron and smuon pair-production searches
 exclusion domains have been computed  in the 
m$_{\tilde{\chi}^0}$ versus m$_{\tilde{\ell}_R}$ 
plane
(Figures~\ref{udd_sle_exc}). 
The M$_2$ value was fixed for each neutralino mass. 
For \mbox{$\Delta$M$ \geq  5$~\GeVcc},
the lower limit on the right-handed selectron mass
was 92~\GeVcc, and    
the lower limit obtained for the right-handed smuon 
was 85~\GeVcc.

\subsection{Squark searches} 

\subsubsection{\BLLE~scenario}
\stitre{Efficiencies and selected events}
\noindent  The selection efficiencies of the analysis
described in~\ref{squa_lle} varied 
with the stop mass and with the mass difference 
between the stop and the lightest neutralino. They lay
around 30\% for most of cases, 
and around 18--20\%  for low neutralino masses. 

After the selection procedure, 35 events remained  with 35.4~$\pm$~0.6 
expected from background contributions, mostly
coming from  \WW\ production.\\

\begin{figure}[htb!]
\begin{center}
\epsfig{file=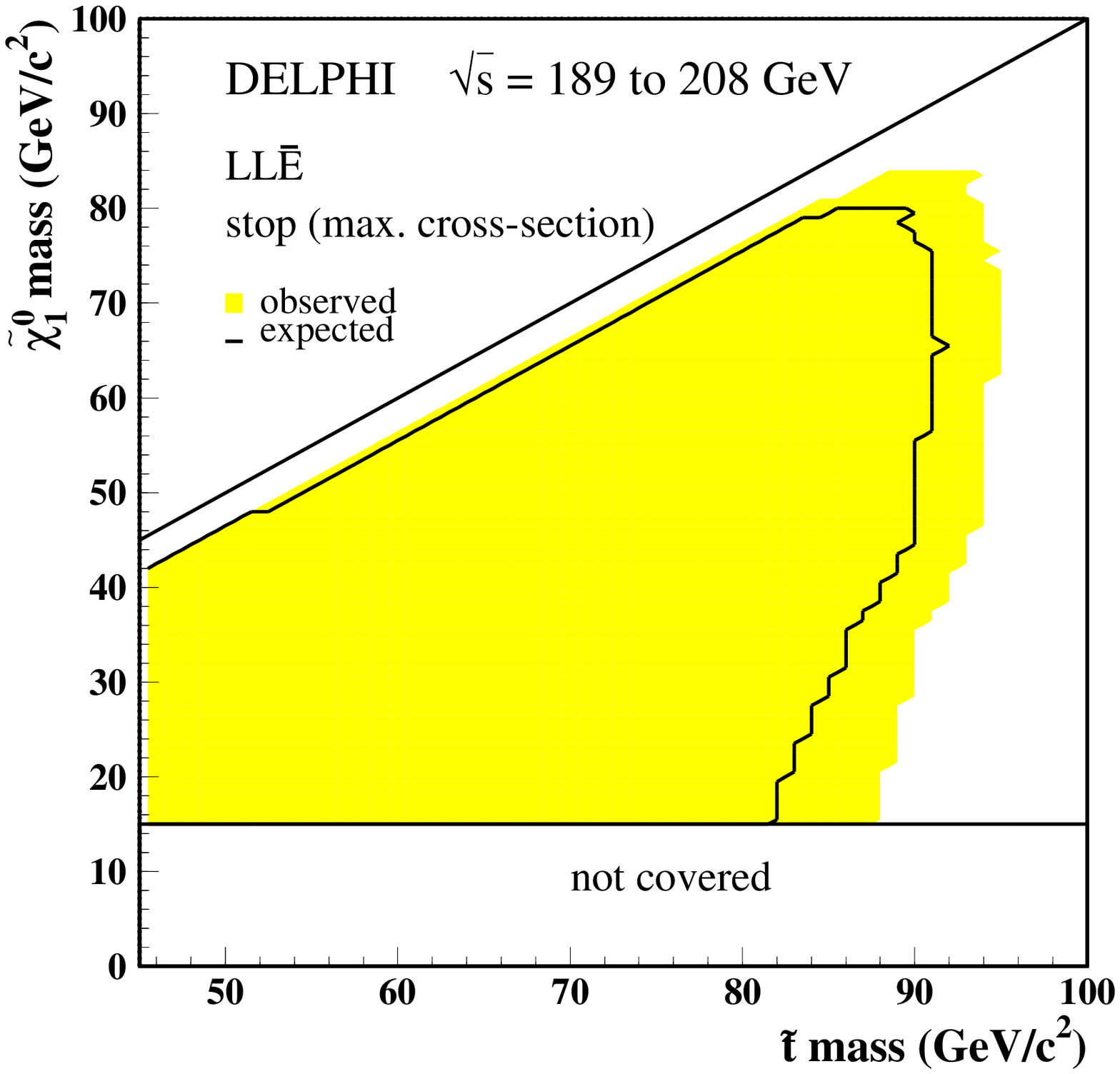,width=.47\linewidth}
\epsfig{file=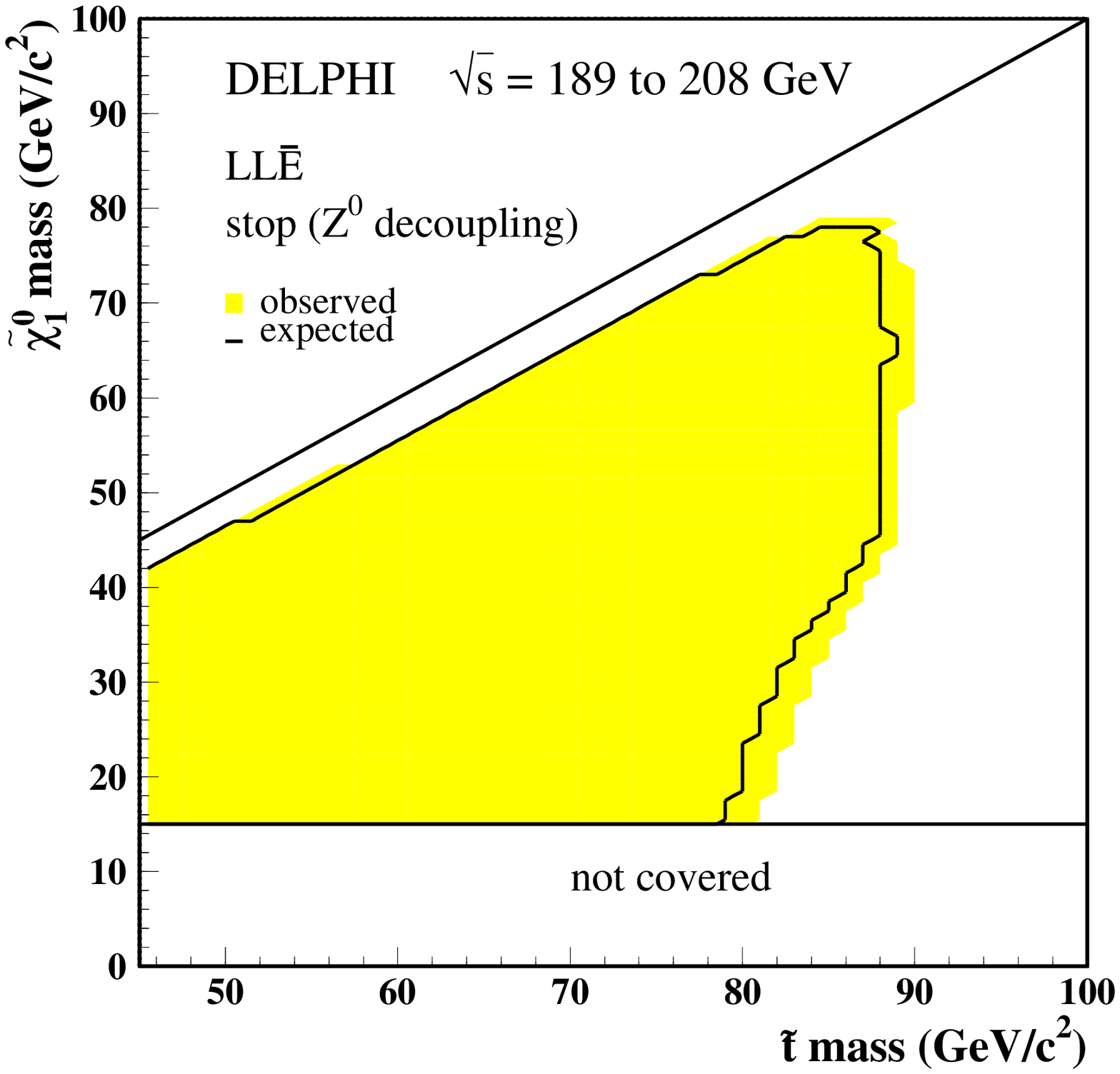,width=.47\linewidth}
\vspace{-0.5cm}
\caption{\LLE: exclusion domains at  95\% CL
in the m$_{\tilde{\chi}^0}$ versus m$_{\tilde{\rm t}}$ plane
for the stop pair-production, with
\mbox{BR(\stp~\Ra~c\XOI)~=~100\%}
and neutralino decay into leptons. The  plots show the exclusion (filled
area) for the lightest stop for no mixing (left) and 
for the mixing leading to the maximal decoupling to the
Z boson (right). The black contour is the corresponding 
expected exclusion at 95\% CL.}
\label{rp_lle_sto}
\end{center}
\end{figure}

\stitre{Limits}
\noindent 
From the study of the stop indirect decay, a
lower limit on the stop pair-production cross-section was derived
as a function of the  stop and neutralino masses. 
Considering two cases of stop mixing (no mixing and
mixing angle~=~56\mydeg), 
the exclusion limit was derived in the 
m$_{\tilde{\chi}^0_1}$ versus m$_{\tilde{\rm t}}$ 
plane, as shown in Figure~\ref{rp_lle_sto}.
With no mixing, the lower bound on the stop mass 
is 88~\GeVcc, valid for \dm~$> 5$~\GeVcc.
If the mixing angle is~56\mydeg,  the lower bound on the stop mass 
is 81~\GeVcc, for \dm~$> 5$~\GeVcc, and
becomes 87~\GeVcc, taking into account the lower limit on the
mass of the lightest neutralino.

\subsubsection{\BUDD~scenario} 

\stitre{Efficiencies and selected events}
\noindent  

\begin{figure}[htb!]
\begin{center}
\vspace{-0.4cm}  
\epsfig{file=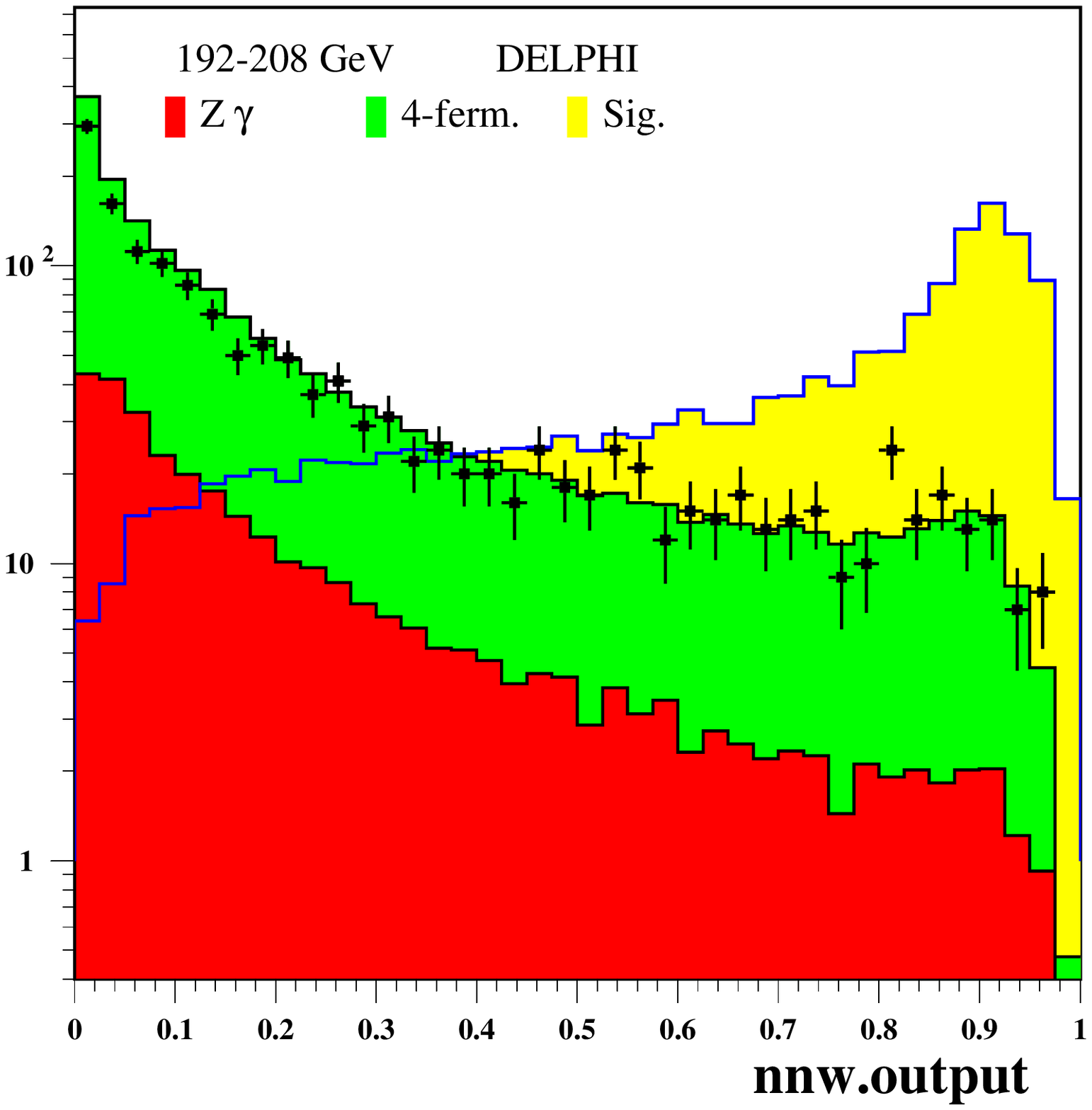,width=.47\linewidth}  
\epsfig{file=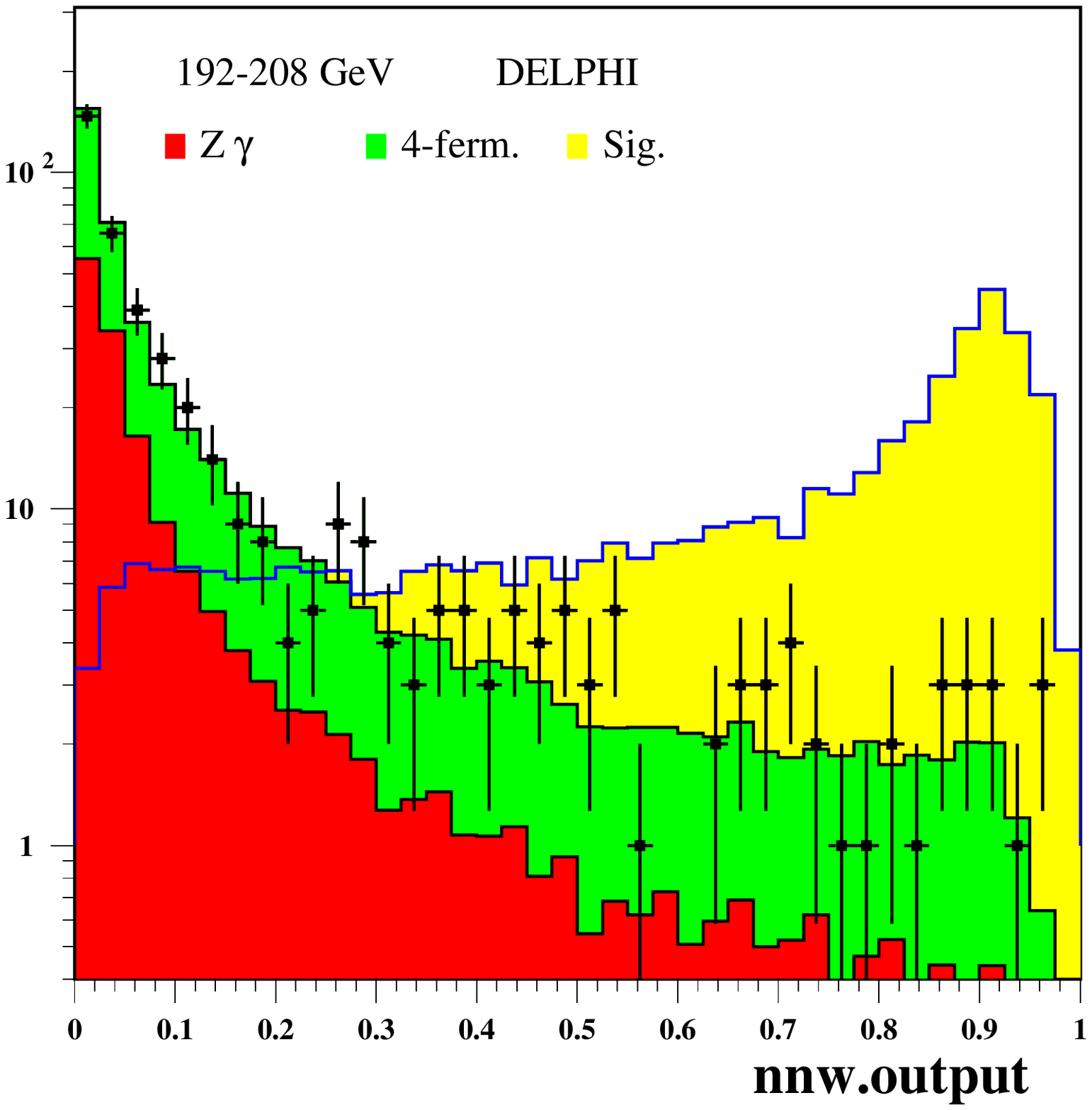,width=.47\linewidth}  
\caption{\UDD: the neural network signal output distributions
for the multi-jet stop (left) and sbottom (right)  window 3 analyses.
The cuts on the neural network output variable 
were chosen for the final selection at 0.86 and 0.75
for stop and sbottom respectively.}
 \label{udd_sqa_nnw}
\end{center}
\end{figure}

The final selection of candidate events was based on the  signal output values of 
the neural networks for the stop and for the sbottom. The
cut on the neural network variable has been relaxed for sbottom due to the
effect of the b-tagging selection.    
The  signal outputs 
of the neural network for the multi-jet stop and sbottom
analyses (window 3) are shown in
Figure~\ref{udd_sqa_nnw}. 
The signal efficiencies of the neural network analyses 
described in~\ref{squa_udd}
were evaluated at each of 
the evenly distributed 
simulated points in the plane of stop (sbottom) and neutralino masses. 
Efficiencies for the signal after the 
final selection were in the range from 10--20\%, for small or large mass
differences between squark and neutralino, up to around 50\% for medium
mass differences.

The numbers of events observed and expected 
from SM processes are shown in Table~\ref{results}.\\

\begin{figure}[htb!]
\begin{center}
\vspace{1.5cm}  
\epsfig{file=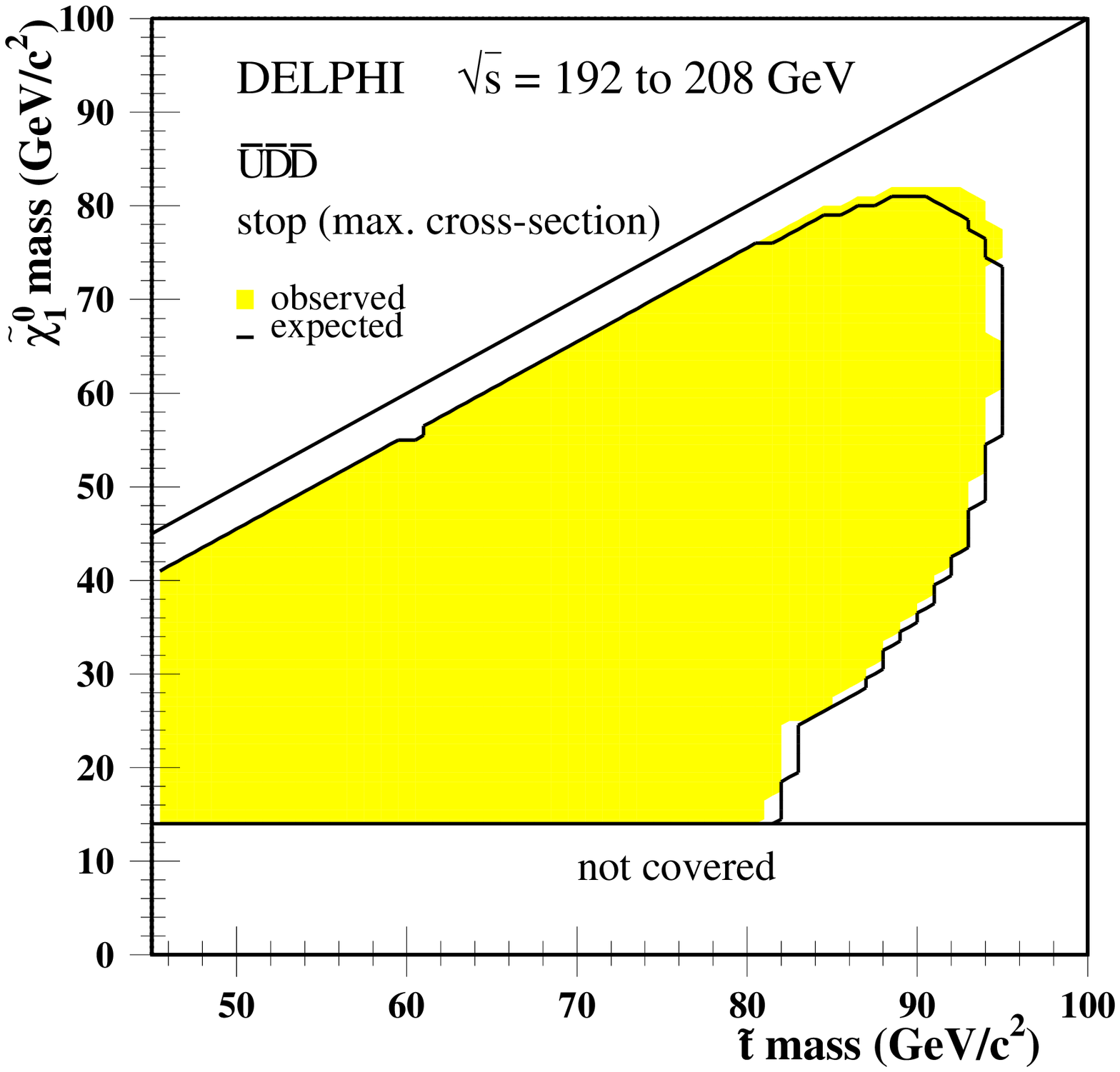,width=.47\linewidth,}
\epsfig{file=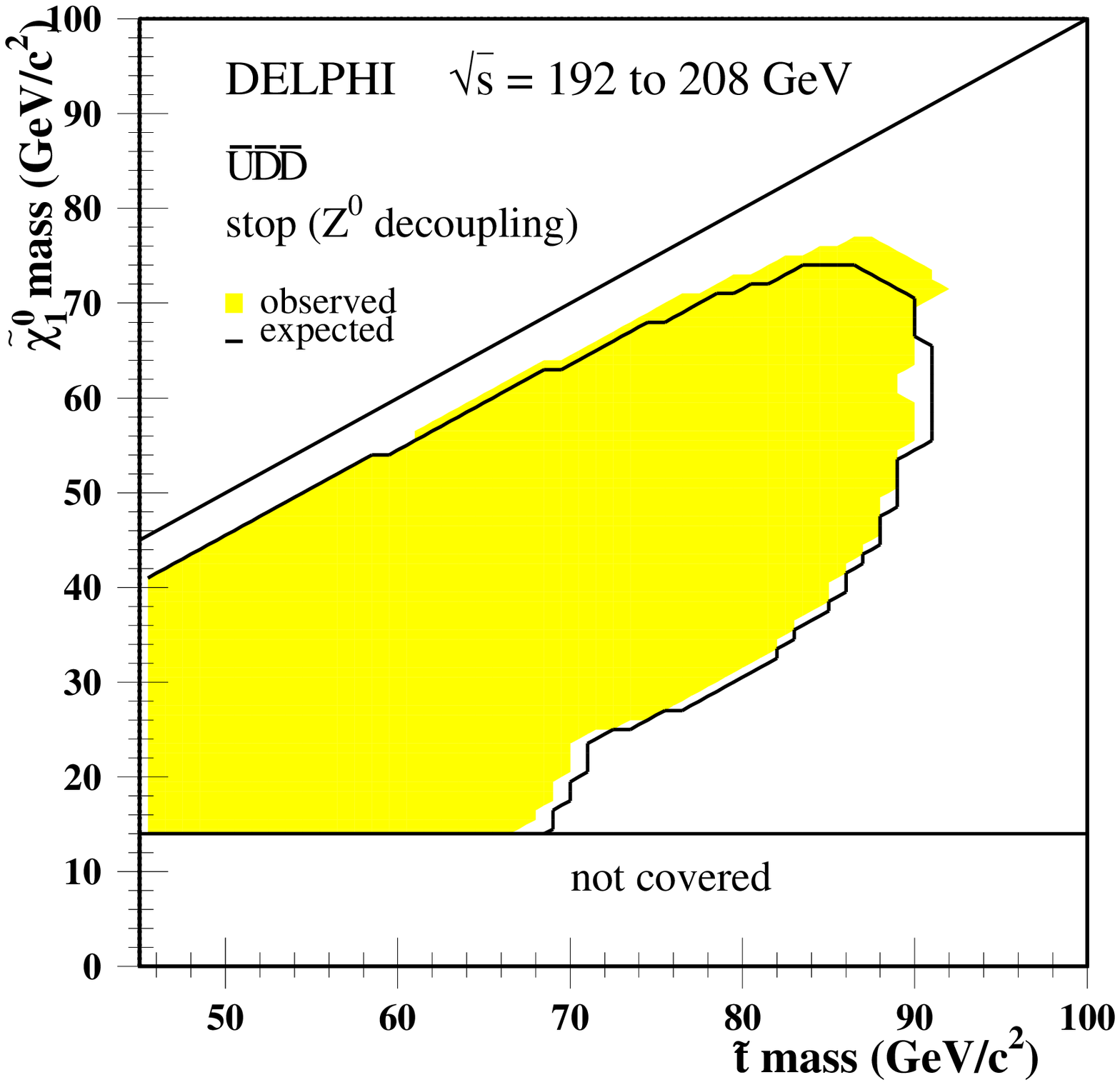,width=0.47\linewidth}
\caption{\UDD: exclusion domains at 95\%~CL in the 
m$_{\tilde{\chi}^0_1}$ versus  m$_{\tilde{\rm t}}$ plane  
for the stop pair-production, with \mbox{BR(\stp~\Ra~c\XOI)~=~100\%} 
and neutralino decay into jets.
The  plots show the exclusion (filled area) for 
the lightest stop for no mixing (left) and for
 the mixing leading to the maximal decoupling to the
Z boson (right). The black contour is the corresponding expected
exclusion at 95\%~CL.}
\label{indexc}

\vspace{1.5cm}  
\epsfig{file=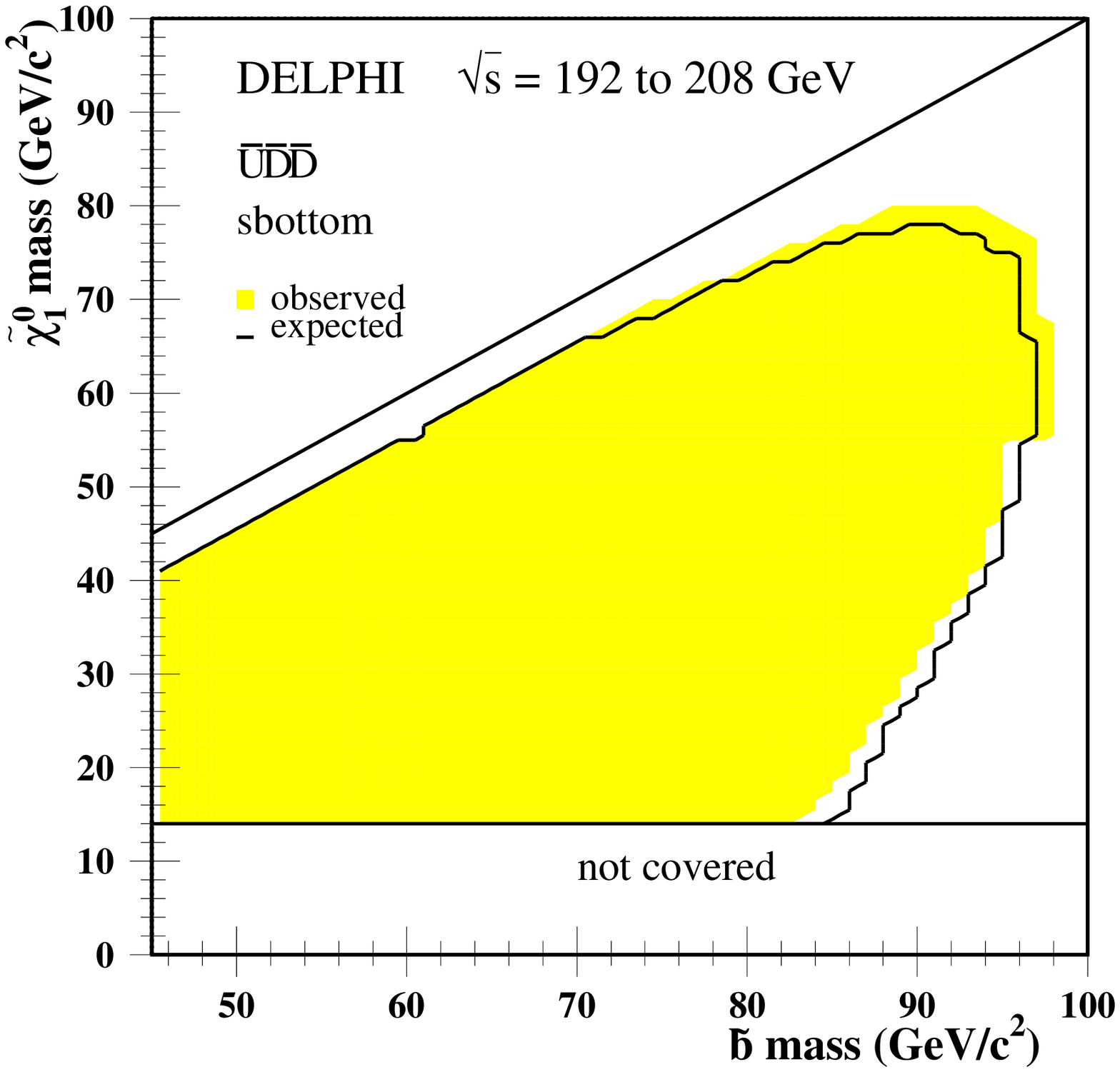,width=.47\linewidth,}
\caption{\UDD: exclusion contours at 95\% CL in the 
m$_{\tilde{\chi}^0_1}$ versus  m$_{\tilde{\rm b}}$ plane
for the sbottom pair-production, with \mbox{BR(\sbt~\Ra~b\XOI)~=~100\%} 
and neutralino decay into jets.
The plot shows the obtained  exclusion (filled area)
for the lightest sbottom in the case of no mixing. 
The black contour is the expected exclusion at 95\% CL.}
\label{indexcbo}
\end{center}
\end{figure}

\stitre{Limits}
\noindent The resulting exclusion contours for stop and sbottom can be 
seen in Figures~\ref{indexc} and~\ref{indexcbo}. 
The lower limit on the mass of 
the left-handed stop, 
assuming the neutralino
mass limit of 38~\GeVcc,   
is  87~\GeVcc\ for $\Delta$M~$ \geq 5$~\GeVcc.  
The lower limit on the stop mass 
with the mixing angle~=~56\mydeg\  
was  77~\GeVcc\ under the same assumptions.  
The lower limit obtained for the left-handed sbottom assuming a neutralino
mass limit of 38~\GeVcc, 
was  78~\GeVcc\ for $\Delta$M~$ \geq  5$~\GeVcc.  
The sbottom pair-production cross-section 
in the case of the mixing angle corresponding to maximum decoupling from the
Z boson (68\mydeg\ for sbottom) was
too low to cover a significant region of the mass 
plane by the excluded cross-section of the sbottom analyses.

\section{Summary}
A large number of different searches for 
supersymmetric particles  
with the assumption of {\rp} violation 
via \Lijk\LiLjEk\ or \Lppijk\UiDjDk\ terms
have been performed on the data
recorded in 1999 and 2000 by the DELPHI experiment,
at centre-of-mass energies between 192 and 208 GeV.
No significant excess has been observed in any of
the channels.
Limits on the pair-production of sparticles have been 
derived. These limits were converted into limits on
sparticle masses and excluded regions in the MSSM parameter
space.
Mass limits are
summarized in Table~\ref{rpv_limits}, together with the assumptions
under which
these limits are valid. 

\newpage
\subsection*{Acknowledgements}
\vskip 3 mm
 We are greatly indebted to our technical 
collaborators, to the members of the CERN-SL Division for the excellent 
performance of the LEP collider, and to the funding agencies for their
support in building and operating the DELPHI detector.\\
We acknowledge in particular the support of \\
Austrian Federal Ministry of Education, Science and Culture,
GZ 616.364/2-III/2a/98, \\
FNRS--FWO, Flanders Institute to encourage scientific and technological 
research in the industry (IWT), Belgium,  \\
FINEP, CNPq, CAPES, FUJB and FAPERJ, Brazil, \\
Czech Ministry of Industry and Trade, GA CR 202/99/1362,\\
Commission of the European Communities (DG XII), \\
Direction des Sciences de la Mati$\grave{\mbox{\rm e}}$re, CEA, France, \\
Bundesministerium f$\ddot{\mbox{\rm u}}$r Bildung, Wissenschaft, Forschung 
und Technologie, Germany,\\
General Secretariat for Research and Technology, Greece, \\
National Science Foundation (NWO) and Foundation for Research on Matter (FOM),
The Netherlands, \\
Norwegian Research Council,  \\
State Committee for Scientific Research, Poland, SPUB-M/CERN/PO3/DZ296/2000,
SPUB-M/CERN/PO3/DZ297/2000, 2P03B 104 19 and 2P03B 69 23(2002-2004)\\
FCT - Funda\c{c}\~ao para a Ci\^encia e Tecnologia, Portugal, \\
Vedecka grantova agentura MS SR, Slovakia, Nr. 95/5195/134, \\
Ministry of Science and Technology of the Republic of Slovenia, \\
CICYT, Spain, AEN99-0950 and AEN99-0761,  \\
The Swedish Natural Science Research Council,      \\
Particle Physics and Astronomy Research Council, UK, \\
Department of Energy, USA, DE-FG02-01ER41155. \\
EEC RTN contract HPRN-CT-00292-2002. \\

\newpage


\begin{table}[htb!]
\begin{center}
\begin{tabular}{l|ccccccc}\hline \hline \strutl
$\sqrt{s}$ (GeV)  & 192   &196   & 200   & 202 &
$<$204.9$>$ & $<$206.6$>$  & $<$206.6$>$    \\ \hline \strutl
 \lum\ (pb$^{-1}$) & 25.1  & 76.0 & 83.3  & 42.5 &
73.7 & 85.4 & 51.8\\  \hline \hline
\end{tabular}
\caption{Data collected by DELPHI in 1999 and 2000: the
integrated luminosities correspond to the data actually used in the present
  analyses after the run selection. The last column refers to the
integrated luminosity collected with one sector of the TPC off.}
\label{lumitable}
\end{center}
\end{table}

\newpage

\vspace{-0.5cm}
\begin{table}[htb!]
\begin{center}
\begin{tabular}{l l l l } 
\multicolumn{4}{c}{\LLE: multi--lepton topologies} \\ \hline \hline \strutl
analysis name  &   final states             &   direct     &   indirect   \\
               &                            & decays of    &    decays of  \\ \hline \strutl
gaugino & 2$\tau$ + n$\ell$ + m$j$ + \Emiss &  \XOI \XOI   & \XPI  \XMI \\
        &  (n$ \geq $2)  (m$ \geq $0)       &              & \XOi\XOj \\ \strutl
slepton &
  2$\tau$ + 2$\ell$ + \Emiss   + p$ \ell$ &            & \snu\asnu, \SLEPP\SLEPM \\ 
        &              (p = 0 or 2)       &              &        \\
        &
  $\tau\tau\tau\tau$ & \snue\asnue  &  \\  \strutl
sneutrino tau  & ee$\tau \tau$              & \snut\asnut  &           \\ \strutl
squark  & 2$\tau$ + 2$\ell$ + \Emiss + $2j$ &              & \SQ \SQB\\ \hline 
                      &                             & \\
\multicolumn{4}{c}{\UDD: multi--jet topologies} \\ \hline  \hline \strutl
analysis name         &   
final states          &     direct      &  indirect  \\
                &      &    decays of &    decays of \\ 
\hline  \strutl
neutralino &$6 j$                   & \XOI \XOI, \XPI \XMI     &            \\ \strutl
chargino   &$10 j$                  &                          & \XPI \XMI \\ \strutl
slepton &$6 j + 2\ell$              &                          &\SLEPP\SLEPM \\ \strutl
squark & $8 j$                   &                          & \SQ \SQB \\ 
\hline  \strutl
\end{tabular}
\caption{The multi-lepton and  multi-jet visible final states which
correspond to the analyses described  in this paper, 
when one \LLE~or \UDD~term is dominant. 
The corresponding pairs of produced sparticles
that may have given rise to them are indicated.
For the \LLE~cases, only topologies
produced with decays via \Lacc\ are considered (see text), 
and for the \UDD~cases, $\ell =$~e or $\mu$.}
\label{topos}
\end{center}
\end{table}

\relax
\newpage

\vspace{-1.5cm}
\begin{table}[htb!]
\begin{center}
\begin{tabular}{l l r r l r r l} 
\multicolumn{8}{c}{\LLE: multi--lepton topologies} \\ \hline \hline
         &   &   &    &    &   \\
\multicolumn{2}{l}{Analysis} & \multicolumn{3}{c}{192--202~GeV} &  \multicolumn{3}{c}{203--208~GeV}\\
name   &     &  observed &  \multicolumn{2}{c}{expected}  &  observed   & \multicolumn{2}{c}{expected} \\ \hline\strutl
gaugino          &   &  15 &  13.4&$\pm$~0.4  &   9 &  10.3&$\pm$~0.4  \\ \strutl
slepton          &   &   4 &   4.2&$\pm$~0.2  &   7 &   3.9&$\pm$~0.2 \\ \strutl
sneutrino $\tau$ &   &   3 &   3.5&$\pm$~0.3  &   3 &   2.8&$\pm$~0.2 \\ \strutl
squark           &   &  19 &  19.9&$\pm$~0.5  &  16 &  15.5&$\pm$~0.4 \\ 
&   &   &    &   &   \\ 
\multicolumn{8}{c}{\UDD: multi--jet topologies} \\ \hline \hline
         &   &   &    &    &   \\
\multicolumn{1}{l}{Analysis} & \multicolumn{1}{l}{Mass windows} &
\multicolumn{3}{c}{192--202~GeV} &  \multicolumn{3}{c}{203--208~GeV}\\
name          & (\GeVcc)  &  observed &  \multicolumn{2}{c}{expected}  &  observed & \multicolumn{2}{c}{expected}\\ \hline\strutl
neutralino&15~$\leq$~m$_{\tilde{\chi}}\leq$~45        &134 & 126.0&$\pm$~13.0 & 121  & 119.3&$\pm$~8.8  \\ 
          &45~$<   $~m$_{\tilde{\chi}}\leq$~75        &192 & 172.5&$\pm$~8.2  & 167  & 164.7&$\pm$~5.7  \\
          &75 $<   $~m$_{\tilde{\chi}}$               &97  & 103.3&$\pm$~3.6  & 82   &  91.7&$\pm$~2.3  \\ \strutl
chargino  &                             \dm~$\leq$~10 &187 & 181.1&$\pm$~5.9  & 156  & 171.7&$\pm$~5.2  \\
          &                             \dm~$>   $~10 &22  &25.6&$\pm$~1.1  & 20   &  23.5&$\pm$~1.0  \\ \strutl
slepton   &  m$_{\tilde{\chi}}>$~55, \dm~$\leq   $~10 & 9 & 5.6&$\pm$~0.2& 1 & 6.2&$\pm$~0.2\\
(\sel)    &  m$_{\tilde{\chi}}>$~55, \dm~$>      $~10 & 1 & 2.0&$\pm$~0.1& 5 & 2.3&$\pm$~0.1\\
          &  15~$\leq$~m$_{\tilde{\chi}}\leq$~55      & 1 & 1.6&$\pm$~0.1& 0 & 1.8&$\pm$~0.1\\ \strutl
slepton   &  m$_{\tilde{\chi}}>$~55, \dm~$\leq   $~10 & 7 & 5.7&$\pm$~0.2& 5 & 6.4&$\pm$~0.2\\
(\smu)    &  m$_{\tilde{\chi}}>$~55, \dm~$>      $~10 & 4 & 3.3&$\pm$~0.2& 1 & 3.5&$\pm$~0.2\\
          &  15~$\leq$~m$_{\tilde{\chi}}\leq$~55      & 2 & 2.0&$\pm$~0.1& 1 & 2.3&$\pm$~0.1\\ \strutl
squark    &  m$_{\tilde{\chi}}>$~55, \dm~$\leq   $~10 &42 & 39.4&$\pm$~0.6&38 &40.4&$\pm$~0.6\\
(\stp)    &  m$_{\tilde{\chi}}>$~55, \dm~$>      $~10 &13 & 10.1&$\pm$~0.3& 8 & 9.5&$\pm$~0.3\\
          &  15~$\leq$~m$_{\tilde{\chi}}\leq$~55      &30 & 26.3&$\pm$~0.5&25 &25.2&$\pm$~0.5\\ \strutl
squark    &  m$_{\tilde{\chi}}>$~55, \dm~$\leq   $~10 &10 & 11.9&$\pm$~0.4&13 &12.0&$\pm$~0.4\\
(\sbt)    &  m$_{\tilde{\chi}}>$~55, \dm~$>      $~10 & 4 &  5.4&$\pm$~0.2& 7 & 4.8&$\pm$~0.2\\
          &  15~$\leq$~m$_{\tilde{\chi}}\leq$~55      & 6 &  4.6&$\pm$~0.2& 8 & 4.2&$\pm$~0.2\\ \hline
\end{tabular}
\caption{Observed and expected numbers of events for all \LLE\ and
\UDD\ analyses. Although not explicitely written, \dm~is always
greater than 5~\GeVcc.}
\label{results}
\end{center}
\end{table}
\newpage 
    
\vspace{-1.5cm}
\begin{table}[htb!]
\begin{center}
\begin{tabular}{c| c| c| c} \hline \hline 
 &  & \multicolumn{2}{|c}{~}\\
SUSY  & Comments about validity conditions& \multicolumn{2}{|c}{Mass
  limit (\GeVcc) } \\
particle  &   & ~~~\LLE ~~~  & ~~~\UDD ~~~ \\ \hline \hline \strutl
     & Validity conditions for gauginos: &  &\\
 \XOI    & 90 $<$ m$_0 < $500~\GeVcc, 0.7 $<$ tan$\beta <$30 , & 39.5  & 38.0 \\
  \XPI   & --200~$< \mu <$~200~\GeVcc\ and 0~$< $M$_2 <$~400~\GeVcc &103.0 
     & 102.5  \\ \hline \hline \strutl 
 & Validity conditions for sfermions: &  &\\
 &  $\mu$~=~--200~\GeVcc\ and tan$\beta$~=~1.5,  &   &\\
 &  BR($\tilde{\rm f}$\Ra f' \XOI\ )=1,   $\Delta$M$ > $5 \GeVcc &   &
     \\ \hline  \strutl
%
$\widetilde{\rm e}_{\rm R}$           &       
\XOI\ mass limit not used
      &  94  & 92  \\ \cline{2-2} \strutl
      & \XOI\ mass limit used & 95 & 92 \\ \hline \strutl
$\widetilde{\mu}_{\rm R}$           &       
\XOI\ mass limit not used
      & 87   & 85  \\ \cline{2-2} \strutl
  & \XOI\ mass limit used & 90 & 87 \\ \hline \strutl
$\widetilde{\tau}_{\rm R}$           &       
\XOI\ mass limit not used
      & 86   &  {--} \\ \cline{2-2} \strutl
  & \XOI\ mass limit used & 90 &  {--} \\ \hline \hline \strutl
%
  & \XOI\ mass limit not used   &96  & {--} \\ \cline{2-2} \strutl
\snue & \XOI\ mass limit used & 98 & {--} \\  \cline{2-2} \strutl
     &   direct decay only   & 96 & $\times$ \\ \hline  \strutl
 & \XOI\ mass limit not used
      & 82 & -- \\ \cline{2-2} \strutl
\snum &\XOI\ mass limit used  & 85 & {--} \\  \cline{2-2} \strutl
      &  direct decay only   & 83 & $\times$ \\ \hline \strutl
& \XOI\ mass limit not used
      & 82  & -- \\ \cline{2-2} \strutl
\snut & \XOI\ mass limit used  & 85  & {--}  \\  \cline{2-2} \strutl
      &   direct decay only   & 91 & $\times$ \\ \hline \hline \strutl
%
     & \XOI\ mass limit not used, no mixing         & 88  & 81 \\ \cline{2-2} \strutl
\stp & \XOI\ mass limit used,   no mixing           & 92  & 87 \\  \cline{2-2} \strutl
     & \XOI\ mass limit not used,   minimal mixing  & 81  & 67 \\ \cline{2-2} \strutl
     & \XOI\ mass limit used, minimal mixing        & 87  & 77 \\  \hline \strutl
\sbt & \XOI\ mass limit not used, no mixing         & {--}& 78 \\ 
     & \XOI\ mass limit used, no mixing             & {--}& 78 \\
\hline \hline 
\end{tabular}
\caption{Sparticle mass limits at 95\% CL from the DELPHI {\rp} violation  
pair-production 
searches of supersymmetric particles;
{$\times$}: the decay channel is not possible; 
{--}: the decay channel is not covered. These results are valid if 
m$_{\widetilde{\chi}^0_{1}} \geq 15$~\GeVcc, except
for the \snu\ mass limits derived  from the \snu\ direct decay
which does not depend on m$_{\widetilde{\chi}^0_{1}}$.}
\label{rpv_limits}
\end{center}
\end{table} 


\newpage

\newpage

\newpage

\newpage

\newpage

\newpage
 
\newpage

\newpage

\newpage
\newpage

\newpage
  
\newpage

\newpage

\newpage

\begin{thebibliography}{99}

\bibitem{mssm} For reviews, see e.g.
H.P. Nilles, {\em Phys. Rep.}{\bf 110} (1984) 1;\\  
H.E. Haber and G.L. Kane, \prep{117}{1985}{75}.
%
\bibitem{fayet}
P.~Fayet, {\em Phys. Lett.} {\bf B69} (1977) 489;\\
G.~Farrar and P. Fayet, \pl{B76}{1978}{575}.
%
\bibitem{lle189}
P.~Abreu {\it et al.} [DELPHI Collaboration], {\em Eur. Phys. J.} {\bf C13} (2000) 591;\\
P.~Abreu {\it et al.} [DELPHI Collaboration], \pl{B487}{2000}{36}. 
%
\bibitem{udd189}
P.~Abreu {\it et al.} [DELPHI Collaboration], \pl{B500}{2001}{22}.
%
\bibitem{lesautres}
A.~Heister {\it et al.}  [ALEPH Collaboration], {\em Eur. Phys. J.} {\bf C31} (2003) 1;\\
P.~Achard {\it et al.}  [L3 Collaboration],
\pl{B524}{2002}{65};\\
G.~Abbiendi {\it et al.}  [OPAL Collaboration], {\em Eur. Phys. J.} {\bf C33} (2004) 149.
%
\bibitem{weinberg}
S. Weinberg, \prev{D26}{1982}{287}.
%
\bibitem{barger89} 
V.~Barger, G.F.~Giudice and T.~Han, \prev{D40}{1989}{2987}.
%
\bibitem{dreiner99}
B.C.~Allanach, A.~Dedes and H.K.~Dreiner, \prev{D60}{1999}{075014}.
%
%
\bibitem{dreiner-ross}
H.~Dreiner and G.G.~Ross, \np{B365}{1991}{597}.
%
\bibitem{dawson} 
S.~Dawson, \np{B261}{1985}{297}.
%
\bibitem{rpc}
J.~Abdallah {\it et al.} [DELPHI Collaboration], CERN-EP-2003-007.
%
\bibitem{displaced}
J.~Abdallah {\it et al.} [DELPHI Collaboration], {\em Eur. Phys. J.} {\bf C27} (2003) 153;\\
P.~Abreu {\it et al.} [DELPHI Collaboration],  \pl{B485}{2000}{95}.
%
\bibitem{singlerpv}
J.~Abdallah {\it et al.} [DELPHI Collaboration], {\em Eur. Phys. J.} {\bf C28} (2003) 15.
%

\bibitem{ellis}
J.~Ellis and S.~Rudaz, \pl{B128}{1983}{248}.
%
\bibitem{bartl}
A.~Bartl et al., \zp{C76}{1997}{549}.
%
\bibitem{drees}
M.~Drees, ``An introduction to supersymmetry''
Lectures given at Inauguration Conference of the Asia Pacific Center
for Theoretical Physics (APCTP), Seoul, Korea, 4-19 Jun 1996; hep-ph/9611409.
%
\bibitem{dreesmartin}
M.~Drees and S.P.~Martin,
``Implications of SUSY model building,"
hep-ph/9504324.
%
\bibitem{dreeshikasa} 
M.~Drees and K.~Hikasa,  \pl{B252}{1990}{127}.
%
\bibitem{delphidet} 
P. Aarnio {\it et al.}, \nim{303}{1991}{233};\\
P. Abreu {\it et al.}, \nim{378}{1996}{57}.
%
\bibitem{bdk}
F.A. Berends, P.H. Daverveldt, R. Kleiss, \cpc{40}{1986}{271, 285 and 309}.
%
%
\bibitem{bhwd}
S. Jadach, W. Placzek, B.F.L. Ward, \pl{B390}{1997}{298}.
%
\bibitem{koralz}
S. Jadach, Z. Was, \cpc{79}{1994}{503}.
%
\bibitem{pythia}
T. Sjostrand, \cpc{82}{1994}{74}.
%
\bibitem{excal} 
F.A. Berends, R. Kleiss, R. Pittau, \cpc{85}{1995}{437}. 
%
\bibitem{grc4f}
J.~Fujimoto {\it et al.},
\cpc{100}{1997}{128}.
%
\bibitem{susygen}
S. Katsanevas, P. Morawitz, \cpc{112}{1998}{227};\\
N. Ghodbane, {\it Proceedings of the Worldwide
Study on Physics and Experiments with Future Linear e$^+$e$^-$ Colliders},
Sitges, Barcelona, Spain, 28~April~--~5~May~1999.

%
\bibitem{Durham}
S.~Catani et al.,   \pl{B269}{1991}{432}.
%
\bibitem{camjet}
Yu.L.~Dokshitzer, G.D.~Leder, S.~Moretti, B.R.~Webber,
{\em J. High Energy Phys.} {8}~(1997)~1.
%
\bibitem{ckern}
S.~Bentvelsen and I.~Meyer, \epj{C4}{1998}{623}.
%



\bibitem{these_VP}
V.~Poireau, DAPNIA/SPP-01-01-T, PhD thesis (2001) in French. 
%
\bibitem{aabtag} 
P.~Abreu et al., \epj{C10}{1999}{415}. 
%
\bibitem{bayesian} 
R.M. Barnett et al., Particle Data Group, \prev{D54}{1996}{1}.
%
\bibitem{aread} A. Read {\it et al, 1st Workshop on Confidence Limits},
CERN 2000-005, 81-103.
\end{thebibliography}
\end{document}